\documentclass[aps,showpacs,prx,twocolumn,superscriptaddress, floatfix]{revtex4-2}

\usepackage[colorlinks,
urlcolor=blue,linkcolor=blue,anchorcolor=blue,citecolor=blue]{hyperref}
\makeatletter

\newcommand{\Rmnum}[1]{\expandafter\@slowromancap\romannumeral #1@}
\makeatother
\usepackage{amsmath}
\usepackage{graphicx}
\usepackage{subfigure}
\usepackage{color}
\usepackage{amsfonts}
\usepackage{tikz}
\usepackage{esint}
\usepackage{mathrsfs}
\usepackage{mathptmx}
\usepackage{lettrine}
\newcommand{\bea}{\begin{eqnarray}}
\newcommand{\eea}{\end{eqnarray}}

\begin{document}
\title{Exploring the topological sector optimization on quantum computers}

\author{Yi-Ming Ding}
\email{dingyiming@westlake.edu.cn}
\affiliation{State Key Laboratory of Surface Physics and Department of Physics, Fudan University, Shanghai 200438, China}
\affiliation{Department of Physics, School of Science and Research Center for Industries of the Future, Westlake University, Hangzhou 310030,  China}
\affiliation{Institute of Natural Sciences, Westlake Institute for Advanced Study, Hangzhou 310024, China}

\author{Yan-Cheng Wang}
\affiliation{Hangzhou International Innovation Institute, Beihang University, Hangzhou 311115, China}
\affiliation{Tianmushan Laboratory, Hangzhou 311115, China}

\author{Shi-Xin Zhang}
\email{shixinzhang@iphy.ac.cn}
\affiliation{Institute of Physics, Chinese Academy of Sciences, Beijing 100190, China}

\author{Zheng Yan}
\email{zhengyan@westlake.edu.cn}
\affiliation{Department of Physics, School of Science and Research Center for Industries of the Future, Westlake University, Hangzhou 310030,  China}
\affiliation{Institute of Natural Sciences, Westlake Institute for Advanced Study, Hangzhou 310024, China}

\begin{abstract}
Optimization problems are the core challenge in many fields of science and engineering, yet general and effective methods for finding optimal solutions remain scarce. Quantum computing has been envisioned to help solve such problems, with methods like quantum annealing (QA), grounded in adiabatic evolution, being extensively explored and successfully implemented on quantum simulators such as D-wave's annealers and some Rydberg arrays.
In this work, we investigate the topological sector optimization (TSO) problem, which attracts particular interests in the quantum simulation and many-body physics community.
We reveal that the topology induced by frustration in the optimization model is an intrinsic obstruction for QA and other traditional methods to approach the ground state. We demonstrate that the difficulties of TSO problem are not restricted to the gaplessness, but are also due to the topological nature which are often ignored for the analysis of optimization problems before.  
To solve TSO problems, we utilize quantum imaginary time evolution (QITE) with a possible realization on quantum computers, which leverages the property of quantum superposition to explore the full Hilbert space and can thus address optimization problems of topological nature. We report the performance of different quantum optimization algorithms on TSO problems and demonstrate that their capability to address optimization problems are distinct even when considering the quantum computational resources required for practical QITE implementations.
\end{abstract}

\maketitle
\section{Introduction}

Optimization problems have a wide range of applications across various fields in science and engineering, which are generally NP-hard since the set of candidate solutions is discrete and expands exponentially with the problem size. As a representative NP-hard problem, the quadratic unconstrained binary optimization problem targets to identify a binary sequence solution for a quadratic cost function~\cite{Hauke_2020, PhysRevResearch.1.033142}. By mapping such a problem onto an Ising-like Hamiltonian, i.e., a spin glass model, the optimal solution of the problem becomes the configuration of the quantum system corresponding to the ground state~\cite{Anderson1986,Virasoro1987,Huse1986,Car2002,Lucas2014}. Based on the quantum adiabatic theorem~\cite{Born1928, berry1984quantal, griffiths_schroeter_2018}, the quantum annealing (QA) method, or adiabatic quantum computation, has turned out to be a promising approach to these problems~\cite{yan2022preparing, Anderson1986,Virasoro1987,Huse1986,Car2002,Lucas2014,kadowaki1998quantum,Car2002,Montanari2009,Heim2015, quant-ph/0001106, RevModPhys.90.015002}. 

Following technological advancements in manufacturing coupled qubit systems recently, QA can be implemented with superconducting flux qubits~\cite{Gildert2011,Boixo2013,Boixo2014}. In the past decades, QA has been considered as a more powerful toolbox to solve the optimization problems than simulated annealing (SA) and other classical methods~\cite{Huse1986,Car2002,perdomo2012finding,novotny2016spanning,qiu2020precise,qiu2020programmable,king2023quantum}. Furthermore, the Rydberg array simulator can also realize the Ising-like encoded Hamiltonian with high tunability and scalability~\cite{Roushan21,Semeghini21}. 
Despite of tremendous potentialities that QA possesses, it is yet an open question on which optimization problems can QA outperform other classical or quantum algorithms. Although the quantum fluctuations during the annealing process gives it more chance to jump out of an local minimum, the tangible limit of QA is hard to quantify. Especially when the ground states at two adjacent time moments are not connected smoothly, i.e., there is a phase transition in between, the validity of QA cannot always be guaranteed.

Recently, topological phases attract numerous attentions and are being simulated via various quantum platforms~\cite{doi:10.1126/science.abi8378, PhysRevD.105.054508, yan2022preparing,semeghini2021probing,yan2022triangular,ZY2023Emergent}. Since the original intention of developing quantum machines is to investigate quantum many-body systems~\cite{Feynman1982}, a natural but long-neglected problem is rising: whether QA can find the ground state of a spin glass model that intrinsically encompasses different topological sectors? For convenience, we call this kind of problems the topological sector optimization (TSO) problems. 

\begin{figure}[ht!] \centering
    \subfigure[] { \label{fig:spin_glass_problem}
    \includegraphics[width=4.2cm]{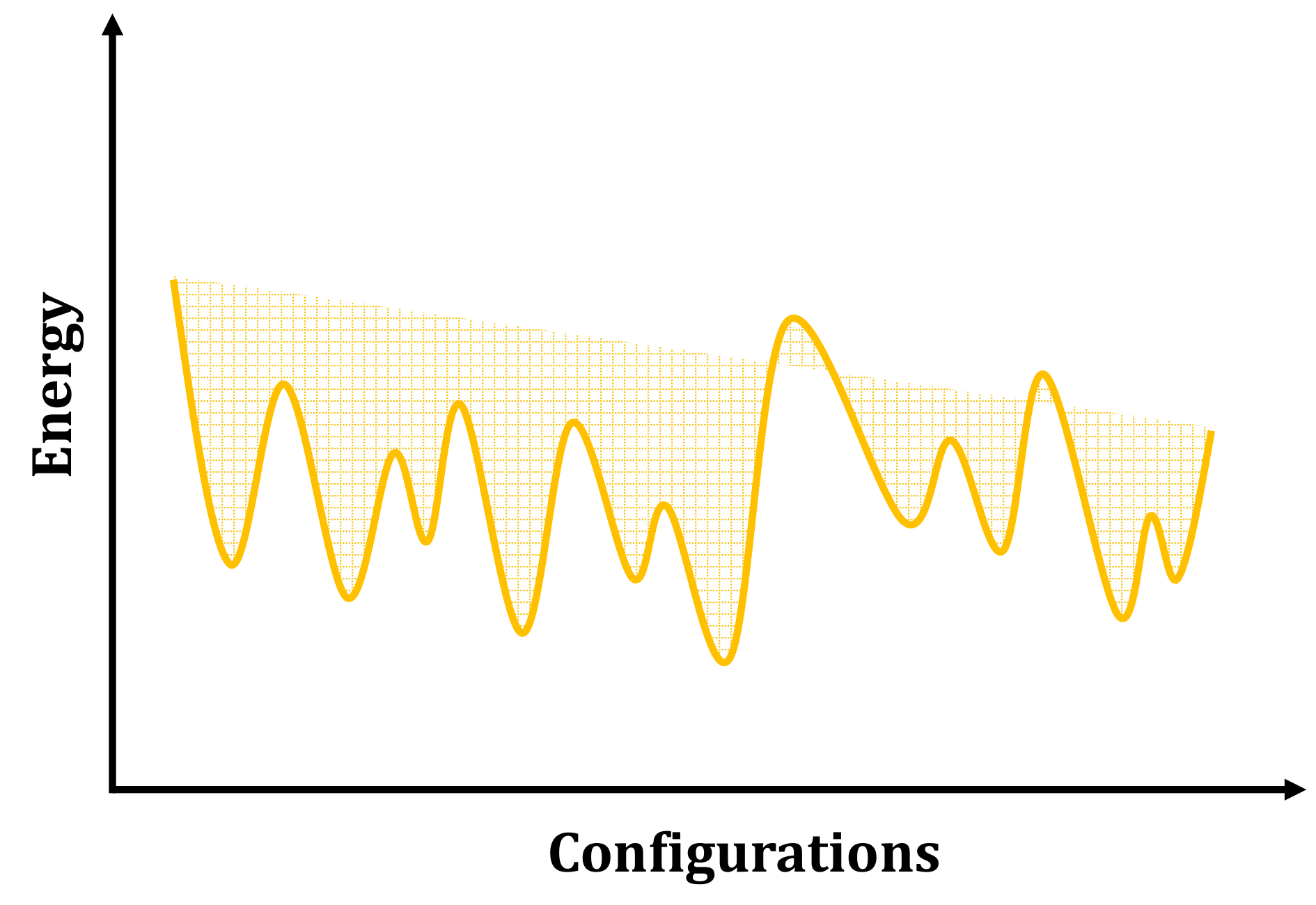}}
    \subfigure[] { \label{fig:tso_problem}
    \includegraphics[width=4.2cm]{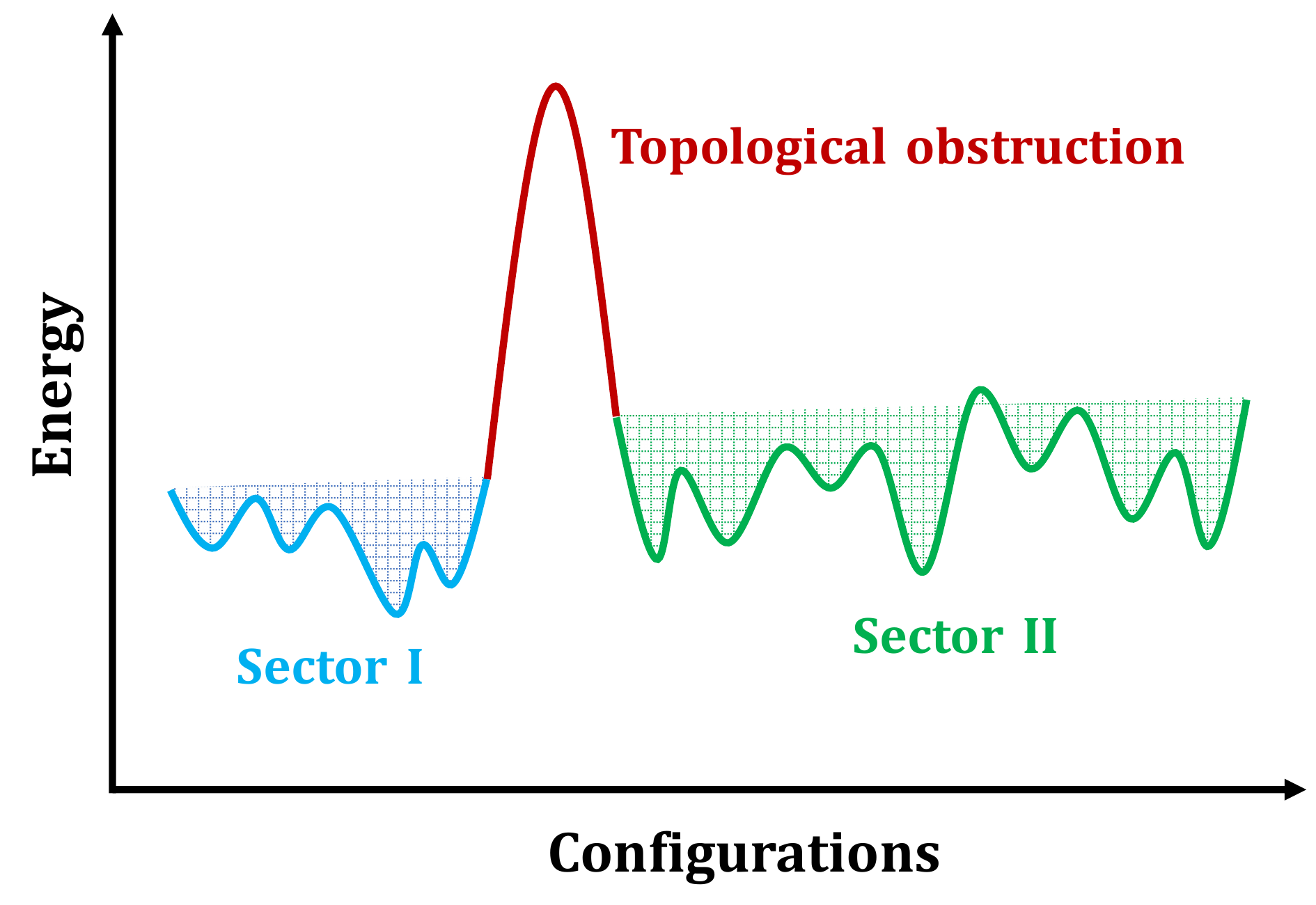}}
    \caption{ 
       (a) Spin glass model with no topological sectors; (b) TSO problem.
        }
        \label{fig:opt_configuration}
\end{figure}

Different from traditional optimization problems which focus on searching the ground state among many local minima in a glass system (see Fig.~\ref{fig:spin_glass_problem}), the TSO problems further introduce several topological sectors in the energy landscape (see Fig.~\ref{fig:tso_problem}). Beside the challenges of local minima, the topological obstructions bring new difficulties for traditional quantum optimization algorithms because each topological sector has been protected and is robust to local quantum fluctuations. Generally speaking, a complex system with frustrations can generate topological properties~\cite{zhou2020quantumstring,zhou2020quantum,zhang01,zhang02,zhang03,YCWang2017,YCWang2018,YCWang2021vestigial} and its optimization problem naturally has the corresponding difficulties. Therefore, the TSO problems are important and urgent to be addressed universally.

In Ref.~\cite{yan2022preparing}, the authors found both SA and QA lose the effectiveness for the TSO problems. QA generally fails since it explores the optimized solution by traversing the valley path of the parameter space, which is of local nature and prone to getting trapped in some local minimum of the wrong topological sector. Furthermore, Ref.~\cite{yan2022preparing} proposed the sweeping quantum annealing (SQA) method to improve the traditional QA, which introduces virtual edges as the annealing terms to overcome topological defects for better results. However, SQA is still inefficient as the searching is taken along the valley path of the energy landscape of the deformed Hamiltonian. Therefore, it is hard to generalize SQA to more complicated topological systems, as the result shown in the supplementary materials of the Ref.~\cite{yan2022preparing}, the SQA method can not reach the ground state strictly even after long time annealing for some TSO problems.

Recently, quantum imaginary time evolution (QITE) has been successfully realized experimentally on quantum computers~\cite{motta2020determining,love2020cooling, Nishi2021, Cao2022, Zhang2023a}, while its advantage for addressing TSO problems has been underestimated. In this work, we apply the quantum imaginary time evolution method to TSO problems, which utilizes the superposition property in quantum mechanics and is able to process all the states of the system simultaneously. As long as the initial state has a non-zero overlap with the ground state (note: all the solutions of a common optimization problem encoded in glass Hamiltonian is a classical state), QITE can decay all the excited states and only the ground state will be preserved after a sufficiently long time. 
Though the performance of QITE also depends on the energy gap between the ground state energy and first excited state energy, it has lower complexity than QA which we will show below. More importantly, QITE evolves the system in a global way such that topological sectors can be mixed and efficiently explored.

In this paper, we elaborate the limitations of QA as well as SQA, and illustrate the advantage of QITE over QA and SQA with two TSO problems. This paper is organized as follows. In Sec.~\ref{sec:algorithm}, we give a brief review of general QA, SQA and QITE methods with one practical implementation of QITE as well as its diagonal approximation on quantum computers, and discuss how to utilize these algorithms to solve TSO problems. Then, in Sec.~\ref{sec:design}, we discuss how a difficult TSO problem should be like. In Sec.~\ref{sec:model_tri} and Sec.~\ref{sec:model_sq}, we introduce two examples of TSO problem and show our numerical simulation results of the forementioned algorithms to solve these two examples. For comparing, we also consider the variational quantum eigensolver (VQE)~\cite{Peruzzo2014, McClean_2016} in this section. Further discussions are in 
Sec.~\ref{sec:discussion}. 

\section{Quantum annealing versus imaginary time evolution}\label{sec:algorithm}
\subsection{Quantum annealing and adiabaticity}
Usually, the archetypal Hamiltonian of QA is 
\begin{equation}\label{eq:adiabatic}
    \hat H(t) = \frac{t}{T}\hat H_0 + (1-\frac{t}{T})\hat H_1,
\end{equation}
where $T$ is the total time of the QA process and $t$ monotonically increases from $0$ to $T$. $\hat H_0$ and $\hat H_1$ are two time-independent Hamiltonian. Obtaining the ground state of $\hat H_0$ is exactly our target and $\hat H_1$ is some auxiliary Hamiltonian which has an accessible ground state. Besides, $[\hat H_0,\hat H_1]\ne 0$ must be satisfied. 

In QA, $\hat H_0$ is typically some spin glass model
\begin{equation}
    \hat H_0 = \sum_{ij}J_{ij}\hat Z_i \hat Z_j
\end{equation}
and $\hat H_1$ is usually chosen to be 
\begin{equation}\label{eq:H_1}
    \hat H_1 = \sum_i \hat X_i.
\end{equation}
Apparently, $\hat H_1$ only has a non-degenerate ground state
\begin{equation}\label{eq:pd_state_minus}
    |\Psi^{-}\rangle=\otimes_j|-\rangle_j = \otimes_j \bigg(\frac{|0\rangle_j-|1\rangle_j}{\sqrt{2}}\bigg),
\end{equation}
which is also a product state and easy to prepare practically.
 
The adiabatic theorem~\cite{Born1928, berry1984quantal, griffiths_schroeter_2018} states that, if the system is initially at the non-degenerate ground state $|\chi_0(0)\rangle$, it will remain at the instantaneous ground state $|\chi_0(t)\rangle$ for any $t>0$ under the following two conditions: (i) the change of the Hamiltonian is sufficiently slow; (ii) there is a non-zero gap $g(t)$ between the ground state energy $E_0(t)$ and the first excited state energy $E_1(t)$. 
The adiabatic approximation requires
\begin{equation}\label{eq:adia_approx}
     \frac{\max_{t\in[0,T]} \langle \chi_n(t) | \dot{\hat{H}} (t) | \chi_0(t)\rangle \hbar}{[\min_{t\in[0,T]} g(t)]^2}\ll 1
\end{equation}
should hold for any $n\ne 0$, where $\{|\chi_n(t)\rangle\}$ are the instantaneous orthogonal eigenvectors of $\hat H(t)$ and $\{E_n(t)\}$ are the corresponding eigenvalues. Let $g_{\text{min}}\equiv \min_{t\in[0,T]} g(t)$ and suppose the numerator achieves its maximum at $t=t^*$. Then the variation in time at $t^*$ should satisfy $\Delta t\gg \hbar \langle \chi_n(t^*)|\Delta \hat H|\chi_0(t^*)\rangle/g_{\text{min}}^2\propto 1/g_{\text{min}}^2$. Therefore, to ensure the adiabacticity, the total evolution time must satisfy $T\gg \tau$, where $\tau \propto 1/g^2_{\min}$.

Based on the adiabatic approximation (\ref{eq:adia_approx}), the main obstacle of QA is accordingly the gaplessness. However, this is not the only factor that influences the efficacy of QA. As we will show later, if the Hilbert space of $\hat H_0$ comprises many topological sectors, the performance of QA will dramatically decrease. Relevant numerical evidence has also been presented in Ref.~\cite{PhysRevD.105.054508}, where the authors numerically demonstrated the performance of QA, equivalently applied via digital quantum simulation, on the toric code model under external magnetic fields~\cite{KITAEV20032}. They found that the true topological sector of the toric code model cannot be achieved from a non-topological limit, which corresponds to the initial Hamiltonian $\hat H_1$. 

\subsection{Real-time simulations of quantum annealing}\label{subsec:qa_tech}
In this subsection, we introduce the method we used to perform our real-time simulations of QA on classical computers. For the QA Hamiltonian $\hat H(s)$, quantum mechanics requires that the immediate quantum state $|\Psi(t)\rangle$ and the state evolved after a small time interval $\Delta t$ satisfy the relation 
\begin{equation}\label{eq:class_qa_sim}
    |\Psi(t+\Delta t)\rangle = e^{-i\hat H(s)\Delta t}
    |\Psi(t)\rangle.
\end{equation}
After this, we change $s$ by $\Delta s$ and perform the next evolution. Without loss of generality, we consider the case that $s$ linearly decreases with $t$. The total number of evolution steps is therefore $N=1/\Delta s$ and the total evolution time is $T=N\Delta t$. 
To ensure the adiabatic approximation, $T$ must be large enough or equivalently, $\Delta s/\Delta t$ is small enough when we use Eq.~(\ref{eq:class_qa_sim}). For the two parameters $\Delta t$ and $\Delta s$, practically, to control the pace of QA, we can fix one of the them and change the other. In our simulations, we set $\Delta t = 0.1$, and utilize different values of $\Delta s$ to adjust the total evolution time $T$. The initial state, as discussed before, is taken to be state~(\ref{eq:pd_state_minus}). 

For the first example and benchmark, we test it on a 16-qubit Ising chain under periodic boundary condition, whose Hamiltonian is written as 
\begin{equation}
    \hat H_{\text{Ising-1d}} = \sum_{\langle j,j+1\rangle}\hat Z_j\hat Z_{j+1}.
\end{equation}

\begin{figure}[ht!] \centering
    \includegraphics[width=6cm]{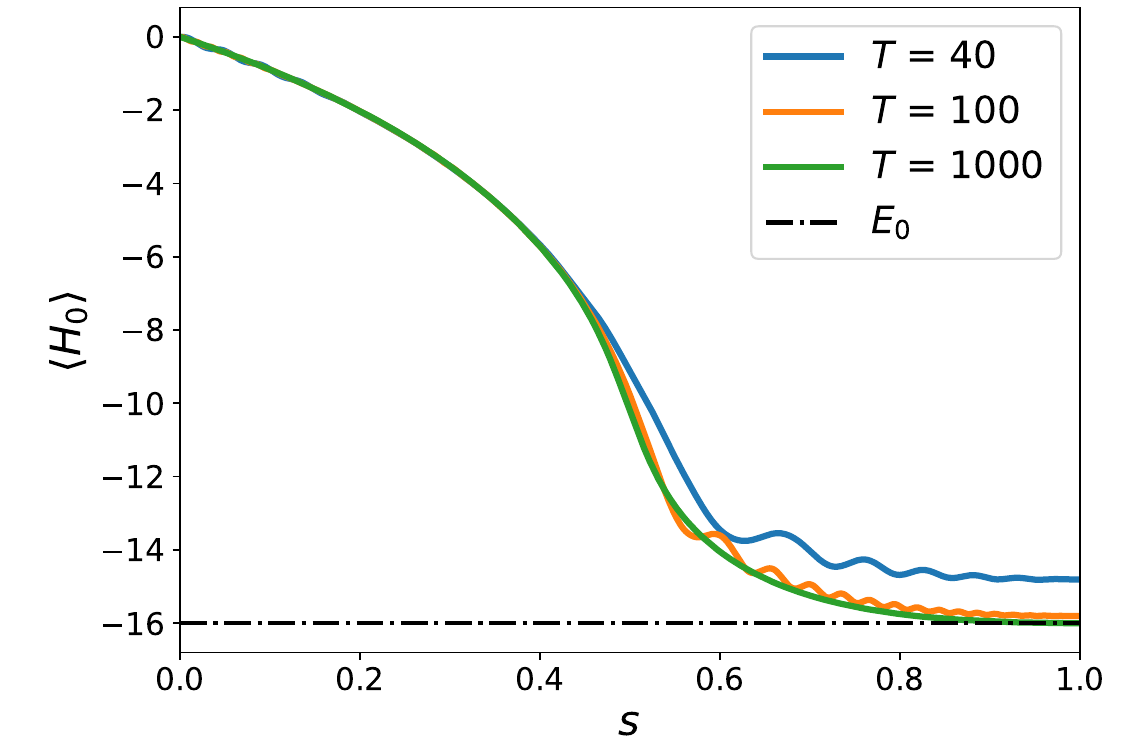}
    \caption{ 
        Real-time simulation of QA for a 16-qubit Ising chain under periodic boundary condition with different evolution time. Here we take $\hat H_0\equiv \hat H_{\text{Ising-1d}}$. The black dashed line denotes the exact ground state energy of the model. 
        }
    \label{fig:QA_isingChain_benchmark}
\end{figure}

As it shown in Fig.~(\ref{fig:QA_isingChain_benchmark}), when $T=40$, the process is non-adiabatic with great fluctuations when $s\to 1$. It turns out that $\Delta s= 10^{-4}$, i.e. $T=10^3$, is sufficient for this model to reach its ground state $E_0$ via QA.

\subsection{Sweeping quantum annealing}
As we will introduce in Sec.~\ref{sec:model_tri} and \ref{sec:model_sq}, different topological sectors are separated and marked by some global topological defects, which in spin systems are usually some domain walls. If we possess prior knowledge of the direction of the defects, we can polarize the spins on a line (called virtual edge) parallel to the defect on the lattice using strong transverse magnetic fields. These fields can disrupt the defect when it overlaps with the virtual edge, enabling the state to escape the confines of that particular topological sector. Since practically we cannot assure the defect can overlap the virtual edge we choose, we adopt the strategy that sweeping all parallel lines and making them to be the virtual edge sequentially in the process of a vanilla QA. This method is called the sweeping quantum annealing (SQA)~\cite{yan2022preparing}, effectively introducing some topological annihilation operators with a high likelihood of eliminating the defect. 

However, since the prior knowledge cannot always be assumed, and furthermore, if the true ground state lies within a topological sector that has many topological defects, the creation of virtual edges would be counterproductive. We also explore this method in our work, employing the same initial state as in QA. More numerical details about the real-time simulation of SQA can be found in Appx.~\ref{appx:sqa}.


\subsection{Quantum imaginary time evolution}\label{sec:ite}
The evolution of the instantaneous ground state in QA can be considered as searching the final ground state along a specific path in the Hilbert space following the adiabatic evolution. On the contrary, QITE offers a more direct approach, capable of discarding high energy states absolutely over time, thereby enabling the exploration of the target state in a global manner through the superposition property inherent in quantum systems.

The QITE acting on a quantum state $|\psi(0)\rangle$ under a given Hamiltonian $\hat H$ after a time interval $t$ is defined by 
\begin{equation}\label{eq:ITE}
    |\psi(t)\rangle \propto e^{- \hat H t} |\psi(0)\rangle,
\end{equation}
where the $t$ represents a imaginary time factor, because there is an additional $i$ in an unitary evolution.

Suppose $\{|\chi_n\rangle\}$ are a set of eigenstates of $\hat H$ with eigenvalues $\{E_n\}$, then $|\psi\rangle$ can be rewritten as
\begin{equation}
    |\psi(0)\rangle = \sum_n a_n |\chi_n\rangle.
\end{equation}
As long as $|\psi(0)\rangle$ has a non-zero overlap with the ground state $|\chi_0\rangle$, by taking $t$ to enough long (infinity), one can always achieve $|\chi_0\rangle$ since 
\begin{equation}\label{eq:to_gs}
    \begin{split}
        &\lim_{t\to\infty} e^{-\hat H t} |\psi(0)\rangle \\ 
        = & \lim_{t\to\infty} e^{-E_nt} a_n |\chi_n\rangle  \\
        = & \lim_{t\to\infty} e^{-E_0 t} 
        \bigg[
            a_0|\chi_0\rangle + \sum_{m\ne 0}  a_m e^{-(E_m-E_0)t} |\chi_m\rangle    
        \bigg] \\ 
        \propto & |\chi_0\rangle.
    \end{split}
\end{equation}
Notice that we have assumed the ground state to be non-degenerate. However, one can similarly discuss the degenerate case and the final state would be the superposition of all degenerate ground states that have non-zero overlaps with $|\psi(0)\rangle$. 

Eq. (\ref{eq:to_gs}) indicates that the minimum evolutionary time $t$ to obtain a sufficiently pure ground state should scale with the reciprocal of the first excited state gap $g$, or $t\gg 1/g $, which offers a quadratic speed up compared to the real-time QA at least. It is worth noting that the $g$ here is the first excited gap in the $H_{t=T}$ while the $g_{\text{min}}$ in QA is the smallest gap during the whole evolution, thus the $g_{\text{min}}\leq g$. 
In most cases, we only need one (classical) solution of an optimization problem, and the gaplessness, implying the degeneracy of solution space, will not harm the application of QITE. 

Moreover, Eq.~(\ref{eq:to_gs}) provides a mathematical representation of the infinite QITE operator ($t \to \infty$), illustrating its role as a projection operation onto the ground state(s) of $\hat H$ by diminishing all other excited states. Therefore, this method is free from problems such as being trapped in a local minimum or lacking ergodicity, since all eigenstates, including our desired solutions, are equally addressed simultaneously. Moreover, this projection imposes no constraints on the specific characteristics of the Hamiltonian, be it short-range or long-range interactions, involving two-body or many-body interactions, or the lattice manifold \cite{LEHTOVAARA2007148, McArdle2019}. Thus, QITE is a more powerful tool for tackling TSO problem theoretically. Since we cannot assume prior knowledge of the ground state, it becomes imperative to consider a superposition of all possible classical states within the Hilbert space as the initial state. In our work, we take
\begin{equation}\label{eq:pd_state_plus}
    |\Psi^{+}\rangle=\otimes_j|+\rangle_j = \otimes_j \bigg(\frac{|0\rangle_j+|1\rangle_j}{\sqrt{2}}\bigg),
\end{equation}
for convenience.


\subsection{Implementation of quantum imaginary time evolution}\label{sec:intro_vqite}
QITE is represented by non-unitary operator which requires exponential resource to implement exactly on quantum computers. 
Non-hermitian physics in open quantum system may be a relevant platform for these non-unitary operations~\cite{doi:10.1080/00018732.2014.933502, doi:10.1080/00018732.2021.1876991}. In terms of gate-based quantum computers, we can expand the QITE operator as a linear combination of some unitary operators exactly or approximately, which enables us to implement QITE on digital quantum computers.

As a demonstration to show the advantage of QITE over QA on TSO problems, we consider an approximate but practical implementation of QITE on quantum computers following variational philosophy \cite{Bharti2021z, Cerezo2021}, dubbed as variational quantum imaginary time evolution (VQITE)~\cite{Li2017bz, Yuan2019theoryofvariational, McArdle2019, Barison2021z, Benedetti2021az, Lee2022z}. This methodology only requires shallower circuits of some appropriate ansatz given by the Hamiltonian form and symmetry \cite{Wiersema2020z} or by quantum architecture search \cite{Zhang2020bz, Du2020az, Zhang2021z}, thus it is suitable for noisy intermediate-scale quantum (NISQ) \cite{Preskill2018z} devices. 

To perform VQITE, we first prepare a trial state 
\begin{equation}
    |\phi(\vec\theta(t))\equiv |\phi(t)\rangle = V(\vec\theta) |\bar 0\rangle, \end{equation}
where $V(\vec\theta)$ is some circuit ansatz, a parameterized quantum circuit (PQC) on a quantum computer; $\vec\theta=[\theta_0,\theta_1,\dots \theta_{p-1}]$ denotes $p$ variational parameters and $|\bar 0\rangle$ is the inital state input into the PQC. Thereafter, we require it to approximately satisfy the Wick-rotated Schrodinger equation under McLachlan's variational principle~\cite{doi:10.1080/00268976400100041}:
\begin{equation}
    \delta\bigg|\bigg| \bigg(
\frac{\partial}{\partial t}+\hat H-E_{t}\bigg) |\phi(\vec\theta(t))\rangle 
\bigg|\bigg| = 0,
\end{equation}
which leads to the following linear differential equations
\begin{equation}\label{eq:linearEq}
    \mathbf{A}\dot{\vec{ \theta}} = \mathbf{C}
\end{equation}
or
\begin{equation}\label{eq:linearEq}
    \sum_k A_{jk}\dot\theta_k = C_j,
\end{equation}
where 
\begin{equation}\label{eq:measure}
    \begin{split}
        &A_{jk} = \text{Re}\bigg\{ 
\frac{\partial \langle\phi(t)|}{\partial\theta_j}\frac{\partial|\phi(t)\rangle}{\partial\theta_k}
\bigg\} \\
&C_j = -\text{Re}\bigg\{ 
\frac{\partial\langle\phi(t)}{\partial\theta_j}\hat H|\phi(t)\rangle
\bigg\}
    \end{split}
\end{equation}
The $A_{jk}$ and $C_j$ in Eq.~(\ref{eq:measure}) can be evaluated on a quantum computer via the Hadamard test, then Eq. (\ref{eq:linearEq}) will be futher solved with the help of a classical computer and the dynamics of $|\phi(t)\rangle$ 
is controlled by $\vec \theta$ as 
\begin{equation}\label{eq:updateTheta}
    \vec\theta(t+\Delta t) \approx\vec\theta(t) + \dot{\vec\theta}(t)\Delta t = \vec\theta(t) +A^{-1}(t)\cdot \vec C(t)\Delta t
\end{equation}
for a small enough $\Delta t$. 

For experimental contemplation, we also consider the diagonal approximation of variational quantum imaginary time evolution (Diag-VQITE)~\cite{Stokes2020quantumnatural}, where we only keep the diagonal elements in $\mathbf{A}$ in Eq. (\ref{eq:linearEq}) to reduce the computational resources. Since the form of VQITE is similar to variational quantum algorithms, we additionally compare it with the gradient-based variational quantum eigensolver (VQE)~\cite{Peruzzo2014, McClean_2016}, which is a frequently-used quantum algorithm for ground states searching.

\section{Designing typical TSO problems}\label{sec:design}
 \begin{figure}[ht!] \centering
    \includegraphics[width=6cm]{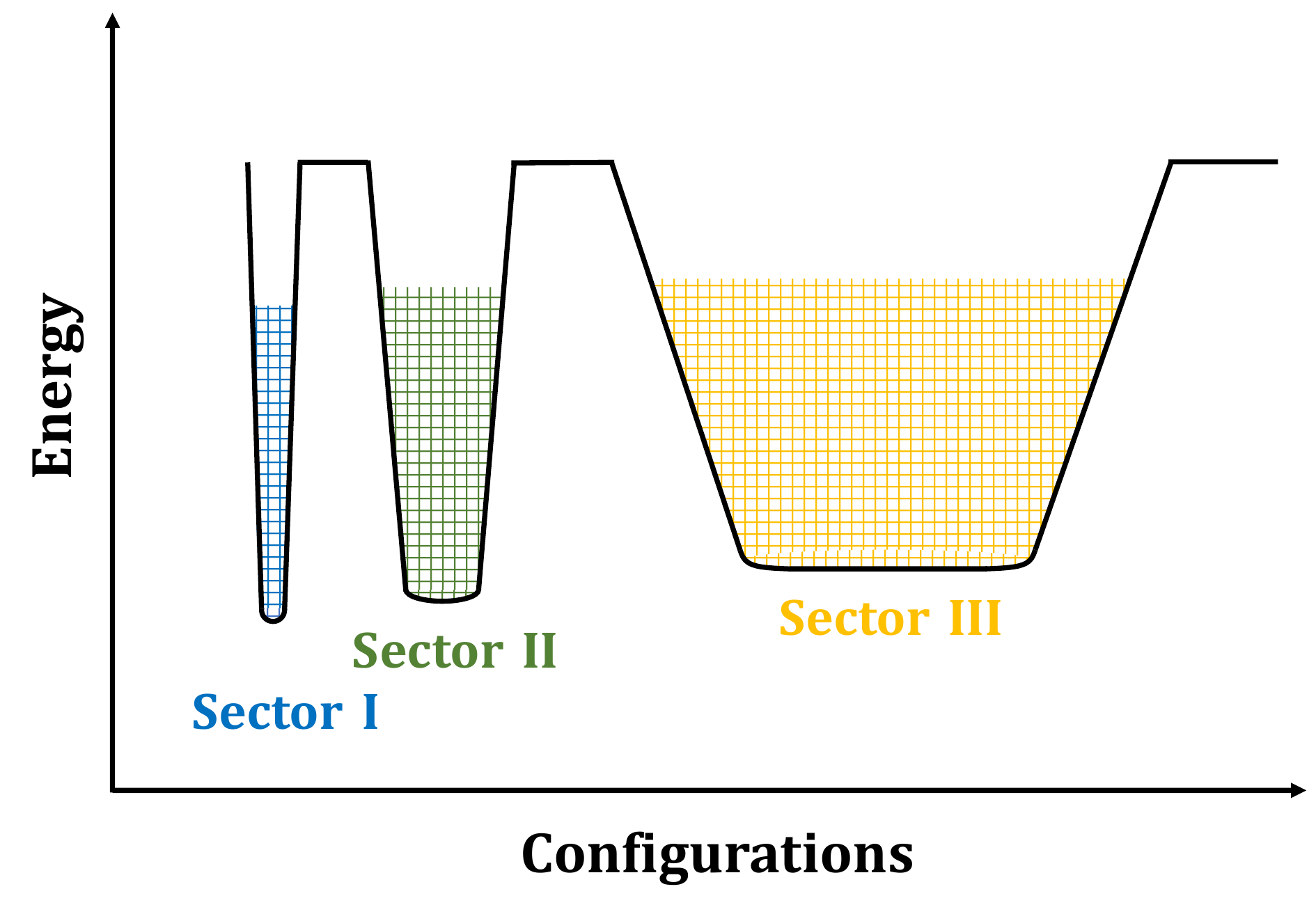}
    \caption{ 
        A schematic diagram of a hardcore TSO problem which satisfy the three difficulty characteristics.
        }
        \label{fig:small_sector}
  \end{figure}

To test the efficiency of the aforementioned quantum optimization algorithms, several representative TSO problems are needed. The TSO model is expected to be simple enough in the Hamiltonian form, but sufficiently difficult to be optimized. Models that are amenable to numerical simulation on classical computers, expected to involve around 16 qubits or more to make topological effects manifest, are rare. We have identified two suitable examples for our study, which will be introduced in Sec.~\ref{sec:model_tri} and Sec.~\ref{sec:model_sq}. Many other models would necessitate greater computational resources, posing challenges for simulating variational algorithms that require substantial memory to store the variational parameters in addition to the qubits. 

Moreover, these two models represent a challenging scenario even within the class of TSO problems. If QITE proves effective in such a scenario where QA struggles, it would underscore the advantage of QITE. In Fig. \ref{fig:small_sector}, we show the three characteristics of this kind of TSO problems: (i) The minimum energies of different topological sectors are nearly equal; (ii) The target topological sector containing the ground state occupies an exponentially small fraction of the Hilbert space thereupon is difficult to be found; (iii) Competing topological sectors have much larger Hilbert space to the benefit of quantum fluctuations. 
In other words, utilizing search methods based on quantum kinetic terms could potentially give rise to a situation where the competing topological sectors become significantly more preferred compared to the desired target sector.

There are three reasons for the above criteria. 1) The case we designed for tests should be a difficult task in the topological optimization problem. Therefore, it is natural to set the degree of freedom of the target sector as small as possible. It can avoid the case that the target can be found easily via random searching. 2) The model needs being simple in understanding, so we do not consider the spin glass problem but focused on the topological optimization itself only. This makes the problem more focused, eliminating the effects of other variables, and the well-known ground state of the model is easy for benchmark. 3) This hardcore mode has been discussed carefully in Ref.\cite{yan2022preparing} by different QA methods, thus it is convenient for comparing the advantages of the QITE with QAs. 
With these characteristics in mind, we design the following two models and test the algorithms by numerical simulations.

\section{Frustrated AFM Ising model on triangular lattice}\label{sec:model_tri} 
\begin{figure}[ht!] \centering
    \subfigure[] { \label{fig:tri_lattice_dimer}
    \includegraphics[width=4.2cm]{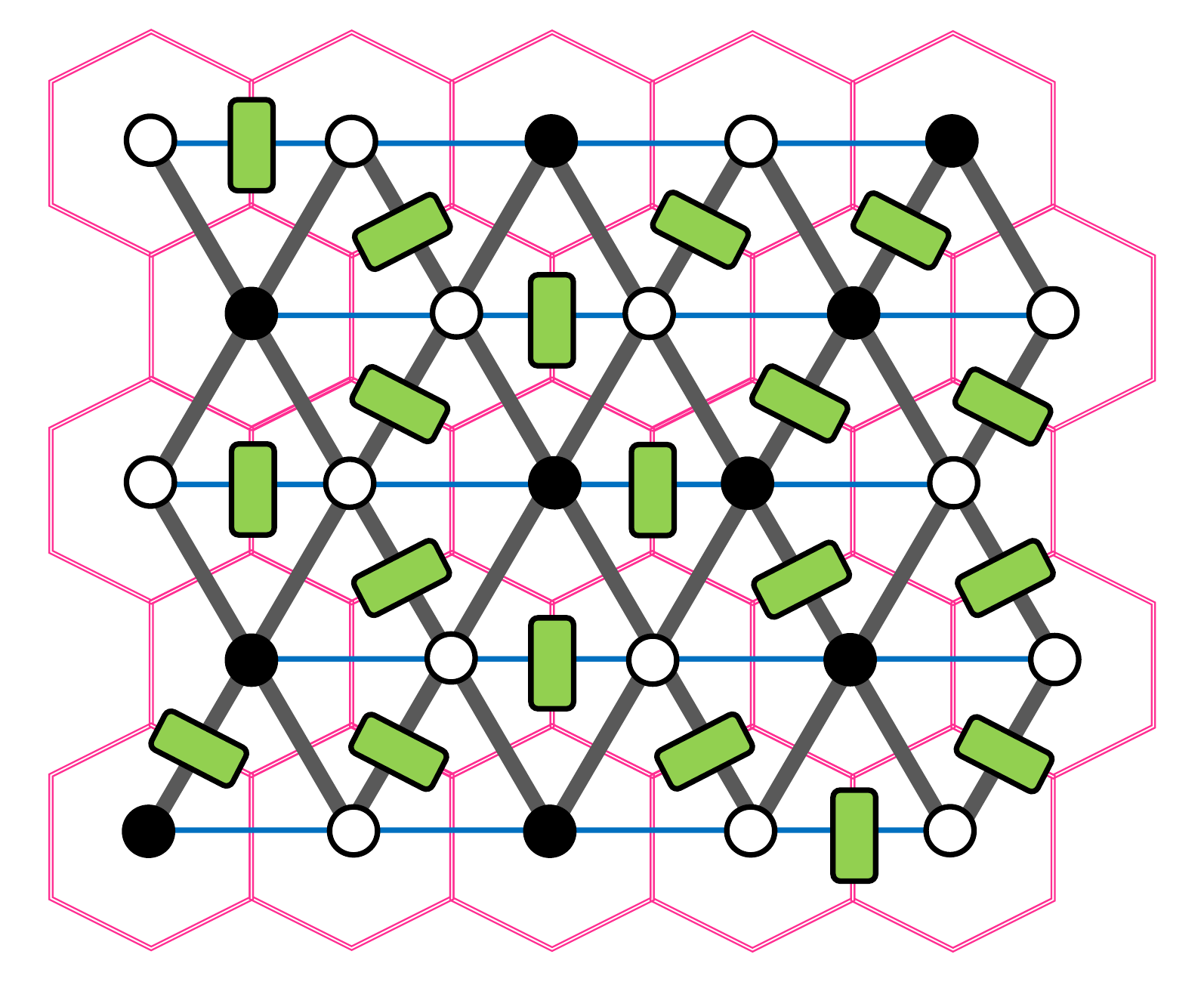}}
    \subfigure[] { \label{fig:tri_lattice_ss}
    \includegraphics[width=4.2cm]{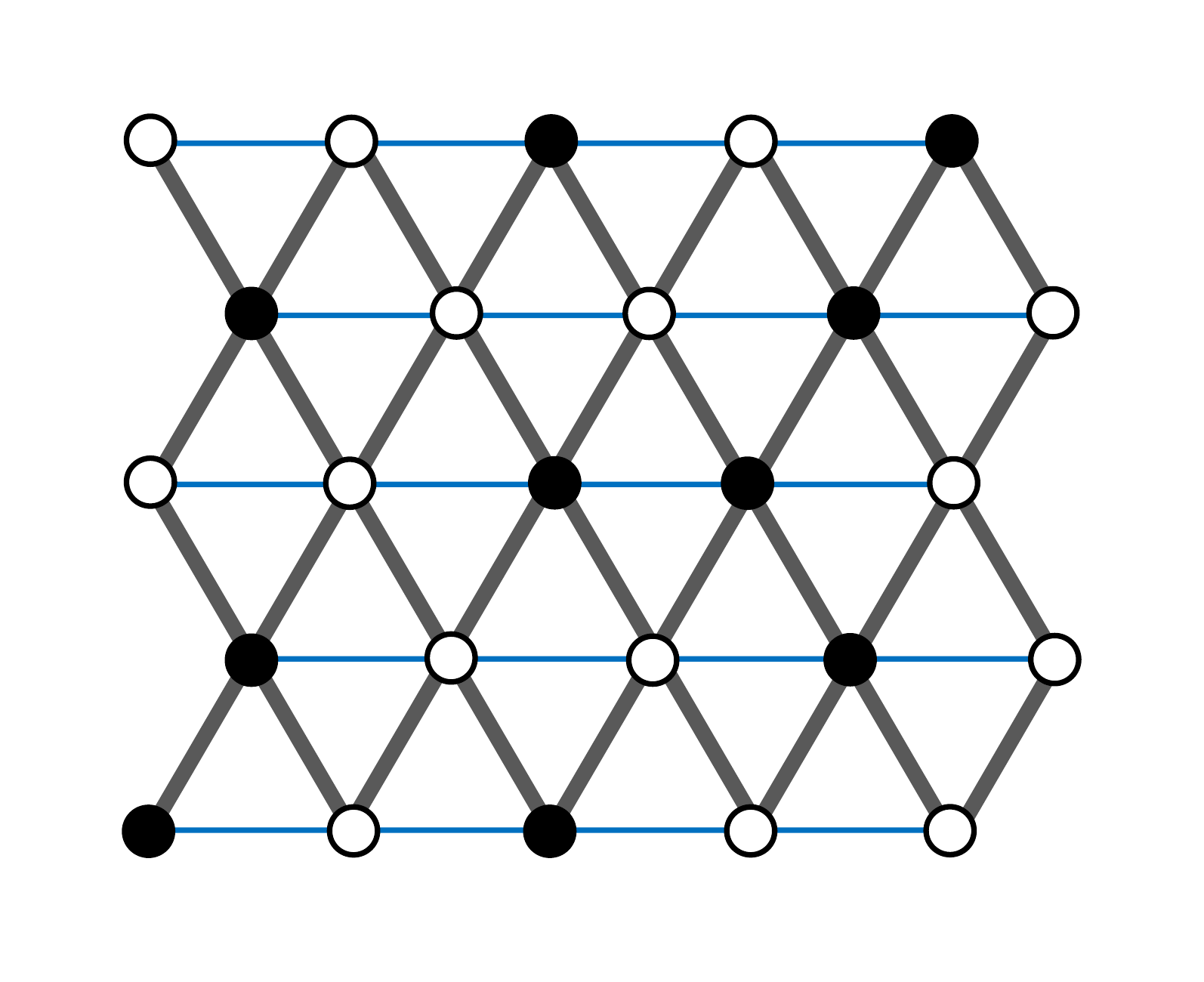}}
    \caption{ 
       Triangular lattice of the AFM Ising model, where the white open circle denotes spin up or $|0\rangle$ and the black solid circle denotes spin down or $|1\rangle$. The blue thick bonds denote the coupling $J_x$ and the gray thick bonds denote the coupling $J_{\wedge}$. (a) shows the mapping between this model and a dimer model on the dual lattice, which is depicted in pink and double-solid link and (b) is an anisotropy triangular lattice AFM Ising model which we used to test different quantum algorithms.
        }
        \label{fig:lattice_tri_diag}
\end{figure}
\subsection{Model}

The first TSO model we used is the nearest neighbor antiferromagnetic (AFM) Ising model on the triangular lattice. The emergent $U(1)$ gauge fields and topological properties in this Ising-like model have been well studied~\cite{Moessner2001,Isakov,YCWang2017,YDLiao2021,zhou2020quantum}. On the other hand, the model Hamiltonian is simple enough formally.

Due to the antiferromagnetic interaction, each triangle in the low-energy Hilbert space must include two spins aligned in parallel and one spin aligned in antiparallel. We refer to this local constraint as the `triangle rule'. The low-energy Hilbert space satisfying this constraint can be precisely mapped onto the well-known dimer models \cite{Moessner2010b, yan2019widely, ZY2019, zhou2020quantumstring,yan2020triangular, yan2020improved}. Fig. \ref{fig:tri_lattice_dimer} shows this mapping between the constrained spin configuration on triangular lattice and the dimer configuration on the dual honeycomb lattice, where the bond with two parallel spins corresponds to a dimer. The dimer density on the honeycomb lattice can be understood as lattice electric field on the dual bond, and the local constraint can be written as the divergenceless condition. There thus emerges an $U(1)$ gauge field in this triangular AFM Ising model~\cite{Moessner2001,zhou2020quantum,ZYhqdm2022}, and the many-body configurations with constraints can be mapped to lattice electromagnetic fields which are naturally categorized into different topological sectors~\cite{Moessner2001}. 

\begin{figure*}[ht!] \centering
    \subfigure[] { \label{fig:tri_lattice_sector0}
    \includegraphics[width=5.8cm]{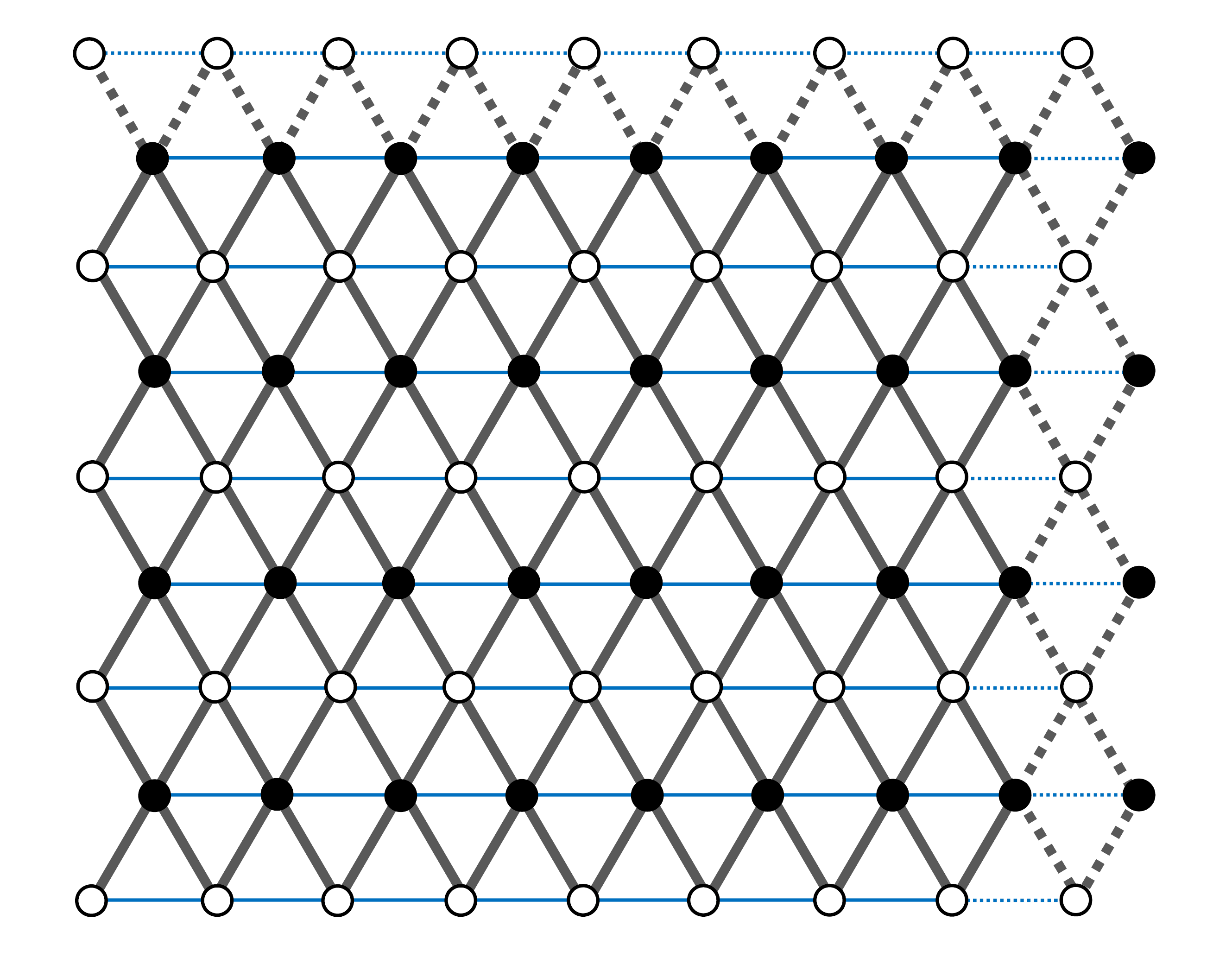}}
    \subfigure[] { \label{fig:tri_lattice_sector1}
    \includegraphics[width=5.8cm]{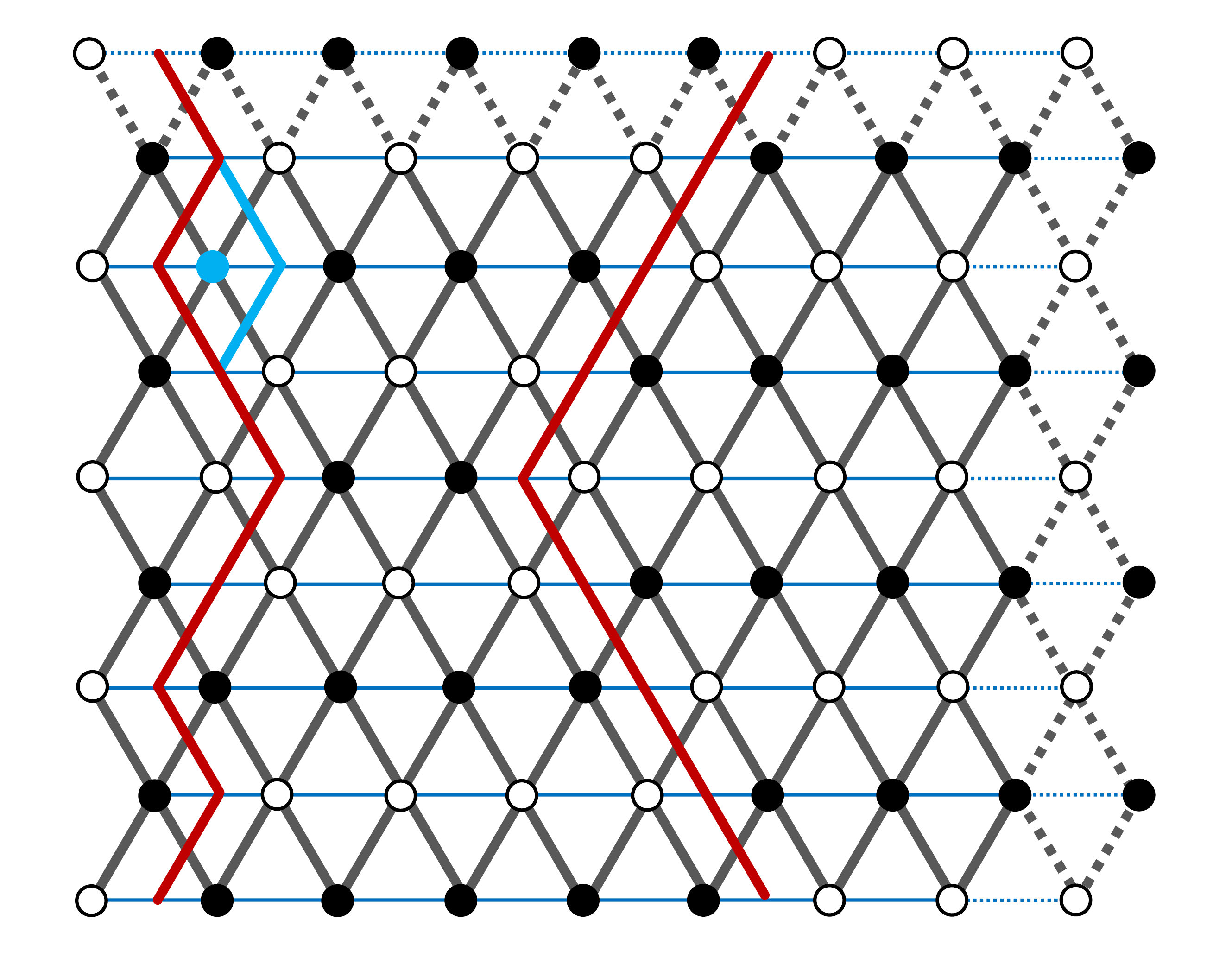}}
    \subfigure[] { \label{fig:tri_lattice_sector2}
    \includegraphics[width=5.8cm]{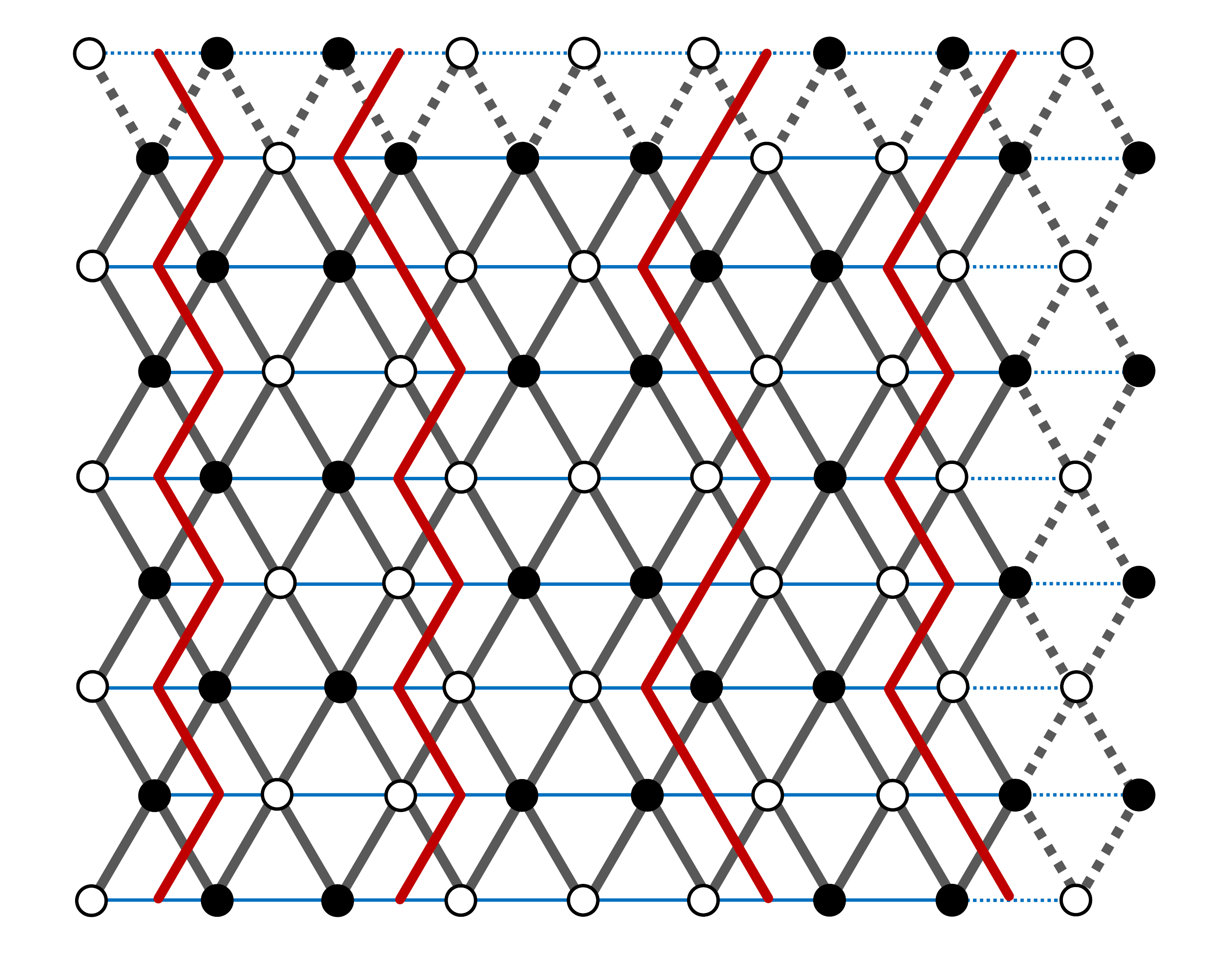}}
    \caption{ 
        (a) Spin configuration for $N_D=0$; (b) spin configuration for $N_D=2$. The red lines are the topological defects on the lattice. We can deform the shape of a defect by flipping a spin in the corner of it without breaking the 'triangular rule'. For example, we can flip the blue spin, then the segment on the left side of the spin will be moved to the right side, which is drew in blue as well; (c) spin configuration for $N_D=4$.
        }
        \label{fig:lattice_tri_config}
\end{figure*}

In the isotropic AFM triangular Ising model, the low-energy effective model is a non-interacting dimer model. It has topologically degenerate ground states which are in different topological sectors.
In order to get a non-trivial TSO problem as Fig.~\ref{fig:small_sector}, we break the degeneracy of the model by adding anisotropy in the coupling.

We finally use the Hamiltonian
\begin{equation}
    \hat H_{\text{tri}} = \sum_{\langle jk\rangle_x} J_x \hat Z_j\hat Z_k +  \sum_{\langle jk\rangle_{\wedge}} J_{\wedge} \hat Z_j\hat Z_k 
\end{equation}
on a triangular lattice with periodic boundary conditions (PBC).
$J_x$ denotes the interaction along the $x$ axis and $J_{\wedge}=1.0$ for the other two directions.

As mentioned above, all these topological sectors are degenerate when $J_x=J_{\wedge}$, therefore we set $J_x=0.9< J_{\wedge} $ to break the degeneracy and favor the stripe configuration [Fig.~\ref{fig:tri_lattice_sector0}] as the ground state. Because putting two parallel spins on the $J_x$ bonds costs least energy. This setting is to satisfy the hardcore three characteristics mentioned in Fig.~\ref{fig:small_sector}. Note that we can also set different configuration with target $N_{\rm D}$ as the ground state by changing the related couplings on bonds.

Through a more pictorial representation, the topological sectors here can be labeled by the number of topological defects $N_{\rm D}$~\cite{zhang01,zhang02,zhang03,zhou2020quantum}. As examples, we show the spin configurations in different topological sectors in Fig.~\ref{fig:tri_lattice_sector0}, \ref{fig:tri_lattice_sector1}, and \ref{fig:tri_lattice_sector2} for $N_{\rm D}=0,\,2,\,4$, respectively. Between two topological defects, the inside configuration is indeed same as the ground state, i.e., the stripe phase. Because the defect goes through the periodic boundary, thus it is topological and can not be removed by local actions.

In fact, the degree of freedom of a topological sector is decided by the corner numbers of defects, because flipping the spins at the corners of defects obeying the `triangle rule' without energy cost, as shown in Fig.~\ref{fig:tri_lattice_sector1}. Therefore, the degeneracy (degree of freedom) increases exponentially with the defect number $N_{\rm D}$~\cite{QDMbook,zhou2020quantum,zhou2020quantumstring,yan2020improved}.

It is worth mentioning that, $J_{\wedge}>J_x$ make the system favor the stripe phase without topological defect while the quantum fluctuation favors more topological defects with flippable corners 
. The sector with many topological defects, which has much more freedom degree as the Sector III of Fig.~\ref{fig:small_sector} shown, is easy to be reached and favored by quantum fluctuations. Of course, we could set Sector III as the sector where the ground state in, but it would be a trivial problem because it is very easy to be explored. 

\subsection{Results}
\begin{figure*}[ht!] \centering
    \subfigure[] { \label{fig:comparison_tri}
    \includegraphics[width=5.8cm]{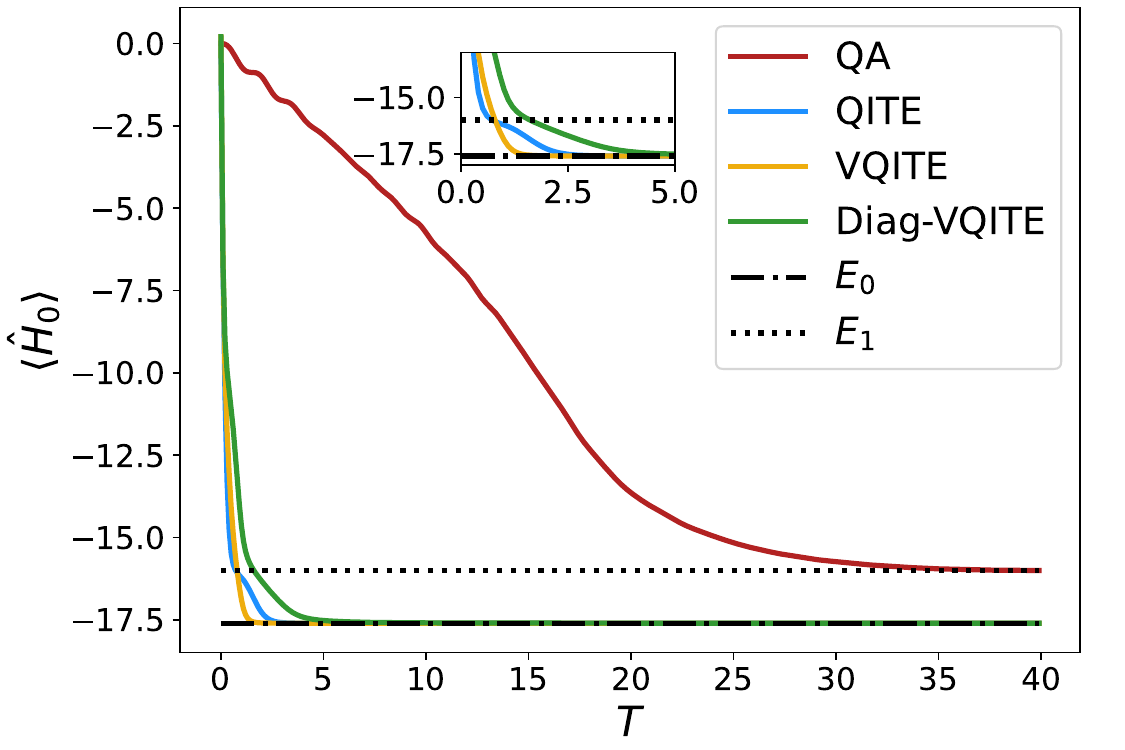}}
    \subfigure[] { \label{fig:QA_SQA_tri_model_44}
    \includegraphics[width=5.8cm]{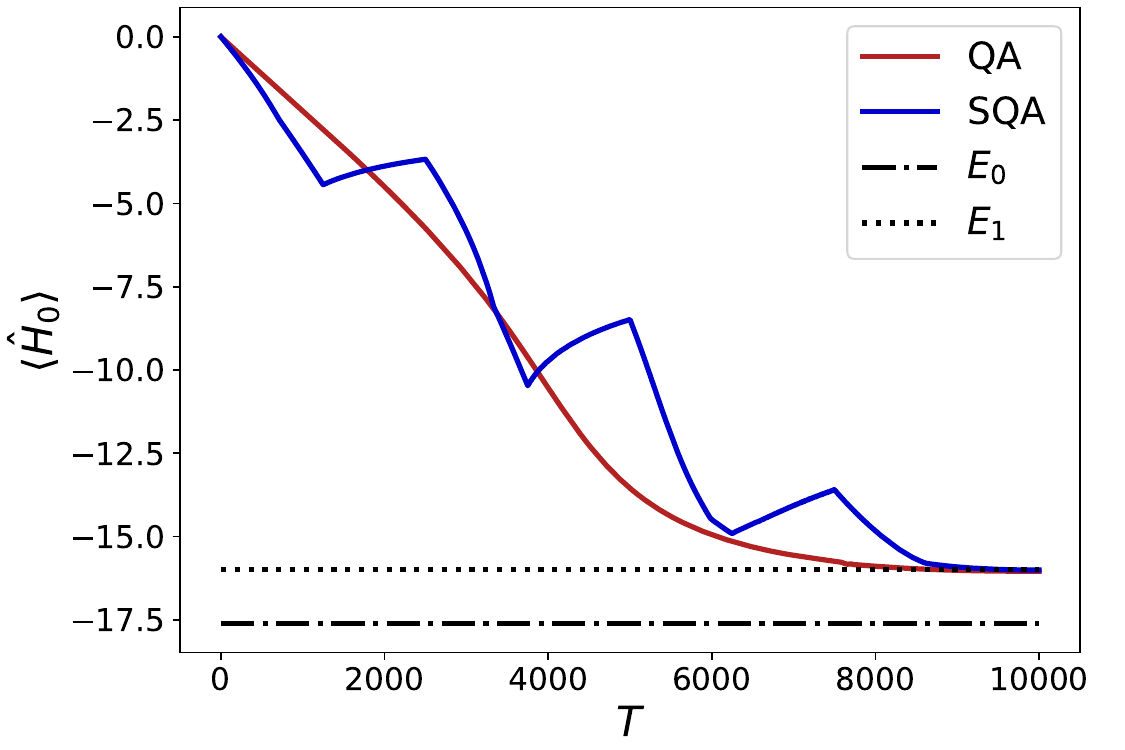}}
    \subfigure[] { \label{fig:VQE_tri_model_44}
    \includegraphics[width=5.8cm]{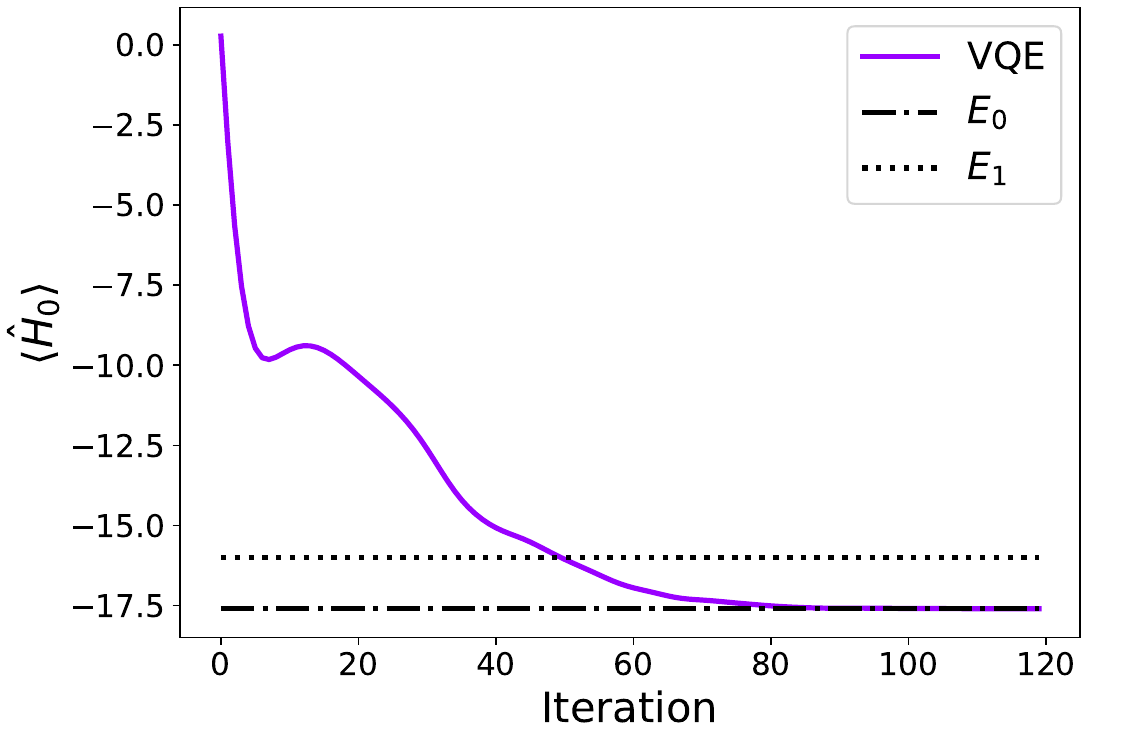}}
    \caption{ 
        (a) For $\hat H_{\text{tri}}$ and $T=40$, the simulation results of QA, QITE and VQITE, respectively; (b) simulation results of QA and SQA for $T=10^4\gg 40$; (c) simulation results of VQE.
        }
        \label{fig:QA_theTri}
\end{figure*}

We perform real-time simulations of QA and SQA numerically, which can provide quantitative estimations for their practical performance on quantum annealers. We simulate the ideal QITE according to the definition Eq. (\ref{eq:ITE}) and on the other side, implement VQITE, Diag-VQITE and VQE using TensorCircuit quantum simulation framework~\cite{Zhang2023tensorcircuit}. We consider a $4\times 4$ lattice with $16$ qubits. More technical details of the simulations can be found in Appx. \ref{appx:sqa} and~\ref{appx:qite_sim}. In addition, all codes for numerical simulations and results shown in this paper are open-source \footnote{\url{https://github.com/msquare1998/SourceCodes-arxiv2310.04291}}.

Fig.~\ref{fig:comparison_tri} shows the simulation results of QA, QITE, VQITE, and Diag-VQITE when the total evolution time $T$ is set to be $40$. Because of the topological obstructions, QA is finally stuck in the first excited state with energy $E_1$, a local minima which is not in the same sector of the ground state with energy $E_0$. By contrast, QITE can find the ground state in a very short time for this model. Besides, though the result of VQITE has some deviations from that of the standard QITE, it can reach $E_0$ as well. We further increase $T$ up to $10^4$, as it shown in Fig.~\ref{fig:QA_SQA_tri_model_44}, and QA still cannot find the ground state. This is an exceedingly long time and the intractability of this model for QA is therefore revealed. Besides, with $T=10^4$, SQA also cannot reach the true ground state, which indicates its inefficiency.

Notice that though the performance of VQITE should be better than Diag-VQITE as expected, it is even better than QITE according to Fig.~\ref{fig:comparison_tri} in this example. As a variational algorithm, the performance of VQITE depends closely on the choice of ansatz as introduced in the Sec.~\ref{sec:intro_vqite}. On the other hand, Diag-VQITE and VQE (see Fig.~\ref{fig:VQE_tri_model_44}) can both find the ground state. The performance advantage of VQITE and the success of Diag-VQITE and VQE may come from the translation symmetry along the $x$-axis of the model, rendering the ground state easier to identify. Therefore, we consider a similar fully frustrated Ising model on square lattice, which breaks the translation symmetry in the next section. 

\section{Fully frustrated Ising model on square lattice}\label{sec:model_sq} 
\subsection{Model}
\begin{figure}[ht!] \centering
    \subfigure[] { \label{fig:lattice_sq_config}
    \includegraphics[width=4.2cm]{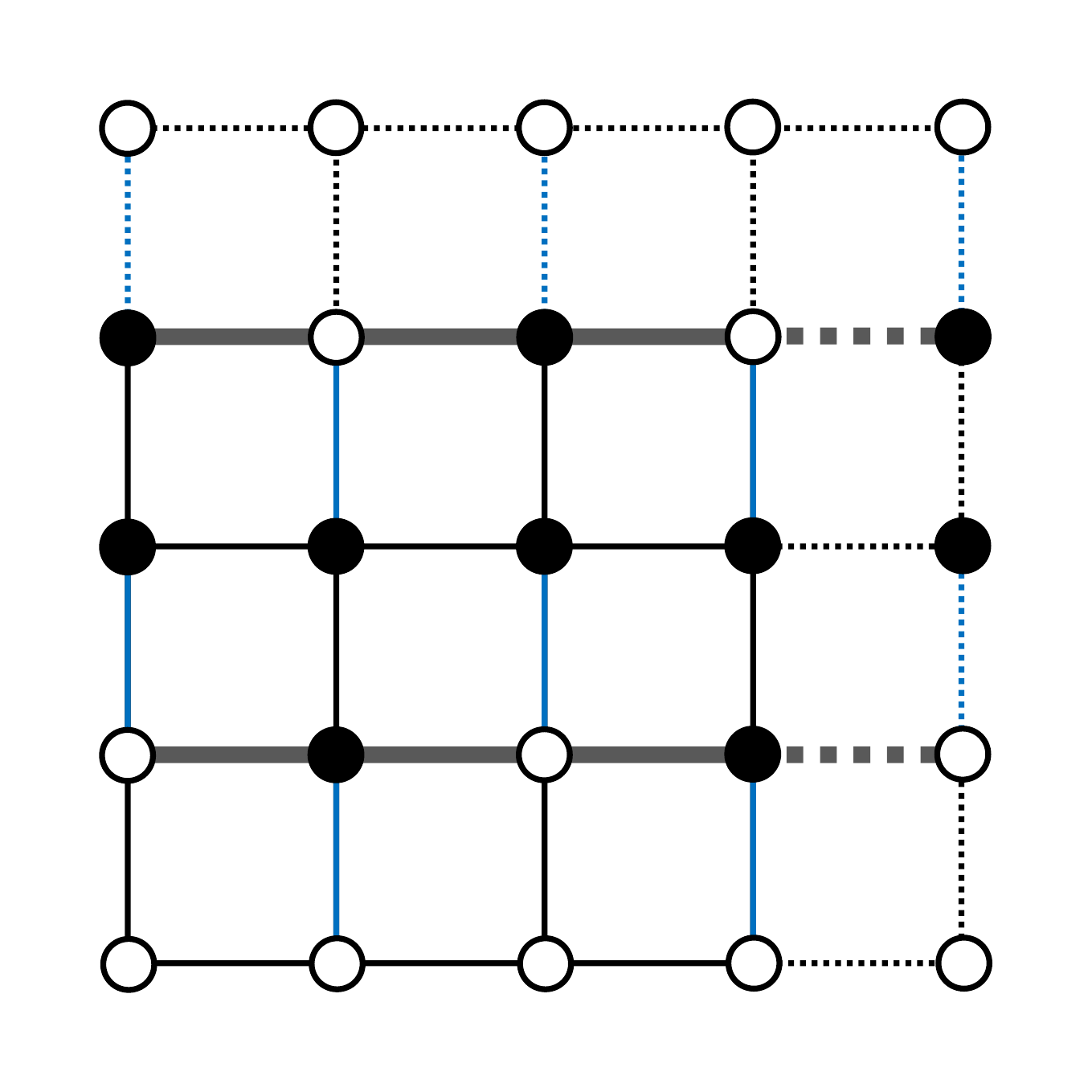}}
    \subfigure[] { \label{fig:lattice_sq}
    \includegraphics[width=4.2cm]{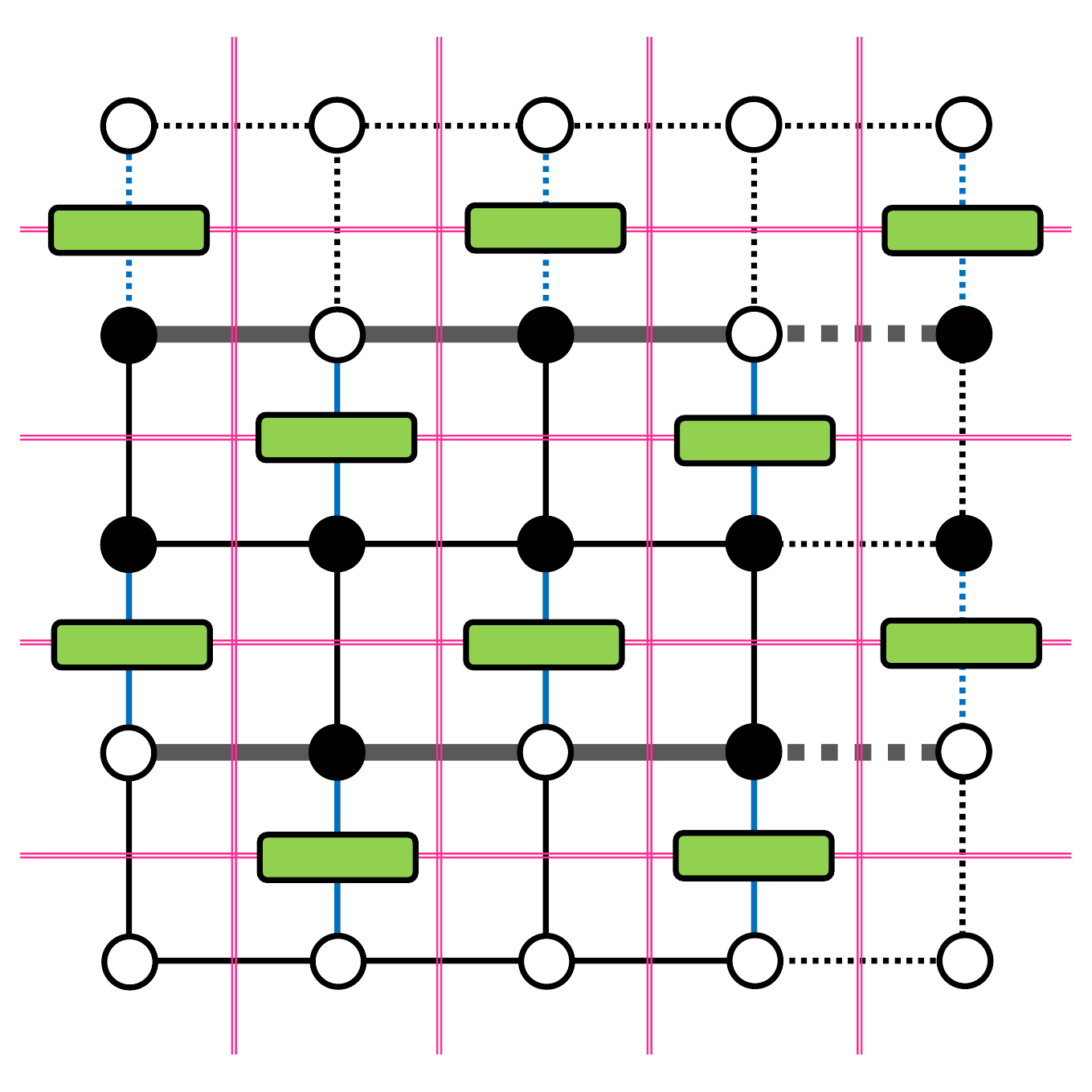}}
    \caption{ 
        (a) Lattice of $\hat H_{\text{sq}}$. The thick gray bonds correspond to the coupling $J$, the think black bonds correspond to the coupling $K$, and the blue think bonds correspond to the coupling $K_p$. (b) the dimer model on the dual lattice which is depicted in pink color with the green links denoting the dimers.
        }
        \label{fig:LATTICE_SQ}
\end{figure}

Similar to the frustrated Ising model on triangular lattice above, we present a fully frustrated Ising model on square lattice [Fig. \ref{fig:lattice_sq_config}] as the other test bed for these quantum optimization algorithms. This model breaks the translation symmetry which makes the solution harder to be reached. The Hamiltonian reads:
\begin{equation}\label{model:sq}
    \hat H_{\text{sq}} = J\sum_{\langle jk\rangle_J} \hat Z_j\hat Z_k + K\sum_{\langle jk\rangle_K}\hat Z_j\hat Z_k + K_p\sum_{\langle jk\rangle_{K_p}}\hat Z_j\hat Z_k.
\end{equation}
As in Fig.~\ref{fig:lattice_sq_config}, the interaction of thick gray bonds are antiferromagnetic (AFM) ($J = 1$), and the thin black bonds are ferromagnetic (FM) (set $K=-1$). Similar to the "triangle rule" mentioned above for the triangular lattice model, here we also have a "square rule" at low temperature where each four bonds in a square plaquette must have one excited bond (e.g., its coupling is FM/AFM, but the two spins on it are antiparallel/parallel). Similarly, each square plaquette must has and only has one excited bond in the low-energy case, thus it can be mapped to a square lattice dimer model with one dimer per site~\cite{yan2019widely}. The dimers live on the dual lattice as Fig. \ref{fig:lattice_sq} shown.

To satisfy the three hardcore characteristics for TSO and break the topological degeneracy, we set the links crossed by dimers a little weaker ($K_p=0.9$ for AFM and $-0.9$ for FM, drew in blue of Fig.~\ref{fig:lattice_sq_config}) to let dimers condensed on the corresponding positions. 
The anisotropy makes the staggered sector with least degree of freedom become the lowest energy one. Therefore, the model has similar topological sector optimization problem as the model in Sec.~\ref{sec:model_tri} but much harder due to the translation symmetry breaking. Even in Ref. \cite{yan2022preparing} of SQA, the SQA works well in the triangular lattice but is not strictly effective in the square lattice.

\subsection{Results}
\begin{figure*}[ht!] \centering
    \subfigure[] { \label{fig:comparison_sq}
    \includegraphics[width=5.8cm]{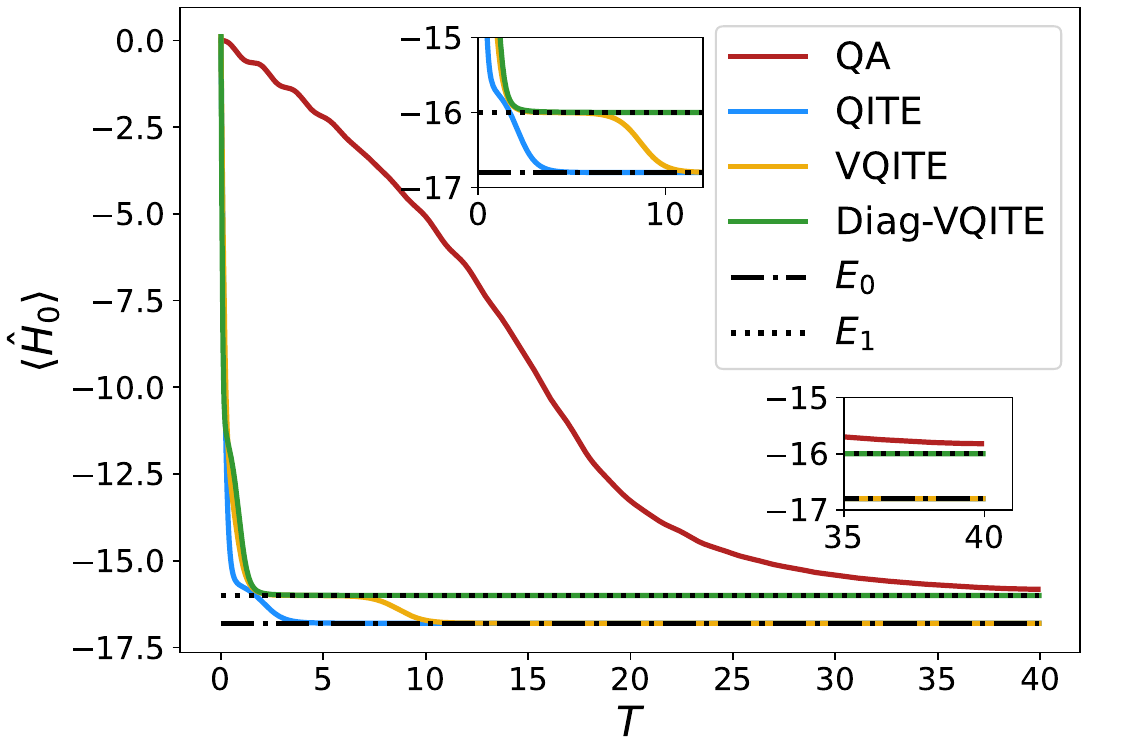}}
    \subfigure[] { \label{fig:qa_sq}
    \includegraphics[width=5.8cm]{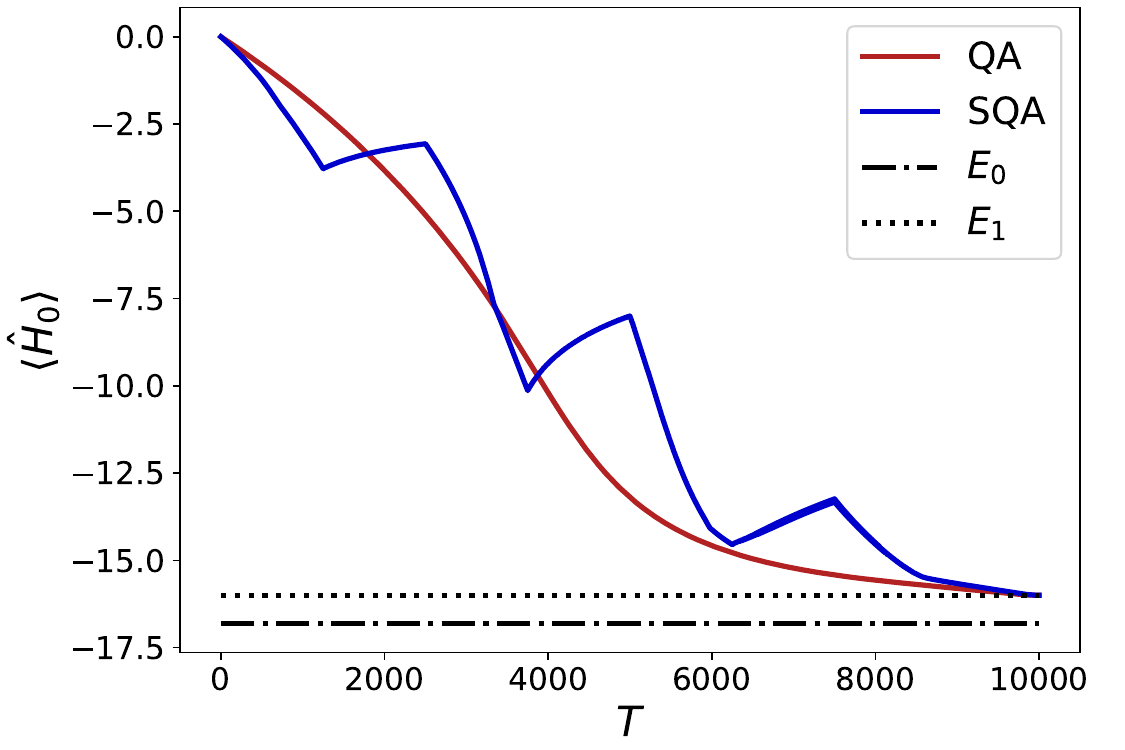}}
    \subfigure[] { \label{fig:vqe_sq}
    \includegraphics[width=5.8cm]{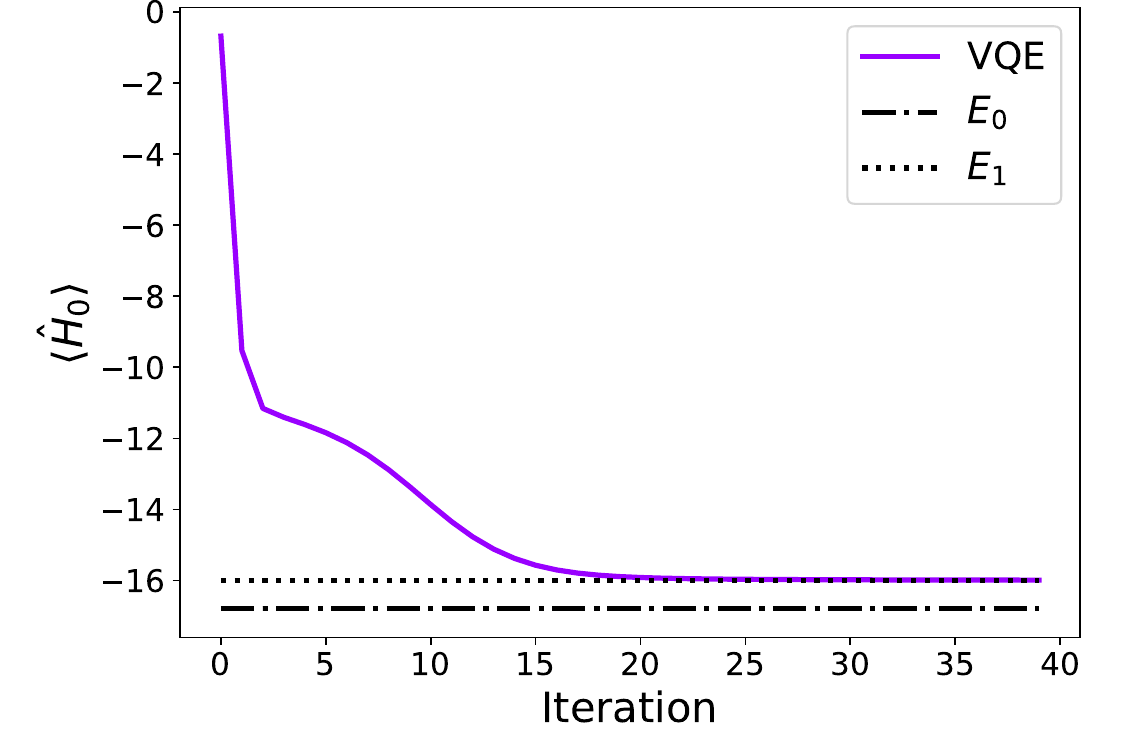}}
    \caption{ 
        (a) For $\hat H_{\text{sq}}$ and $T=40$, the simulation results of QA, QITE and VQITE, respectively; (b) simulation results of QA and SQA for $T=10^4\gg 40$; (c) simulation results of VQE.
        }
        \label{fig:square_sim}
\end{figure*}

We use the same optimization schemes with $\hat H_{\text{tri}}$. As it is shown in Fig.~\ref{fig:comparison_sq} and \ref{fig:qa_sq}, QA and SQA also fails in this TSO problem. However, different from $\hat H_{\text{tri}}$, $T=40$ is not even sufficient to reach the first excited state $E_1$ for QA. On the other hand, QITE needs $T\approx 4$ while VQITE needs $T< 20$ to reach the ground state. In this example, both Diag-VQITE and VQE fail (see Fig.~\ref{fig:VQE_tri_model_44}). Therefore the complete version of VQITE is necessary in this more difficult TSO problem. Besides, since the imaginary time evolution relates to the quantum natural gradient descent~\cite{Stokes2020quantumnatural}, its advantage over the stochastic gradient gradient by VQE is also showed in this model.

\begin{figure}[ht!] \centering
    \subfigure[] { \label{fig:multi_Ising}
    \includegraphics[width=4.1cm]{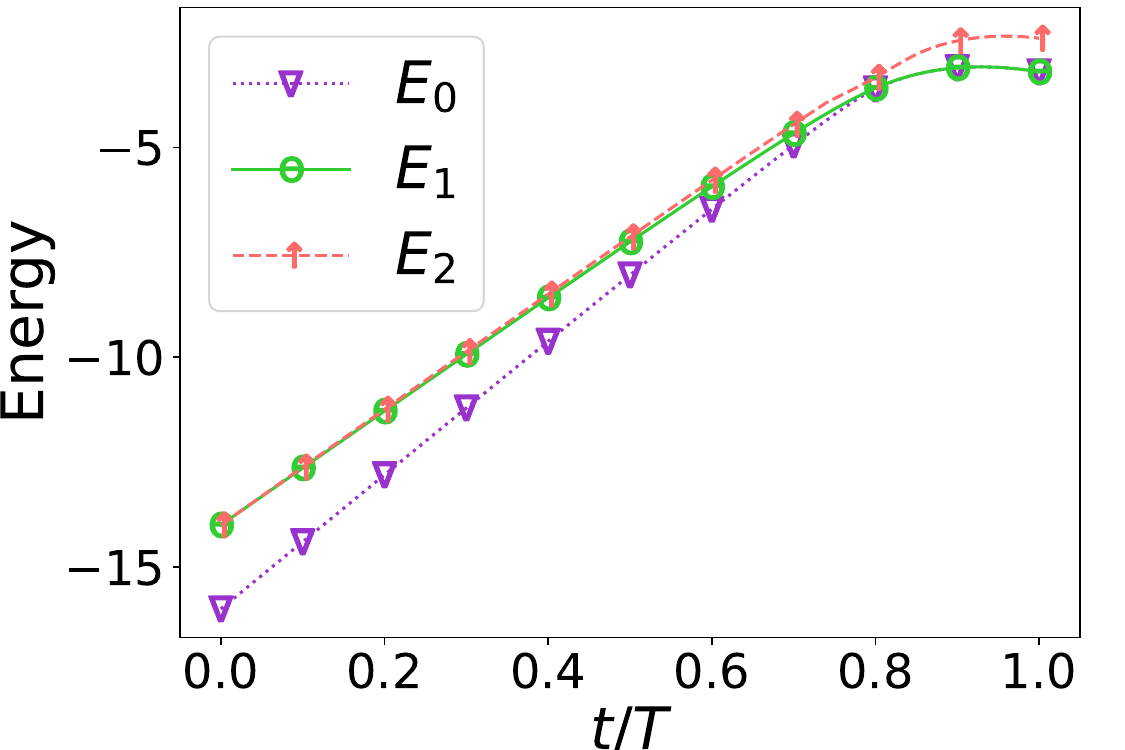}
    }
    \subfigure[] { \label{fig:multi_square}
    \includegraphics[width=4.1cm]
    {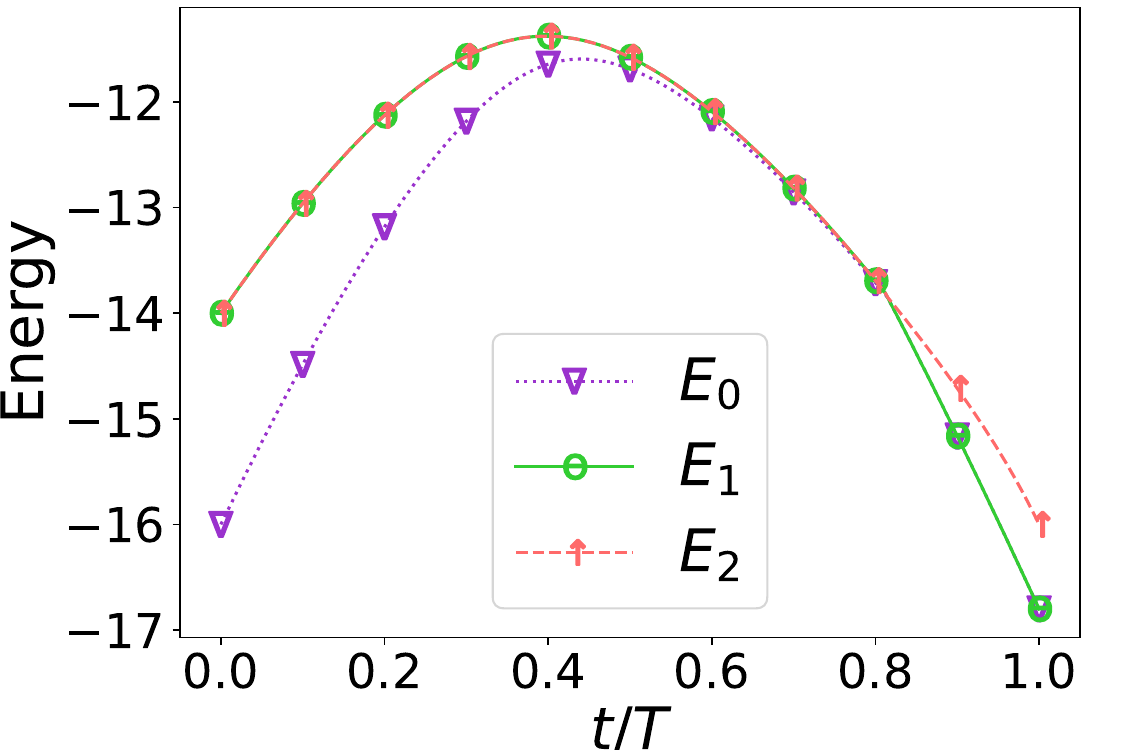}
    }
    \caption{ 
        Lower energies of a $4\times 4$ model for (a) $H_{\text{Ising-2d}}$, (b) $H_{\text{sq}}$.
        }
        \label{fig:energySpectrum}
  \end{figure}

\subsection{The role of topology}
  
To better show the interplay between topology and QA, we compare $\hat H_{\text{sq}}$ with a spin model of no topological character, which is the isotropic 2D AFM Ising model on a square lattice (PBC)
\begin{equation}
    \hat H_{\text{Ising-2d}} = J\sum_{\langle i,j\rangle} \hat Z_i\hat Z_j.
\end{equation}
We take $J=0.1$ to make sure that $\hat H_{\text{Ising}}$ shares the same gap with $\hat H_{\text{sq}}$ in the QA Hamiltonian Eq. (\ref{eq:adiabatic}). Consequently, the lower energies of $\hat H_{\text{eq}}$ and $\hat H_{\text{Ising}}$ computed by the exact diagonalization method show similar behaviors: both of the spectra have a closed gap at some point around $s=0.8$ (not the same one), and $E_0$ merges with $E_1$ after the gapless point, as it shown in Fig. \ref{fig:energySpectrum}.

\begin{figure}[ht!] \centering
    \includegraphics[width=6cm]{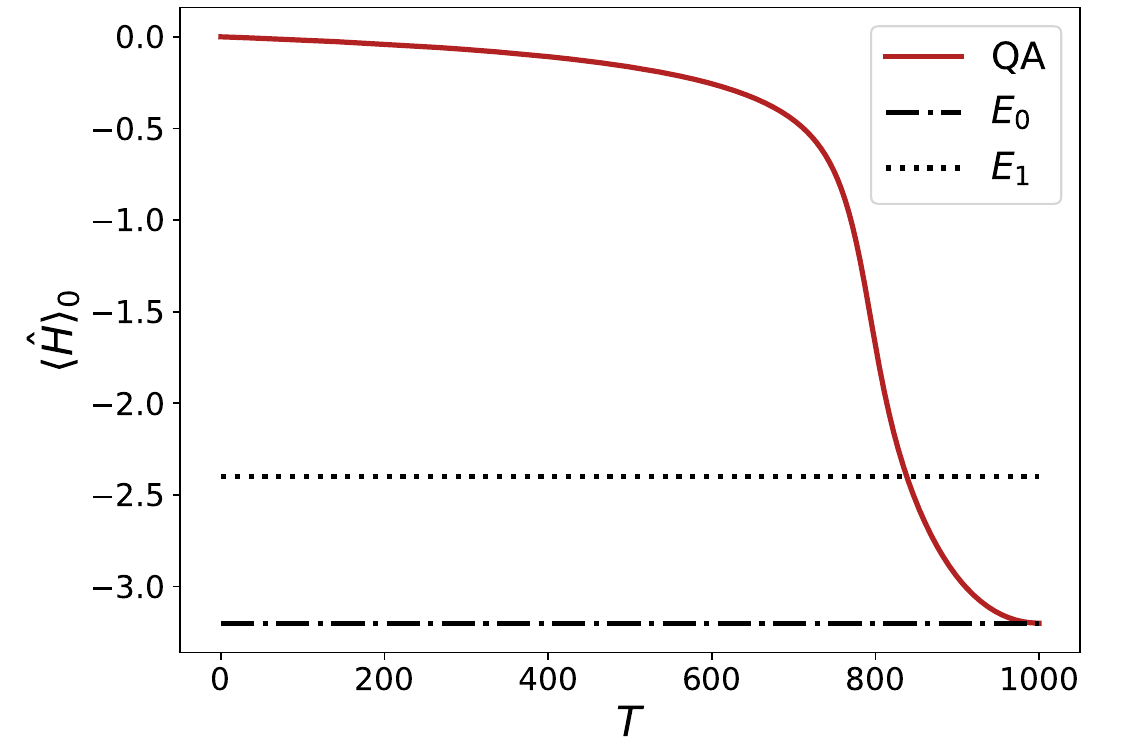}
    \caption{ 
        Real-time simulations of QA for $\hat H_{\text{Ising-2d}}$ on a $4\times 4$ lattice under PBC.
        }
        \label{fig:qa_2d_ising}
  \end{figure}

Fig.~\ref{fig:qa_2d_ising} shows that when $T=10^3$, QA can find the ground state of $\hat H_{\text{Ising-2d}}$ exactly. Although there is a point that gap closes in the annealing path, the second order phase transition nature renders the critical dynamics following Kibble-Zurek mechanism, which secures the large fidelity with the ground state at the end~\cite{T_W_B_Kibble_1976, PhysRevLett.95.105701}. 
In $\hat H_{\text{sq}}$, the states before and after the gapless point are in different topological sectors, thus they are not connected smoothly hence it is hard for QA to remain correct.

QA with the Hamiltonian in Eq.(\ref{eq:adiabatic}) leverages quantum fluctuations induced by magnetic fields along the $x$ direction to seek the true ground state. Since the magnetic fields are local (i.e. each $\hat X_i$ only acts on a single site $i$), they have little chance to induce a collective fluctuation that aligns precisely with the global topological defect to be bypassed for an instaneous state. As the system size increases, the length of the defect grows, decreasing the likelihood of such collective fluctuation. We can also understand this point through general topological phases~\cite{Wen2017topoZoo, Wen2013topoorder}, in which the topological sectors are robust to local magnetic perturbations.

\section{Discussions and Conclusions}\label{sec:discussion}
In this work, we explore the role played by topology in optimization problems. We report the results of different quantum optimization methods on the so-called TSO problem which is of topological nature. We find that the QA method fails to address TSO problem as it is intractable to explore the Hilbert space composed of different topological sectors. QA fails to transform between two states that are not smoothly connected and are separated in different topological sectors. Though the two states have close energies, the topological protection leads to a large Hamming distance (also a large energy barrier) and requires some specific global operations that can wind the lattice to break it. Consequently, the probability to generate such a global operation via quantum fluctuation is exponentially small, thus QA loses its efficacy in TSO problems. Although the improved QA method, SQA, can go across different topological sectors, but it is still not sufficiently efficient to address these optimization problems. 

It is worth of emphasizing that for some TSO models, it is possible to reach the true ground state via QA if we have some preknowledge of the model and are able to prepare the initial ground state in the right topological sector. Depending on the number of sectors $\mathfrak{g}$, at most $\mathfrak{g}$ times can be tried. This is generally hard because even though we know the right sector, it does not mean the corresponding ground state can be efficiently prepared. 

On the contrary, QITE algorithm and its NISQ approximate implementation can both successfully address TSO problems demonstrated in our paper. Not only QITE has quadratic advantage on time complexity over QA, but its mechanism which utilizes the property of quantum superposition prevents it from being trapped in some local minima. Topological obstructions have no effects on QITE and thus QITE is a better algorithm for tackling TSO problem than QA. We have to emphasize that the two TSO problems we consider in this work do not even introduce external magnetic fields and it is yet difficult for QA and SQA. If some small random external fields are added, the energy landscape in each sector will be much more glassy and the problems will be more difficult. Unquestionably, QITE can still find the solution as we discussed. The fate of VQITE and VQE in this case is an interesting problem to explore in the future.

\section*{Acknowledgements.}
We wish to thank Xiaopeng Li and Shuai Yin for helpful discussions. 
Y.M.D. and Z.Y. thank the start-up fund of Westlake University. The authors acknowledge Beijng PARATERA Tech Co.,Ltd. (\url{https://www.paratera.com/}) for providing HPC resources that have contributed to the research results reported within this paper. Y.C.W. acknowledges  support from Zhejiang Provincial Natural Science Foundation of China (Grant Nos. LZ23A040003). S.X.Z. is supported by a
startup grant in IOP-CAS.

\appendix
\section{Simulation of SQA}\label{appx:sqa}
For $\hat H_1$ in the QA Hamiltonian (\ref{eq:adiabatic}), QA always applys uniform magnetic fields on all sites. Different from QA, the SQA method introduces virtual edges to try overcoming the topological obstructions, where a virtual edge along a specified direction will be created imposing strong magnetic fields on some spins. After the spins are polarized, the virtual edge will be glued by decreasing the intensity of the magnetic fields~\cite{yan2022preparing}. 

\begin{figure}[ht!] \centering
    \subfigure[] { \label{fig:sqa_edges}
    \includegraphics[width=4.2cm]{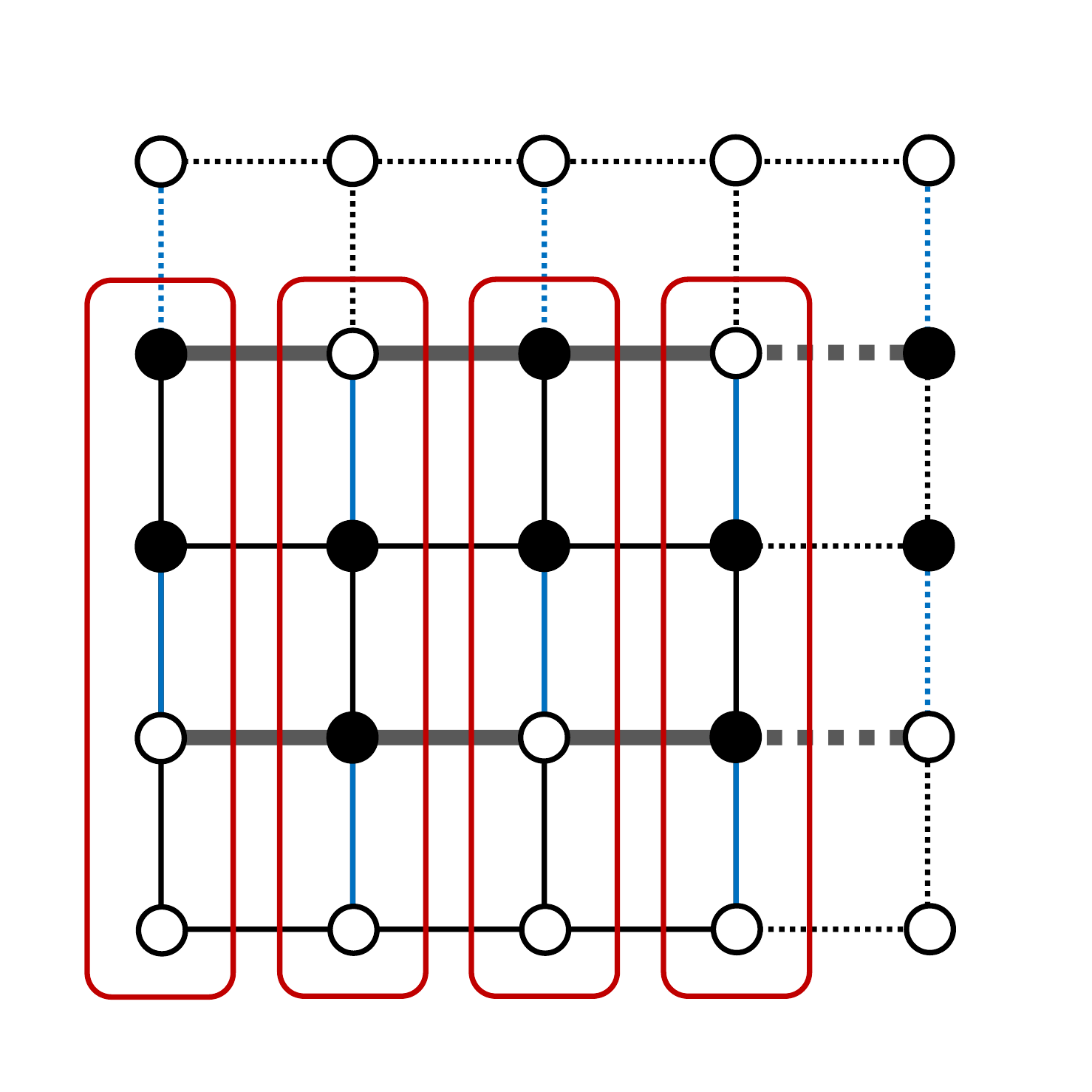}}
    \subfigure[] { \label{fig:sqa_schematic}
    \includegraphics[width=4.2cm]{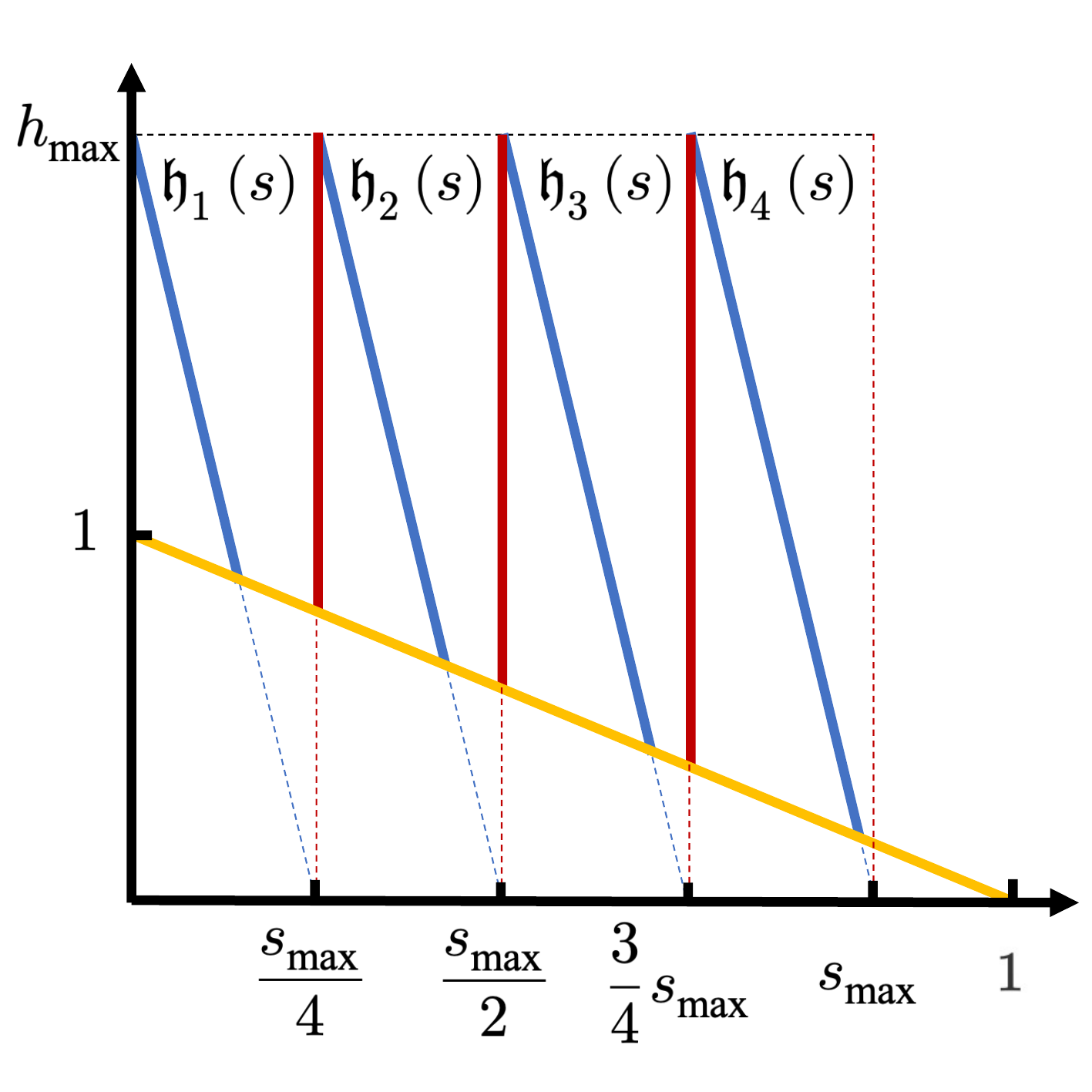}}
    \caption{ 
        For $\hat H_{\text{sq}}$ on a $4\times 4$ lattice, (a) the four rounded rectangles represent the four edges on the lattice; (b) Schematic diagram of the SQA scheme for $\hat H_{\text{sq}}$ on a $4\times 4$ lattice. The orange line is the standard QA which corresponds to the function $\mathfrak{f}_(s)=1-s$ and the four blue solid lines denote $\mathfrak{h}_j(t)$ ($j=1,2,3,4$) when glueing edges. The red lines represent the process when opening the edges in which $s$ is fixed.
        }
        \label{fig:SQA_sch}
\end{figure}

Here we take the $\hat H_{\text{sq}}$ on a $4\times 4$ lattice as an example. Along the vertical direction, there are four virtual edges to be created (See Fig.~\ref{fig:sqa_edges}). Here we let $s=t/T$ for convenience. We use $\mathfrak{h}_{j}(s)$ to denote the intensities of magnetic fields acting on the $j$th virtual edge and $\overline{\mathfrak{h}}_j(s)$ to denote intensities on those sites that are not included in the $j$th edge at time $s$. Furthermore, we define $s_{\text{max}}<1$ as the time at which we consider all virtual edges have been glued well and $h_{\text{max}}>1$ is the maximum value of $\mathfrak{h}_j(s)$. 

As shown in Fig.~\ref{fig:sqa_schematic}, setting $\mathfrak{h}_1(0)=h_{\text{max}}$ signifies the creation of the first virtual edge at the beginning of the SQA process. The gluing process of the first edge corresponds to $\mathfrak{h}_1(s)=-4h_{\text{max}}s/s_{\text{max}} + h_{\text{max}}$ when $\overline{\mathfrak{h}}_1(s)=1-s$, where $s\in[0,(1-h_{\text{max}})/(1-4h_{\text{max}}/s_{\text{max}})]$. Then, a standard QA will be appiled to all sites until $s=s_{\text{max}}/4$. After that, we moves to the second virtual edge by increasing $\mathfrak{h}_2(s_{\text{max}}/4)$ from $1-s_{\text{max}}/4$ to $h_{\text{max}}$. It is important note that the standard QA is paused until $\mathfrak{h}_2(s_{\text{max}})=h_{\text{max}}$. Subsequent to this, we similarly glue the second edge akin to what we do to the first edge. When $s=- 4h_{\text{max}}s / s_{\text{max}} + 4h_{\text{max}}$, a standard QA will be performed until $s=1$.

Taking into account the time $T_{\text{sweeping}}$ for sequentially creating the virtual edges, the total evolution time should be $T=T_{\text{evo}}+T_{\text{sweeping}}$, where $T_{\text{evo}}$ stands for the time when $s$ is changed. In our simulations for both $\hat H_{\text{tri}}$ and $\hat H_{\text{sq}}$, we take $h_{\text{max}}=2$ and $s_{\text{max}}=0.8$, which are same as that in Ref.~\cite{yan2022preparing}. Besides, the choice of  $\Delta t$ and $\Delta s$ follows Appx.~\ref{subsec:qa_tech}.

\section{Details of VQITE, Diag-VQITE and VQE}\label{appx:qite_sim}
In VQITE, Diag-VQITE and VQE, we use similar PQC for both $\hat H_{\text{tri}}$ and $\hat H_{\text{sq}}$ which encompasses single- and two- qubit gates. We write $V(\vec{\theta})$ as 
\begin{equation}
    V(\vec{\theta}) = V_2(\vec{\alpha},\vec{\beta})V_1(\vec{\omega})
\end{equation}
for convenience, which means $\vec{\theta} \equiv (\vec\omega,\vec\alpha,\vec\beta)$, where $V_1(\vec\omega)$ denotes all single-qubit gates and $V_2(\vec\alpha,\vec\beta)$ denotes all double-qubit gates.

Denote the number of qubits to be $n_q$, and label the qubits by $\{q\ |\ q\in[0,n_q-1] \text{ and } q\in\mathrm{Z}\}$. We consider the ZXZ decomposition, a set of universal single-qubit operations, for each qubit $q$, therefore
\begin{equation}
    V_1(\vec{\omega}) = \prod_q R_z^{q}(\omega_{q,2})R_x^{q}(\omega_{q,1})R_z^{q}(\omega_{q,0}).
\end{equation}

Our choice of $V_2(\vec\alpha,\vec\beta)$ comes from the 2d structure of the lattice. Specifically, for each bond $\overline{jk}$ connecting qubits $j$ and $k$ on the lattice, apply two Pauli gadgets (double-qubit rotation gates) to it, and 
\begin{equation}
    V_2(\vec\alpha,\vec\beta) = \prod_{\overline{jk}}e^{-i\beta_{\overline{jk}} \hat Y_{j}\hat Z_k/2}
    e^{-i\alpha_{\overline{jk}} \hat Z_{j}\hat Y_k/2}.
\end{equation}

Since the solutions to an optimization problem must be some classical states, the most straightforward way is to prepare the initial state to be a superposition of all possible classical configurations. Therefore, we can set $|\bar 0\rangle$ as
\begin{equation}\label{eq:sp_state}
    |\bar 0\rangle  = H^{\otimes n_q} |0\rangle^{\otimes n_q},
\end{equation}
which is exactly the $\Psi^{+}$ in (\ref{eq:pd_state_plus}), where $H$ denotes Hadamard gate. The variational parameters of the PQC are initially set to follow Gaussian distribution ($\mu=0,\sigma=0.05$). For updating the parameters, we add a small quantity $\epsilon=10^{-4}$ to all the digonal elements $A_{jj}$ in Eq. (\ref{eq:linearEq}) for the purpose of regularization when solving the linear equations in VQITE and Diag-VQITE. In VQE, we use the stochastic gradient descent to minimize $\langle \phi(\vec\theta)|\hat H|\phi(\vec\theta)\rangle$. 

In principle, a smaller $\Delta t$ in Eq. (\ref{eq:updateTheta}) would make VQITE a better approximation to the standard QITE. However, our simulation results show that $\Delta t=0.1$ is sufficient for VQITE to reach the ground state while for Diag-VQITE we need $\Delta t=0.05$. This comes from the specific ansatz we use. The learning rate we choose in VQE is $\eta = 0.02$ for $\hat H_{\text{tri}}$ and $\eta=0.05$ for $\hat H_{\text{sq}}$. We have simulated 200 times for each of the algorithm with different initial parameters, and the same results occur 200 times.

\newpage 
\bibliography{itqc}

\begin{thebibliography}{82}%
\makeatletter
\providecommand \@ifxundefined [1]{%
 \@ifx{#1\undefined}
}%
\providecommand \@ifnum [1]{%
 \ifnum #1\expandafter \@firstoftwo
 \else \expandafter \@secondoftwo
 \fi
}%
\providecommand \@ifx [1]{%
 \ifx #1\expandafter \@firstoftwo
 \else \expandafter \@secondoftwo
 \fi
}%
\providecommand \natexlab [1]{#1}%
\providecommand \enquote  [1]{``#1''}%
\providecommand \bibnamefont  [1]{#1}%
\providecommand \bibfnamefont [1]{#1}%
\providecommand \citenamefont [1]{#1}%
\providecommand \href@noop [0]{\@secondoftwo}%
\providecommand \href [0]{\begingroup \@sanitize@url \@href}%
\providecommand \@href[1]{\@@startlink{#1}\@@href}%
\providecommand \@@href[1]{\endgroup#1\@@endlink}%
\providecommand \@sanitize@url [0]{\catcode `\\12\catcode `\$12\catcode `\&12\catcode `\#12\catcode `\^12\catcode `\_12\catcode `\%12\relax}%
\providecommand \@@startlink[1]{}%
\providecommand \@@endlink[0]{}%
\providecommand \url  [0]{\begingroup\@sanitize@url \@url }%
\providecommand \@url [1]{\endgroup\@href {#1}{\urlprefix }}%
\providecommand \urlprefix  [0]{URL }%
\providecommand \Eprint [0]{\href }%
\providecommand \doibase [0]{https://doi.org/}%
\providecommand \selectlanguage [0]{\@gobble}%
\providecommand \bibinfo  [0]{\@secondoftwo}%
\providecommand \bibfield  [0]{\@secondoftwo}%
\providecommand \translation [1]{[#1]}%
\providecommand \BibitemOpen [0]{}%
\providecommand \bibitemStop [0]{}%
\providecommand \bibitemNoStop [0]{.\EOS\space}%
\providecommand \EOS [0]{\spacefactor3000\relax}%
\providecommand \BibitemShut  [1]{\csname bibitem#1\endcsname}%
\let\auto@bib@innerbib\@empty
\bibitem [{\citenamefont {Hauke}\ \emph {et~al.}(2020)\citenamefont {Hauke}, \citenamefont {Katzgraber}, \citenamefont {Lechner}, \citenamefont {Nishimori},\ and\ \citenamefont {Oliver}}]{Hauke_2020}%
  \BibitemOpen
  \bibfield  {author} {\bibinfo {author} {\bibfnamefont {P.}~\bibnamefont {Hauke}}, \bibinfo {author} {\bibfnamefont {H.~G.}\ \bibnamefont {Katzgraber}}, \bibinfo {author} {\bibfnamefont {W.}~\bibnamefont {Lechner}}, \bibinfo {author} {\bibfnamefont {H.}~\bibnamefont {Nishimori}},\ and\ \bibinfo {author} {\bibfnamefont {W.~D.}\ \bibnamefont {Oliver}},\ }\bibfield  {title} {\bibinfo {title} {Perspectives of quantum annealing: methods and implementations},\ }\href {https://doi.org/10.1088/1361-6633/ab85b8} {\bibfield  {journal} {\bibinfo  {journal} {Reports on Progress in Physics}\ }\textbf {\bibinfo {volume} {83}},\ \bibinfo {pages} {054401} (\bibinfo {year} {2020})}\BibitemShut {NoStop}%
\bibitem [{\citenamefont {Boettcher}(2019)}]{PhysRevResearch.1.033142}%
  \BibitemOpen
  \bibfield  {author} {\bibinfo {author} {\bibfnamefont {S.}~\bibnamefont {Boettcher}},\ }\bibfield  {title} {\bibinfo {title} {Analysis of the relation between quadratic unconstrained binary optimization and the spin-glass ground-state problem},\ }\href {https://doi.org/10.1103/PhysRevResearch.1.033142} {\bibfield  {journal} {\bibinfo  {journal} {Phys. Rev. Res.}\ }\textbf {\bibinfo {volume} {1}},\ \bibinfo {pages} {033142} (\bibinfo {year} {2019})}\BibitemShut {NoStop}%
\bibitem [{\citenamefont {Fu}\ and\ \citenamefont {Anderson}(1986)}]{Anderson1986}%
  \BibitemOpen
  \bibfield  {author} {\bibinfo {author} {\bibfnamefont {Y.}~\bibnamefont {Fu}}\ and\ \bibinfo {author} {\bibfnamefont {P.~W.}\ \bibnamefont {Anderson}},\ }\bibfield  {title} {\bibinfo {title} {Application of statistical mechanics to {NP}-complete problems in combinatorial optimisation},\ }\href {https://doi.org/10.1088/0305-4470/19/9/033} {\bibfield  {journal} {\bibinfo  {journal} {J. Phys. A}\ }\textbf {\bibinfo {volume} {19}},\ \bibinfo {pages} {1605} (\bibinfo {year} {1986})}\BibitemShut {NoStop}%
\bibitem [{\citenamefont {Mezard}\ \emph {et~al.}(1986)\citenamefont {Mezard}, \citenamefont {Parisi},\ and\ \citenamefont {Virasoro}}]{Virasoro1987}%
  \BibitemOpen
  \bibfield  {author} {\bibinfo {author} {\bibfnamefont {M.}~\bibnamefont {Mezard}}, \bibinfo {author} {\bibfnamefont {G.}~\bibnamefont {Parisi}},\ and\ \bibinfo {author} {\bibfnamefont {M.}~\bibnamefont {Virasoro}},\ }\href {https://doi.org/10.1142/0271} {\emph {\bibinfo {title} {Spin Glass Theory and Beyond}}}\ (\bibinfo  {publisher} {World Scientific},\ \bibinfo {year} {1986})\BibitemShut {NoStop}%
\bibitem [{\citenamefont {Huse}\ and\ \citenamefont {Fisher}(1986)}]{Huse1986}%
  \BibitemOpen
  \bibfield  {author} {\bibinfo {author} {\bibfnamefont {D.~A.}\ \bibnamefont {Huse}}\ and\ \bibinfo {author} {\bibfnamefont {D.~S.}\ \bibnamefont {Fisher}},\ }\bibfield  {title} {\bibinfo {title} {Residual energies after slow cooling of disordered systems},\ }\href {https://doi.org/10.1103/PhysRevLett.57.2203} {\bibfield  {journal} {\bibinfo  {journal} {Phys. Rev. Lett.}\ }\textbf {\bibinfo {volume} {57}},\ \bibinfo {pages} {2203} (\bibinfo {year} {1986})}\BibitemShut {NoStop}%
\bibitem [{\citenamefont {Santoro}\ \emph {et~al.}(2002)\citenamefont {Santoro}, \citenamefont {Marto{\v n}{\'a}k}, \citenamefont {Tosatti},\ and\ \citenamefont {Car}}]{Car2002}%
  \BibitemOpen
  \bibfield  {author} {\bibinfo {author} {\bibfnamefont {G.~E.}\ \bibnamefont {Santoro}}, \bibinfo {author} {\bibfnamefont {R.}~\bibnamefont {Marto{\v n}{\'a}k}}, \bibinfo {author} {\bibfnamefont {E.}~\bibnamefont {Tosatti}},\ and\ \bibinfo {author} {\bibfnamefont {R.}~\bibnamefont {Car}},\ }\bibfield  {title} {\bibinfo {title} {Theory of quantum annealing of an {I}sing spin glass},\ }\href {https://doi.org/10.1126/science.1068774} {\bibfield  {journal} {\bibinfo  {journal} {Science}\ }\textbf {\bibinfo {volume} {295}},\ \bibinfo {pages} {2427} (\bibinfo {year} {2002})}\BibitemShut {NoStop}%
\bibitem [{\citenamefont {Lucas}(2014)}]{Lucas2014}%
  \BibitemOpen
  \bibfield  {author} {\bibinfo {author} {\bibfnamefont {A.}~\bibnamefont {Lucas}},\ }\bibfield  {title} {\bibinfo {title} {{I}sing formulations of many {NP} problems},\ }\href {https://doi.org/10.3389/fphy.2014.00005} {\bibfield  {journal} {\bibinfo  {journal} {Front. Phys.}\ }\textbf {\bibinfo {volume} {2}},\ \bibinfo {pages} {5} (\bibinfo {year} {2014})}\BibitemShut {NoStop}%
\bibitem [{\citenamefont {Born}\ and\ \citenamefont {Fock}(1928)}]{Born1928}%
  \BibitemOpen
  \bibfield  {author} {\bibinfo {author} {\bibfnamefont {M.}~\bibnamefont {Born}}\ and\ \bibinfo {author} {\bibfnamefont {V.}~\bibnamefont {Fock}},\ }\bibfield  {title} {\bibinfo {title} {Beweis des adiabatensatzes},\ }\href {https://doi.org/10.1007/BF01343193} {\bibfield  {journal} {\bibinfo  {journal} {Zeitschrift f{\"u}r Physik}\ }\textbf {\bibinfo {volume} {51}},\ \bibinfo {pages} {165} (\bibinfo {year} {1928})}\BibitemShut {NoStop}%
\bibitem [{\citenamefont {Berry}(1984)}]{berry1984quantal}%
  \BibitemOpen
  \bibfield  {author} {\bibinfo {author} {\bibfnamefont {M.~V.}\ \bibnamefont {Berry}},\ }\bibfield  {title} {\bibinfo {title} {Quantal phase factors accompanying adiabatic changes},\ }\href@noop {} {\bibfield  {journal} {\bibinfo  {journal} {Proceedings of the Royal Society of London. A. Mathematical and Physical Sciences}\ }\textbf {\bibinfo {volume} {392}},\ \bibinfo {pages} {45} (\bibinfo {year} {1984})}\BibitemShut {NoStop}%
\bibitem [{\citenamefont {Griffiths}\ and\ \citenamefont {Schroeter}(2018)}]{griffiths_schroeter_2018}%
  \BibitemOpen
  \bibfield  {author} {\bibinfo {author} {\bibfnamefont {D.~J.}\ \bibnamefont {Griffiths}}\ and\ \bibinfo {author} {\bibfnamefont {D.~F.}\ \bibnamefont {Schroeter}},\ }\href {https://doi.org/10.1017/9781316995433} {\emph {\bibinfo {title} {Introduction to Quantum Mechanics}}},\ \bibinfo {edition} {3rd}\ ed.\ (\bibinfo  {publisher} {Cambridge University Press},\ \bibinfo {year} {2018})\BibitemShut {NoStop}%
\bibitem [{\citenamefont {Yan}\ \emph {et~al.}(2023{\natexlab{a}})\citenamefont {Yan}, \citenamefont {Zhou}, \citenamefont {Zhou}, \citenamefont {Wang}, \citenamefont {Qiu}, \citenamefont {Meng},\ and\ \citenamefont {Zhang}}]{yan2022preparing}%
  \BibitemOpen
  \bibfield  {author} {\bibinfo {author} {\bibfnamefont {Z.}~\bibnamefont {Yan}}, \bibinfo {author} {\bibfnamefont {Z.}~\bibnamefont {Zhou}}, \bibinfo {author} {\bibfnamefont {Y.-H.}\ \bibnamefont {Zhou}}, \bibinfo {author} {\bibfnamefont {Y.-C.}\ \bibnamefont {Wang}}, \bibinfo {author} {\bibfnamefont {X.}~\bibnamefont {Qiu}}, \bibinfo {author} {\bibfnamefont {Z.~Y.}\ \bibnamefont {Meng}},\ and\ \bibinfo {author} {\bibfnamefont {X.-F.}\ \bibnamefont {Zhang}},\ }\bibfield  {title} {\bibinfo {title} {Quantum optimization within lattice gauge theory model on a quantum simulator},\ }\href {https://doi.org/10.1038/s41534-023-00755-z} {\bibfield  {journal} {\bibinfo  {journal} {npj Quantum Information}\ }\textbf {\bibinfo {volume} {9}},\ \bibinfo {pages} {89} (\bibinfo {year} {2023}{\natexlab{a}})}\BibitemShut {NoStop}%
\bibitem [{\citenamefont {Kadowaki}\ and\ \citenamefont {Nishimori}(1998)}]{kadowaki1998quantum}%
  \BibitemOpen
  \bibfield  {author} {\bibinfo {author} {\bibfnamefont {T.}~\bibnamefont {Kadowaki}}\ and\ \bibinfo {author} {\bibfnamefont {H.}~\bibnamefont {Nishimori}},\ }\bibfield  {title} {\bibinfo {title} {Quantum annealing in the transverse ising model},\ }\href@noop {} {\bibfield  {journal} {\bibinfo  {journal} {Physical Review E}\ }\textbf {\bibinfo {volume} {58}},\ \bibinfo {pages} {5355} (\bibinfo {year} {1998})}\BibitemShut {NoStop}%
\bibitem [{\citenamefont {Mezard}\ and\ \citenamefont {Montanari}(2009)}]{Montanari2009}%
  \BibitemOpen
  \bibfield  {author} {\bibinfo {author} {\bibfnamefont {M.}~\bibnamefont {Mezard}}\ and\ \bibinfo {author} {\bibfnamefont {A.}~\bibnamefont {Montanari}},\ }\href@noop {} {\emph {\bibinfo {title} {Information, physics, and computation}}}\ (\bibinfo  {publisher} {Oxford University Press},\ \bibinfo {year} {2009})\BibitemShut {NoStop}%
\bibitem [{\citenamefont {Heim}\ \emph {et~al.}(2015)\citenamefont {Heim}, \citenamefont {R{\o}nnow}, \citenamefont {Isakov},\ and\ \citenamefont {Troyer}}]{Heim2015}%
  \BibitemOpen
  \bibfield  {author} {\bibinfo {author} {\bibfnamefont {B.}~\bibnamefont {Heim}}, \bibinfo {author} {\bibfnamefont {T.~F.}\ \bibnamefont {R{\o}nnow}}, \bibinfo {author} {\bibfnamefont {S.~V.}\ \bibnamefont {Isakov}},\ and\ \bibinfo {author} {\bibfnamefont {M.}~\bibnamefont {Troyer}},\ }\bibfield  {title} {\bibinfo {title} {Quantum versus classical annealing of {I}sing spin glasses},\ }\href {https://doi.org/10.1126/science.aaa4170} {\bibfield  {journal} {\bibinfo  {journal} {Science}\ }\textbf {\bibinfo {volume} {348}},\ \bibinfo {pages} {215} (\bibinfo {year} {2015})}\BibitemShut {NoStop}%
\bibitem [{\citenamefont {Farhi}\ \emph {et~al.}(2000)\citenamefont {Farhi}, \citenamefont {Goldstone}, \citenamefont {Gutmann},\ and\ \citenamefont {Sipser}}]{quant-ph/0001106}%
  \BibitemOpen
  \bibfield  {author} {\bibinfo {author} {\bibfnamefont {E.}~\bibnamefont {Farhi}}, \bibinfo {author} {\bibfnamefont {J.}~\bibnamefont {Goldstone}}, \bibinfo {author} {\bibfnamefont {S.}~\bibnamefont {Gutmann}},\ and\ \bibinfo {author} {\bibfnamefont {M.}~\bibnamefont {Sipser}},\ }\href@noop {} {\bibinfo {title} {Quantum computation by adiabatic evolution}} (\bibinfo {year} {2000}),\ \Eprint {https://arxiv.org/abs/arXiv:quant-ph/0001106} {arXiv:quant-ph/0001106} \BibitemShut {NoStop}%
\bibitem [{\citenamefont {Albash}\ and\ \citenamefont {Lidar}(2018)}]{RevModPhys.90.015002}%
  \BibitemOpen
  \bibfield  {author} {\bibinfo {author} {\bibfnamefont {T.}~\bibnamefont {Albash}}\ and\ \bibinfo {author} {\bibfnamefont {D.~A.}\ \bibnamefont {Lidar}},\ }\bibfield  {title} {\bibinfo {title} {Adiabatic quantum computation},\ }\href {https://doi.org/10.1103/RevModPhys.90.015002} {\bibfield  {journal} {\bibinfo  {journal} {Rev. Mod. Phys.}\ }\textbf {\bibinfo {volume} {90}},\ \bibinfo {pages} {015002} (\bibinfo {year} {2018})}\BibitemShut {NoStop}%
\bibitem [{\citenamefont {Johnson}\ \emph {et~al.}(2011)\citenamefont {Johnson}, \citenamefont {Amin}, \citenamefont {Gildert}, \citenamefont {Lanting}, \citenamefont {Hamze}, \citenamefont {Dickson}, \citenamefont {Harris}, \citenamefont {Berkley}, \citenamefont {Johansson}, \citenamefont {Bunyk} \emph {et~al.}}]{Gildert2011}%
  \BibitemOpen
  \bibfield  {author} {\bibinfo {author} {\bibfnamefont {M.~W.}\ \bibnamefont {Johnson}}, \bibinfo {author} {\bibfnamefont {M.~H.}\ \bibnamefont {Amin}}, \bibinfo {author} {\bibfnamefont {S.}~\bibnamefont {Gildert}}, \bibinfo {author} {\bibfnamefont {T.}~\bibnamefont {Lanting}}, \bibinfo {author} {\bibfnamefont {F.}~\bibnamefont {Hamze}}, \bibinfo {author} {\bibfnamefont {N.}~\bibnamefont {Dickson}}, \bibinfo {author} {\bibfnamefont {R.}~\bibnamefont {Harris}}, \bibinfo {author} {\bibfnamefont {A.~J.}\ \bibnamefont {Berkley}}, \bibinfo {author} {\bibfnamefont {J.}~\bibnamefont {Johansson}}, \bibinfo {author} {\bibfnamefont {P.}~\bibnamefont {Bunyk}}, \emph {et~al.},\ }\bibfield  {title} {\bibinfo {title} {Quantum annealing with manufactured spins},\ }\href {https://doi.org/10.1038/nature10012} {\bibfield  {journal} {\bibinfo  {journal} {Nature}\ }\textbf {\bibinfo {volume} {473}},\ \bibinfo {pages} {194} (\bibinfo {year} {2011})}\BibitemShut {NoStop}%
\bibitem [{\citenamefont {Boixo}\ \emph {et~al.}(2013)\citenamefont {Boixo}, \citenamefont {Albash}, \citenamefont {Spedalieri}, \citenamefont {Chancellor},\ and\ \citenamefont {Lidar}}]{Boixo2013}%
  \BibitemOpen
  \bibfield  {author} {\bibinfo {author} {\bibfnamefont {S.}~\bibnamefont {Boixo}}, \bibinfo {author} {\bibfnamefont {T.}~\bibnamefont {Albash}}, \bibinfo {author} {\bibfnamefont {F.~M.}\ \bibnamefont {Spedalieri}}, \bibinfo {author} {\bibfnamefont {N.}~\bibnamefont {Chancellor}},\ and\ \bibinfo {author} {\bibfnamefont {D.~A.}\ \bibnamefont {Lidar}},\ }\bibfield  {title} {\bibinfo {title} {Experimental signature of programmable quantum annealing},\ }\href {https://doi.org/10.1038/ncomms3067} {\bibfield  {journal} {\bibinfo  {journal} {Nat. Commun.}\ }\textbf {\bibinfo {volume} {4}},\ \bibinfo {pages} {1} (\bibinfo {year} {2013})}\BibitemShut {NoStop}%
\bibitem [{\citenamefont {Boixo}\ \emph {et~al.}(2014)\citenamefont {Boixo}, \citenamefont {R{\o}nnow}, \citenamefont {Isakov}, \citenamefont {Wang}, \citenamefont {Wecker}, \citenamefont {Lidar}, \citenamefont {Martinis},\ and\ \citenamefont {Troyer}}]{Boixo2014}%
  \BibitemOpen
  \bibfield  {author} {\bibinfo {author} {\bibfnamefont {S.}~\bibnamefont {Boixo}}, \bibinfo {author} {\bibfnamefont {T.~F.}\ \bibnamefont {R{\o}nnow}}, \bibinfo {author} {\bibfnamefont {S.~V.}\ \bibnamefont {Isakov}}, \bibinfo {author} {\bibfnamefont {Z.}~\bibnamefont {Wang}}, \bibinfo {author} {\bibfnamefont {D.}~\bibnamefont {Wecker}}, \bibinfo {author} {\bibfnamefont {D.~A.}\ \bibnamefont {Lidar}}, \bibinfo {author} {\bibfnamefont {J.~M.}\ \bibnamefont {Martinis}},\ and\ \bibinfo {author} {\bibfnamefont {M.}~\bibnamefont {Troyer}},\ }\bibfield  {title} {\bibinfo {title} {Evidence for quantum annealing with more than one hundred qubits},\ }\href {https://doi.org/10.1038/nphys2900} {\bibfield  {journal} {\bibinfo  {journal} {Nature Phys.}\ }\textbf {\bibinfo {volume} {10}},\ \bibinfo {pages} {218} (\bibinfo {year} {2014})}\BibitemShut {NoStop}%
\bibitem [{\citenamefont {Perdomo-Ortiz}\ \emph {et~al.}(2012)\citenamefont {Perdomo-Ortiz}, \citenamefont {Dickson}, \citenamefont {Drew-Brook}, \citenamefont {Rose},\ and\ \citenamefont {Aspuru-Guzik}}]{perdomo2012finding}%
  \BibitemOpen
  \bibfield  {author} {\bibinfo {author} {\bibfnamefont {A.}~\bibnamefont {Perdomo-Ortiz}}, \bibinfo {author} {\bibfnamefont {N.}~\bibnamefont {Dickson}}, \bibinfo {author} {\bibfnamefont {M.}~\bibnamefont {Drew-Brook}}, \bibinfo {author} {\bibfnamefont {G.}~\bibnamefont {Rose}},\ and\ \bibinfo {author} {\bibfnamefont {A.}~\bibnamefont {Aspuru-Guzik}},\ }\bibfield  {title} {\bibinfo {title} {Finding low-energy conformations of lattice protein models by quantum annealing},\ }\href {https://doi.org/10.1038/srep00571} {\bibfield  {journal} {\bibinfo  {journal} {Sci. Rep.}\ }\textbf {\bibinfo {volume} {2}},\ \bibinfo {pages} {571} (\bibinfo {year} {2012})}\BibitemShut {NoStop}%
\bibitem [{\citenamefont {Novotny}\ \emph {et~al.}(2016)\citenamefont {Novotny}, \citenamefont {Hobl}, \citenamefont {Hall},\ and\ \citenamefont {Michielsen}}]{novotny2016spanning}%
  \BibitemOpen
  \bibfield  {author} {\bibinfo {author} {\bibfnamefont {M.}~\bibnamefont {Novotny}}, \bibinfo {author} {\bibfnamefont {Q.~L.}\ \bibnamefont {Hobl}}, \bibinfo {author} {\bibfnamefont {J.}~\bibnamefont {Hall}},\ and\ \bibinfo {author} {\bibfnamefont {K.}~\bibnamefont {Michielsen}},\ }\bibfield  {title} {\bibinfo {title} {Spanning tree calculations on d-wave 2 machines},\ }\href {https://doi.org/10.1088/1742-6596/681/1/012005} {\bibfield  {journal} {\bibinfo  {journal} {J. Phys.: Conf. Ser.}\ }\textbf {\bibinfo {volume} {681}},\ \bibinfo {pages} {012005} (\bibinfo {year} {2016})}\BibitemShut {NoStop}%
\bibitem [{\citenamefont {Qiu}\ \emph {et~al.}(2020{\natexlab{a}})\citenamefont {Qiu}, \citenamefont {Zou}, \citenamefont {Qi},\ and\ \citenamefont {Li}}]{qiu2020precise}%
  \BibitemOpen
  \bibfield  {author} {\bibinfo {author} {\bibfnamefont {X.}~\bibnamefont {Qiu}}, \bibinfo {author} {\bibfnamefont {J.}~\bibnamefont {Zou}}, \bibinfo {author} {\bibfnamefont {X.}~\bibnamefont {Qi}},\ and\ \bibinfo {author} {\bibfnamefont {X.}~\bibnamefont {Li}},\ }\bibfield  {title} {\bibinfo {title} {Precise programmable quantum simulations with optical lattices},\ }\href {https://doi.org/10.1038/s41534-020-00315-9} {\bibfield  {journal} {\bibinfo  {journal} {npj Quantum Inf.}\ }\textbf {\bibinfo {volume} {6}},\ \bibinfo {pages} {87} (\bibinfo {year} {2020}{\natexlab{a}})}\BibitemShut {NoStop}%
\bibitem [{\citenamefont {Qiu}\ \emph {et~al.}(2020{\natexlab{b}})\citenamefont {Qiu}, \citenamefont {Zoller},\ and\ \citenamefont {Li}}]{qiu2020programmable}%
  \BibitemOpen
  \bibfield  {author} {\bibinfo {author} {\bibfnamefont {X.}~\bibnamefont {Qiu}}, \bibinfo {author} {\bibfnamefont {P.}~\bibnamefont {Zoller}},\ and\ \bibinfo {author} {\bibfnamefont {X.}~\bibnamefont {Li}},\ }\bibfield  {title} {\bibinfo {title} {Programmable quantum annealing architectures with {I}sing quantum wires},\ }\href {https://doi.org/10.1103/PRXQuantum.1.020311} {\bibfield  {journal} {\bibinfo  {journal} {PRX Quantum}\ }\textbf {\bibinfo {volume} {1}},\ \bibinfo {pages} {020311} (\bibinfo {year} {2020}{\natexlab{b}})}\BibitemShut {NoStop}%
\bibitem [{\citenamefont {King}\ \emph {et~al.}(2023)\citenamefont {King}, \citenamefont {Raymond}, \citenamefont {Lanting}, \citenamefont {Harris}, \citenamefont {Zucca}, \citenamefont {Altomare}, \citenamefont {Berkley}, \citenamefont {Boothby}, \citenamefont {Ejtemaee}, \citenamefont {Enderud} \emph {et~al.}}]{king2023quantum}%
  \BibitemOpen
  \bibfield  {author} {\bibinfo {author} {\bibfnamefont {A.~D.}\ \bibnamefont {King}}, \bibinfo {author} {\bibfnamefont {J.}~\bibnamefont {Raymond}}, \bibinfo {author} {\bibfnamefont {T.}~\bibnamefont {Lanting}}, \bibinfo {author} {\bibfnamefont {R.}~\bibnamefont {Harris}}, \bibinfo {author} {\bibfnamefont {A.}~\bibnamefont {Zucca}}, \bibinfo {author} {\bibfnamefont {F.}~\bibnamefont {Altomare}}, \bibinfo {author} {\bibfnamefont {A.~J.}\ \bibnamefont {Berkley}}, \bibinfo {author} {\bibfnamefont {K.}~\bibnamefont {Boothby}}, \bibinfo {author} {\bibfnamefont {S.}~\bibnamefont {Ejtemaee}}, \bibinfo {author} {\bibfnamefont {C.}~\bibnamefont {Enderud}}, \emph {et~al.},\ }\bibfield  {title} {\bibinfo {title} {Quantum critical dynamics in a 5,000-qubit programmable spin glass},\ }\href@noop {} {\bibfield  {journal} {\bibinfo  {journal} {Nature}\ ,\ \bibinfo {pages} {1}} (\bibinfo {year} {2023})}\BibitemShut {NoStop}%
\bibitem [{\citenamefont {{Satzinger}}\ \emph {et~al.}(2021)\citenamefont {{Satzinger}}, \citenamefont {{Liu}}, \citenamefont {{Smith}}, \citenamefont {{Knapp}}, \citenamefont {{Newman}}, \citenamefont {{Jones}}, \citenamefont {{Chen}}, \citenamefont {{Quintana}}, \citenamefont {{Mi}}, \citenamefont {{Dunsworth}}, \citenamefont {{Gidney}}, \citenamefont {{Aleiner}}, \citenamefont {{Arute}}, \citenamefont {{Arya}}, \citenamefont {{Atalaya}}, \citenamefont {{Babbush}}, \citenamefont {{Bardin}}, \citenamefont {{Barends}}, \citenamefont {{Basso}}, \citenamefont {{Bengtsson}}, \citenamefont {{Bilmes}}, \citenamefont {{Broughton}}, \citenamefont {{Buckley}}, \citenamefont {{Buell}}, \citenamefont {{Burkett}}, \citenamefont {{Bushnell}}, \citenamefont {{Chiaro}}, \citenamefont {{Collins}}, \citenamefont {{Courtney}}, \citenamefont {{Demura}}, \citenamefont {{Derk}}, \citenamefont {{Eppens}}, \citenamefont {{Erickson}}, \citenamefont {{Faoro}}, \citenamefont {{Farhi}}, \citenamefont {{Fowler}}, \citenamefont
  {{Foxen}}, \citenamefont {{Giustina}}, \citenamefont {{Greene}}, \citenamefont {{Gross}}, \citenamefont {{Harrigan}}, \citenamefont {{Harrington}}, \citenamefont {{Hilton}}, \citenamefont {{Hong}}, \citenamefont {{Huang}}, \citenamefont {{Huggins}}, \citenamefont {{Ioffe}}, \citenamefont {{Isakov}}, \citenamefont {{Jeffrey}}, \citenamefont {{Jiang}}, \citenamefont {{Kafri}}, \citenamefont {{Kechedzhi}}, \citenamefont {{Khattar}}, \citenamefont {{Kim}}, \citenamefont {{Klimov}}, \citenamefont {{Korotkov}}, \citenamefont {{Kostritsa}}, \citenamefont {{Landhuis}}, \citenamefont {{Laptev}}, \citenamefont {{Locharla}}, \citenamefont {{Lucero}}, \citenamefont {{Martin}}, \citenamefont {{McClean}}, \citenamefont {{McEwen}}, \citenamefont {{Miao}}, \citenamefont {{Mohseni}}, \citenamefont {{Montazeri}}, \citenamefont {{Mruczkiewicz}}, \citenamefont {{Mutus}}, \citenamefont {{Naaman}}, \citenamefont {{Neeley}}, \citenamefont {{Neill}}, \citenamefont {{Niu}}, \citenamefont {{O{\textquoteright}Brien}}, \citenamefont
  {{Opremcak}}, \citenamefont {{Pat{\'o}}}, \citenamefont {{Petukhov}}, \citenamefont {{Rubin}}, \citenamefont {{Sank}}, \citenamefont {{Shvarts}}, \citenamefont {{Strain}}, \citenamefont {{Szalay}}, \citenamefont {{Villalonga}}, \citenamefont {{White}}, \citenamefont {{Yao}}, \citenamefont {{Yeh}}, \citenamefont {{Yoo}}, \citenamefont {{Zalcman}}, \citenamefont {{Neven}}, \citenamefont {{Boixo}}, \citenamefont {{Megrant}}, \citenamefont {{Chen}}, \citenamefont {{Kelly}}, \citenamefont {{Smelyanskiy}}, \citenamefont {{Kitaev}}, \citenamefont {{Knap}}, \citenamefont {{Pollmann}},\ and\ \citenamefont {{Roushan}}}]{Roushan21}%
  \BibitemOpen
  \bibfield  {author} {\bibinfo {author} {\bibfnamefont {K.~J.}\ \bibnamefont {{Satzinger}}}, \bibinfo {author} {\bibfnamefont {Y.~J.}\ \bibnamefont {{Liu}}}, \bibinfo {author} {\bibfnamefont {A.}~\bibnamefont {{Smith}}}, \bibinfo {author} {\bibfnamefont {C.}~\bibnamefont {{Knapp}}}, \bibinfo {author} {\bibfnamefont {M.}~\bibnamefont {{Newman}}}, \bibinfo {author} {\bibfnamefont {C.}~\bibnamefont {{Jones}}}, \bibinfo {author} {\bibfnamefont {Z.}~\bibnamefont {{Chen}}}, \bibinfo {author} {\bibfnamefont {C.}~\bibnamefont {{Quintana}}}, \bibinfo {author} {\bibfnamefont {X.}~\bibnamefont {{Mi}}}, \bibinfo {author} {\bibfnamefont {A.}~\bibnamefont {{Dunsworth}}}, \bibinfo {author} {\bibfnamefont {C.}~\bibnamefont {{Gidney}}}, \bibinfo {author} {\bibfnamefont {I.}~\bibnamefont {{Aleiner}}}, \bibinfo {author} {\bibfnamefont {F.}~\bibnamefont {{Arute}}}, \bibinfo {author} {\bibfnamefont {K.}~\bibnamefont {{Arya}}}, \bibinfo {author} {\bibfnamefont {J.}~\bibnamefont {{Atalaya}}}, \bibinfo {author} {\bibfnamefont
  {R.}~\bibnamefont {{Babbush}}}, \bibinfo {author} {\bibfnamefont {J.~C.}\ \bibnamefont {{Bardin}}}, \bibinfo {author} {\bibfnamefont {R.}~\bibnamefont {{Barends}}}, \bibinfo {author} {\bibfnamefont {J.}~\bibnamefont {{Basso}}}, \bibinfo {author} {\bibfnamefont {A.}~\bibnamefont {{Bengtsson}}}, \bibinfo {author} {\bibfnamefont {A.}~\bibnamefont {{Bilmes}}}, \bibinfo {author} {\bibfnamefont {M.}~\bibnamefont {{Broughton}}}, \bibinfo {author} {\bibfnamefont {B.~B.}\ \bibnamefont {{Buckley}}}, \bibinfo {author} {\bibfnamefont {D.~A.}\ \bibnamefont {{Buell}}}, \bibinfo {author} {\bibfnamefont {B.}~\bibnamefont {{Burkett}}}, \bibinfo {author} {\bibfnamefont {N.}~\bibnamefont {{Bushnell}}}, \bibinfo {author} {\bibfnamefont {B.}~\bibnamefont {{Chiaro}}}, \bibinfo {author} {\bibfnamefont {R.}~\bibnamefont {{Collins}}}, \bibinfo {author} {\bibfnamefont {W.}~\bibnamefont {{Courtney}}}, \bibinfo {author} {\bibfnamefont {S.}~\bibnamefont {{Demura}}}, \bibinfo {author} {\bibfnamefont {A.~R.}\ \bibnamefont {{Derk}}},
  \bibinfo {author} {\bibfnamefont {D.}~\bibnamefont {{Eppens}}}, \bibinfo {author} {\bibfnamefont {C.}~\bibnamefont {{Erickson}}}, \bibinfo {author} {\bibfnamefont {L.}~\bibnamefont {{Faoro}}}, \bibinfo {author} {\bibfnamefont {E.}~\bibnamefont {{Farhi}}}, \bibinfo {author} {\bibfnamefont {A.~G.}\ \bibnamefont {{Fowler}}}, \bibinfo {author} {\bibfnamefont {B.}~\bibnamefont {{Foxen}}}, \bibinfo {author} {\bibfnamefont {M.}~\bibnamefont {{Giustina}}}, \bibinfo {author} {\bibfnamefont {A.}~\bibnamefont {{Greene}}}, \bibinfo {author} {\bibfnamefont {J.~A.}\ \bibnamefont {{Gross}}}, \bibinfo {author} {\bibfnamefont {M.~P.}\ \bibnamefont {{Harrigan}}}, \bibinfo {author} {\bibfnamefont {S.~D.}\ \bibnamefont {{Harrington}}}, \bibinfo {author} {\bibfnamefont {J.}~\bibnamefont {{Hilton}}}, \bibinfo {author} {\bibfnamefont {S.}~\bibnamefont {{Hong}}}, \bibinfo {author} {\bibfnamefont {T.}~\bibnamefont {{Huang}}}, \bibinfo {author} {\bibfnamefont {W.~J.}\ \bibnamefont {{Huggins}}}, \bibinfo {author} {\bibfnamefont
  {L.~B.}\ \bibnamefont {{Ioffe}}}, \bibinfo {author} {\bibfnamefont {S.~V.}\ \bibnamefont {{Isakov}}}, \bibinfo {author} {\bibfnamefont {E.}~\bibnamefont {{Jeffrey}}}, \bibinfo {author} {\bibfnamefont {Z.}~\bibnamefont {{Jiang}}}, \bibinfo {author} {\bibfnamefont {D.}~\bibnamefont {{Kafri}}}, \bibinfo {author} {\bibfnamefont {K.}~\bibnamefont {{Kechedzhi}}}, \bibinfo {author} {\bibfnamefont {T.}~\bibnamefont {{Khattar}}}, \bibinfo {author} {\bibfnamefont {S.}~\bibnamefont {{Kim}}}, \bibinfo {author} {\bibfnamefont {P.~V.}\ \bibnamefont {{Klimov}}}, \bibinfo {author} {\bibfnamefont {A.~N.}\ \bibnamefont {{Korotkov}}}, \bibinfo {author} {\bibfnamefont {F.}~\bibnamefont {{Kostritsa}}}, \bibinfo {author} {\bibfnamefont {D.}~\bibnamefont {{Landhuis}}}, \bibinfo {author} {\bibfnamefont {P.}~\bibnamefont {{Laptev}}}, \bibinfo {author} {\bibfnamefont {A.}~\bibnamefont {{Locharla}}}, \bibinfo {author} {\bibfnamefont {E.}~\bibnamefont {{Lucero}}}, \bibinfo {author} {\bibfnamefont {O.}~\bibnamefont {{Martin}}},
  \bibinfo {author} {\bibfnamefont {J.~R.}\ \bibnamefont {{McClean}}}, \bibinfo {author} {\bibfnamefont {M.}~\bibnamefont {{McEwen}}}, \bibinfo {author} {\bibfnamefont {K.~C.}\ \bibnamefont {{Miao}}}, \bibinfo {author} {\bibfnamefont {M.}~\bibnamefont {{Mohseni}}}, \bibinfo {author} {\bibfnamefont {S.}~\bibnamefont {{Montazeri}}}, \bibinfo {author} {\bibfnamefont {W.}~\bibnamefont {{Mruczkiewicz}}}, \bibinfo {author} {\bibfnamefont {J.}~\bibnamefont {{Mutus}}}, \bibinfo {author} {\bibfnamefont {O.}~\bibnamefont {{Naaman}}}, \bibinfo {author} {\bibfnamefont {M.}~\bibnamefont {{Neeley}}}, \bibinfo {author} {\bibfnamefont {C.}~\bibnamefont {{Neill}}}, \bibinfo {author} {\bibfnamefont {M.~Y.}\ \bibnamefont {{Niu}}}, \bibinfo {author} {\bibfnamefont {T.~E.}\ \bibnamefont {{O{\textquoteright}Brien}}}, \bibinfo {author} {\bibfnamefont {A.}~\bibnamefont {{Opremcak}}}, \bibinfo {author} {\bibfnamefont {B.}~\bibnamefont {{Pat{\'o}}}}, \bibinfo {author} {\bibfnamefont {A.}~\bibnamefont {{Petukhov}}}, \bibinfo {author}
  {\bibfnamefont {N.~C.}\ \bibnamefont {{Rubin}}}, \bibinfo {author} {\bibfnamefont {D.}~\bibnamefont {{Sank}}}, \bibinfo {author} {\bibfnamefont {V.}~\bibnamefont {{Shvarts}}}, \bibinfo {author} {\bibfnamefont {D.}~\bibnamefont {{Strain}}}, \bibinfo {author} {\bibfnamefont {M.}~\bibnamefont {{Szalay}}}, \bibinfo {author} {\bibfnamefont {B.}~\bibnamefont {{Villalonga}}}, \bibinfo {author} {\bibfnamefont {T.~C.}\ \bibnamefont {{White}}}, \bibinfo {author} {\bibfnamefont {Z.}~\bibnamefont {{Yao}}}, \bibinfo {author} {\bibfnamefont {P.}~\bibnamefont {{Yeh}}}, \bibinfo {author} {\bibfnamefont {J.}~\bibnamefont {{Yoo}}}, \bibinfo {author} {\bibfnamefont {A.}~\bibnamefont {{Zalcman}}}, \bibinfo {author} {\bibfnamefont {H.}~\bibnamefont {{Neven}}}, \bibinfo {author} {\bibfnamefont {S.}~\bibnamefont {{Boixo}}}, \bibinfo {author} {\bibfnamefont {A.}~\bibnamefont {{Megrant}}}, \bibinfo {author} {\bibfnamefont {Y.}~\bibnamefont {{Chen}}}, \bibinfo {author} {\bibfnamefont {J.}~\bibnamefont {{Kelly}}}, \bibinfo {author}
  {\bibfnamefont {V.}~\bibnamefont {{Smelyanskiy}}}, \bibinfo {author} {\bibfnamefont {A.}~\bibnamefont {{Kitaev}}}, \bibinfo {author} {\bibfnamefont {M.}~\bibnamefont {{Knap}}}, \bibinfo {author} {\bibfnamefont {F.}~\bibnamefont {{Pollmann}}},\ and\ \bibinfo {author} {\bibfnamefont {P.}~\bibnamefont {{Roushan}}},\ }\bibfield  {title} {\bibinfo {title} {{Realizing topologically ordered states on a quantum processor}},\ }\href {https://doi.org/10.1126/science.abi8378} {\bibfield  {journal} {\bibinfo  {journal} {Science}\ }\textbf {\bibinfo {volume} {374}},\ \bibinfo {pages} {1237} (\bibinfo {year} {2021})}\BibitemShut {NoStop}%
\bibitem [{\citenamefont {{Semeghini}}\ \emph {et~al.}(2021)\citenamefont {{Semeghini}}, \citenamefont {{Levine}}, \citenamefont {{Keesling}}, \citenamefont {{Ebadi}}, \citenamefont {{Wang}}, \citenamefont {{Bluvstein}}, \citenamefont {{Verresen}}, \citenamefont {{Pichler}}, \citenamefont {{Kalinowski}}, \citenamefont {{Samajdar}}, \citenamefont {{Omran}}, \citenamefont {{Sachdev}}, \citenamefont {{Vishwanath}}, \citenamefont {{Greiner}}, \citenamefont {{Vuleti{\'c}}},\ and\ \citenamefont {{Lukin}}}]{Semeghini21}%
  \BibitemOpen
  \bibfield  {author} {\bibinfo {author} {\bibfnamefont {G.}~\bibnamefont {{Semeghini}}}, \bibinfo {author} {\bibfnamefont {H.}~\bibnamefont {{Levine}}}, \bibinfo {author} {\bibfnamefont {A.}~\bibnamefont {{Keesling}}}, \bibinfo {author} {\bibfnamefont {S.}~\bibnamefont {{Ebadi}}}, \bibinfo {author} {\bibfnamefont {T.~T.}\ \bibnamefont {{Wang}}}, \bibinfo {author} {\bibfnamefont {D.}~\bibnamefont {{Bluvstein}}}, \bibinfo {author} {\bibfnamefont {R.}~\bibnamefont {{Verresen}}}, \bibinfo {author} {\bibfnamefont {H.}~\bibnamefont {{Pichler}}}, \bibinfo {author} {\bibfnamefont {M.}~\bibnamefont {{Kalinowski}}}, \bibinfo {author} {\bibfnamefont {R.}~\bibnamefont {{Samajdar}}}, \bibinfo {author} {\bibfnamefont {A.}~\bibnamefont {{Omran}}}, \bibinfo {author} {\bibfnamefont {S.}~\bibnamefont {{Sachdev}}}, \bibinfo {author} {\bibfnamefont {A.}~\bibnamefont {{Vishwanath}}}, \bibinfo {author} {\bibfnamefont {M.}~\bibnamefont {{Greiner}}}, \bibinfo {author} {\bibfnamefont {V.}~\bibnamefont {{Vuleti{\'c}}}},\ and\
  \bibinfo {author} {\bibfnamefont {M.~D.}\ \bibnamefont {{Lukin}}},\ }\bibfield  {title} {\bibinfo {title} {{Probing topological spin liquids on a programmable quantum simulator}},\ }\href {https://doi.org/10.1126/science.abi8794} {\bibfield  {journal} {\bibinfo  {journal} {Science}\ }\textbf {\bibinfo {volume} {374}},\ \bibinfo {pages} {1242} (\bibinfo {year} {2021})}\BibitemShut {NoStop}%
\bibitem [{\citenamefont {Satzinger}\ \emph {et~al.}(2021)\citenamefont {Satzinger}, \citenamefont {Liu}, \citenamefont {Smith}, \citenamefont {Knapp}, \citenamefont {Newman}, \citenamefont {Jones}, \citenamefont {Chen}, \citenamefont {Quintana}, \citenamefont {Mi}, \citenamefont {Dunsworth}, \citenamefont {Gidney}, \citenamefont {Aleiner}, \citenamefont {Arute}, \citenamefont {Arya}, \citenamefont {Atalaya}, \citenamefont {Babbush}, \citenamefont {Bardin}, \citenamefont {Barends}, \citenamefont {Basso}, \citenamefont {Bengtsson}, \citenamefont {Bilmes}, \citenamefont {Broughton}, \citenamefont {Buckley}, \citenamefont {Buell}, \citenamefont {Burkett}, \citenamefont {Bushnell}, \citenamefont {Chiaro}, \citenamefont {Collins}, \citenamefont {Courtney}, \citenamefont {Demura}, \citenamefont {Derk}, \citenamefont {Eppens}, \citenamefont {Erickson}, \citenamefont {Faoro}, \citenamefont {Farhi}, \citenamefont {Fowler}, \citenamefont {Foxen}, \citenamefont {Giustina}, \citenamefont {Greene}, \citenamefont {Gross},
  \citenamefont {Harrigan}, \citenamefont {Harrington}, \citenamefont {Hilton}, \citenamefont {Hong}, \citenamefont {Huang}, \citenamefont {Huggins}, \citenamefont {Ioffe}, \citenamefont {Isakov}, \citenamefont {Jeffrey}, \citenamefont {Jiang}, \citenamefont {Kafri}, \citenamefont {Kechedzhi}, \citenamefont {Khattar}, \citenamefont {Kim}, \citenamefont {Klimov}, \citenamefont {Korotkov}, \citenamefont {Kostritsa}, \citenamefont {Landhuis}, \citenamefont {Laptev}, \citenamefont {Locharla}, \citenamefont {Lucero}, \citenamefont {Martin}, \citenamefont {McClean}, \citenamefont {McEwen}, \citenamefont {Miao}, \citenamefont {Mohseni}, \citenamefont {Montazeri}, \citenamefont {Mruczkiewicz}, \citenamefont {Mutus}, \citenamefont {Naaman}, \citenamefont {Neeley}, \citenamefont {Neill}, \citenamefont {Niu}, \citenamefont {O’Brien}, \citenamefont {Opremcak}, \citenamefont {Pató}, \citenamefont {Petukhov}, \citenamefont {Rubin}, \citenamefont {Sank}, \citenamefont {Shvarts}, \citenamefont {Strain}, \citenamefont
  {Szalay}, \citenamefont {Villalonga}, \citenamefont {White}, \citenamefont {Yao}, \citenamefont {Yeh}, \citenamefont {Yoo}, \citenamefont {Zalcman}, \citenamefont {Neven}, \citenamefont {Boixo}, \citenamefont {Megrant}, \citenamefont {Chen}, \citenamefont {Kelly}, \citenamefont {Smelyanskiy}, \citenamefont {Kitaev}, \citenamefont {Knap}, \citenamefont {Pollmann},\ and\ \citenamefont {Roushan}}]{doi:10.1126/science.abi8378}%
  \BibitemOpen
  \bibfield  {author} {\bibinfo {author} {\bibfnamefont {K.~J.}\ \bibnamefont {Satzinger}}, \bibinfo {author} {\bibfnamefont {Y.-J.}\ \bibnamefont {Liu}}, \bibinfo {author} {\bibfnamefont {A.}~\bibnamefont {Smith}}, \bibinfo {author} {\bibfnamefont {C.}~\bibnamefont {Knapp}}, \bibinfo {author} {\bibfnamefont {M.}~\bibnamefont {Newman}}, \bibinfo {author} {\bibfnamefont {C.}~\bibnamefont {Jones}}, \bibinfo {author} {\bibfnamefont {Z.}~\bibnamefont {Chen}}, \bibinfo {author} {\bibfnamefont {C.}~\bibnamefont {Quintana}}, \bibinfo {author} {\bibfnamefont {X.}~\bibnamefont {Mi}}, \bibinfo {author} {\bibfnamefont {A.}~\bibnamefont {Dunsworth}}, \bibinfo {author} {\bibfnamefont {C.}~\bibnamefont {Gidney}}, \bibinfo {author} {\bibfnamefont {I.}~\bibnamefont {Aleiner}}, \bibinfo {author} {\bibfnamefont {F.}~\bibnamefont {Arute}}, \bibinfo {author} {\bibfnamefont {K.}~\bibnamefont {Arya}}, \bibinfo {author} {\bibfnamefont {J.}~\bibnamefont {Atalaya}}, \bibinfo {author} {\bibfnamefont {R.}~\bibnamefont {Babbush}},
  \bibinfo {author} {\bibfnamefont {J.~C.}\ \bibnamefont {Bardin}}, \bibinfo {author} {\bibfnamefont {R.}~\bibnamefont {Barends}}, \bibinfo {author} {\bibfnamefont {J.}~\bibnamefont {Basso}}, \bibinfo {author} {\bibfnamefont {A.}~\bibnamefont {Bengtsson}}, \bibinfo {author} {\bibfnamefont {A.}~\bibnamefont {Bilmes}}, \bibinfo {author} {\bibfnamefont {M.}~\bibnamefont {Broughton}}, \bibinfo {author} {\bibfnamefont {B.~B.}\ \bibnamefont {Buckley}}, \bibinfo {author} {\bibfnamefont {D.~A.}\ \bibnamefont {Buell}}, \bibinfo {author} {\bibfnamefont {B.}~\bibnamefont {Burkett}}, \bibinfo {author} {\bibfnamefont {N.}~\bibnamefont {Bushnell}}, \bibinfo {author} {\bibfnamefont {B.}~\bibnamefont {Chiaro}}, \bibinfo {author} {\bibfnamefont {R.}~\bibnamefont {Collins}}, \bibinfo {author} {\bibfnamefont {W.}~\bibnamefont {Courtney}}, \bibinfo {author} {\bibfnamefont {S.}~\bibnamefont {Demura}}, \bibinfo {author} {\bibfnamefont {A.~R.}\ \bibnamefont {Derk}}, \bibinfo {author} {\bibfnamefont {D.}~\bibnamefont {Eppens}},
  \bibinfo {author} {\bibfnamefont {C.}~\bibnamefont {Erickson}}, \bibinfo {author} {\bibfnamefont {L.}~\bibnamefont {Faoro}}, \bibinfo {author} {\bibfnamefont {E.}~\bibnamefont {Farhi}}, \bibinfo {author} {\bibfnamefont {A.~G.}\ \bibnamefont {Fowler}}, \bibinfo {author} {\bibfnamefont {B.}~\bibnamefont {Foxen}}, \bibinfo {author} {\bibfnamefont {M.}~\bibnamefont {Giustina}}, \bibinfo {author} {\bibfnamefont {A.}~\bibnamefont {Greene}}, \bibinfo {author} {\bibfnamefont {J.~A.}\ \bibnamefont {Gross}}, \bibinfo {author} {\bibfnamefont {M.~P.}\ \bibnamefont {Harrigan}}, \bibinfo {author} {\bibfnamefont {S.~D.}\ \bibnamefont {Harrington}}, \bibinfo {author} {\bibfnamefont {J.}~\bibnamefont {Hilton}}, \bibinfo {author} {\bibfnamefont {S.}~\bibnamefont {Hong}}, \bibinfo {author} {\bibfnamefont {T.}~\bibnamefont {Huang}}, \bibinfo {author} {\bibfnamefont {W.~J.}\ \bibnamefont {Huggins}}, \bibinfo {author} {\bibfnamefont {L.~B.}\ \bibnamefont {Ioffe}}, \bibinfo {author} {\bibfnamefont {S.~V.}\ \bibnamefont {Isakov}},
  \bibinfo {author} {\bibfnamefont {E.}~\bibnamefont {Jeffrey}}, \bibinfo {author} {\bibfnamefont {Z.}~\bibnamefont {Jiang}}, \bibinfo {author} {\bibfnamefont {D.}~\bibnamefont {Kafri}}, \bibinfo {author} {\bibfnamefont {K.}~\bibnamefont {Kechedzhi}}, \bibinfo {author} {\bibfnamefont {T.}~\bibnamefont {Khattar}}, \bibinfo {author} {\bibfnamefont {S.}~\bibnamefont {Kim}}, \bibinfo {author} {\bibfnamefont {P.~V.}\ \bibnamefont {Klimov}}, \bibinfo {author} {\bibfnamefont {A.~N.}\ \bibnamefont {Korotkov}}, \bibinfo {author} {\bibfnamefont {F.}~\bibnamefont {Kostritsa}}, \bibinfo {author} {\bibfnamefont {D.}~\bibnamefont {Landhuis}}, \bibinfo {author} {\bibfnamefont {P.}~\bibnamefont {Laptev}}, \bibinfo {author} {\bibfnamefont {A.}~\bibnamefont {Locharla}}, \bibinfo {author} {\bibfnamefont {E.}~\bibnamefont {Lucero}}, \bibinfo {author} {\bibfnamefont {O.}~\bibnamefont {Martin}}, \bibinfo {author} {\bibfnamefont {J.~R.}\ \bibnamefont {McClean}}, \bibinfo {author} {\bibfnamefont {M.}~\bibnamefont {McEwen}}, \bibinfo
  {author} {\bibfnamefont {K.~C.}\ \bibnamefont {Miao}}, \bibinfo {author} {\bibfnamefont {M.}~\bibnamefont {Mohseni}}, \bibinfo {author} {\bibfnamefont {S.}~\bibnamefont {Montazeri}}, \bibinfo {author} {\bibfnamefont {W.}~\bibnamefont {Mruczkiewicz}}, \bibinfo {author} {\bibfnamefont {J.}~\bibnamefont {Mutus}}, \bibinfo {author} {\bibfnamefont {O.}~\bibnamefont {Naaman}}, \bibinfo {author} {\bibfnamefont {M.}~\bibnamefont {Neeley}}, \bibinfo {author} {\bibfnamefont {C.}~\bibnamefont {Neill}}, \bibinfo {author} {\bibfnamefont {M.~Y.}\ \bibnamefont {Niu}}, \bibinfo {author} {\bibfnamefont {T.~E.}\ \bibnamefont {O’Brien}}, \bibinfo {author} {\bibfnamefont {A.}~\bibnamefont {Opremcak}}, \bibinfo {author} {\bibfnamefont {B.}~\bibnamefont {Pató}}, \bibinfo {author} {\bibfnamefont {A.}~\bibnamefont {Petukhov}}, \bibinfo {author} {\bibfnamefont {N.~C.}\ \bibnamefont {Rubin}}, \bibinfo {author} {\bibfnamefont {D.}~\bibnamefont {Sank}}, \bibinfo {author} {\bibfnamefont {V.}~\bibnamefont {Shvarts}}, \bibinfo
  {author} {\bibfnamefont {D.}~\bibnamefont {Strain}}, \bibinfo {author} {\bibfnamefont {M.}~\bibnamefont {Szalay}}, \bibinfo {author} {\bibfnamefont {B.}~\bibnamefont {Villalonga}}, \bibinfo {author} {\bibfnamefont {T.~C.}\ \bibnamefont {White}}, \bibinfo {author} {\bibfnamefont {Z.}~\bibnamefont {Yao}}, \bibinfo {author} {\bibfnamefont {P.}~\bibnamefont {Yeh}}, \bibinfo {author} {\bibfnamefont {J.}~\bibnamefont {Yoo}}, \bibinfo {author} {\bibfnamefont {A.}~\bibnamefont {Zalcman}}, \bibinfo {author} {\bibfnamefont {H.}~\bibnamefont {Neven}}, \bibinfo {author} {\bibfnamefont {S.}~\bibnamefont {Boixo}}, \bibinfo {author} {\bibfnamefont {A.}~\bibnamefont {Megrant}}, \bibinfo {author} {\bibfnamefont {Y.}~\bibnamefont {Chen}}, \bibinfo {author} {\bibfnamefont {J.}~\bibnamefont {Kelly}}, \bibinfo {author} {\bibfnamefont {V.}~\bibnamefont {Smelyanskiy}}, \bibinfo {author} {\bibfnamefont {A.}~\bibnamefont {Kitaev}}, \bibinfo {author} {\bibfnamefont {M.}~\bibnamefont {Knap}}, \bibinfo {author} {\bibfnamefont
  {F.}~\bibnamefont {Pollmann}},\ and\ \bibinfo {author} {\bibfnamefont {P.}~\bibnamefont {Roushan}},\ }\bibfield  {title} {\bibinfo {title} {Realizing topologically ordered states on a quantum processor},\ }\href {https://doi.org/10.1126/science.abi8378} {\bibfield  {journal} {\bibinfo  {journal} {Science}\ }\textbf {\bibinfo {volume} {374}},\ \bibinfo {pages} {1237} (\bibinfo {year} {2021})},\ \Eprint {https://arxiv.org/abs/https://www.science.org/doi/pdf/10.1126/science.abi8378} {https://www.science.org/doi/pdf/10.1126/science.abi8378} \BibitemShut {NoStop}%
\bibitem [{\citenamefont {Ding}\ \emph {et~al.}(2022)\citenamefont {Ding}, \citenamefont {Cui},\ and\ \citenamefont {Shi}}]{PhysRevD.105.054508}%
  \BibitemOpen
  \bibfield  {author} {\bibinfo {author} {\bibfnamefont {Y.}~\bibnamefont {Ding}}, \bibinfo {author} {\bibfnamefont {X.}~\bibnamefont {Cui}},\ and\ \bibinfo {author} {\bibfnamefont {Y.}~\bibnamefont {Shi}},\ }\bibfield  {title} {\bibinfo {title} {Digital quantum simulation and pseudoquantum simulation of the $\mathbf{Z}_2$ gauge--higgs model},\ }\href {https://doi.org/10.1103/PhysRevD.105.054508} {\bibfield  {journal} {\bibinfo  {journal} {Phys. Rev. D}\ }\textbf {\bibinfo {volume} {105}},\ \bibinfo {pages} {054508} (\bibinfo {year} {2022})}\BibitemShut {NoStop}%
\bibitem [{\citenamefont {Semeghini}\ \emph {et~al.}(2021)\citenamefont {Semeghini}, \citenamefont {Levine}, \citenamefont {Keesling}, \citenamefont {Ebadi}, \citenamefont {Wang}, \citenamefont {Bluvstein}, \citenamefont {Verresen}, \citenamefont {Pichler}, \citenamefont {Kalinowski}, \citenamefont {Samajdar}, \citenamefont {Omran}, \citenamefont {Sachdev}, \citenamefont {Vishwanath}, \citenamefont {Greiner}, \citenamefont {Vuletić},\ and\ \citenamefont {Lukin}}]{semeghini2021probing}%
  \BibitemOpen
  \bibfield  {author} {\bibinfo {author} {\bibfnamefont {G.}~\bibnamefont {Semeghini}}, \bibinfo {author} {\bibfnamefont {H.}~\bibnamefont {Levine}}, \bibinfo {author} {\bibfnamefont {A.}~\bibnamefont {Keesling}}, \bibinfo {author} {\bibfnamefont {S.}~\bibnamefont {Ebadi}}, \bibinfo {author} {\bibfnamefont {T.~T.}\ \bibnamefont {Wang}}, \bibinfo {author} {\bibfnamefont {D.}~\bibnamefont {Bluvstein}}, \bibinfo {author} {\bibfnamefont {R.}~\bibnamefont {Verresen}}, \bibinfo {author} {\bibfnamefont {H.}~\bibnamefont {Pichler}}, \bibinfo {author} {\bibfnamefont {M.}~\bibnamefont {Kalinowski}}, \bibinfo {author} {\bibfnamefont {R.}~\bibnamefont {Samajdar}}, \bibinfo {author} {\bibfnamefont {A.}~\bibnamefont {Omran}}, \bibinfo {author} {\bibfnamefont {S.}~\bibnamefont {Sachdev}}, \bibinfo {author} {\bibfnamefont {A.}~\bibnamefont {Vishwanath}}, \bibinfo {author} {\bibfnamefont {M.}~\bibnamefont {Greiner}}, \bibinfo {author} {\bibfnamefont {V.}~\bibnamefont {Vuletić}},\ and\ \bibinfo {author} {\bibfnamefont
  {M.~D.}\ \bibnamefont {Lukin}},\ }\bibfield  {title} {\bibinfo {title} {Probing topological spin liquids on a programmable quantum simulator},\ }\href {https://doi.org/10.1126/science.abi8794} {\bibfield  {journal} {\bibinfo  {journal} {Science}\ }\textbf {\bibinfo {volume} {374}},\ \bibinfo {pages} {1242} (\bibinfo {year} {2021})}\BibitemShut {NoStop}%
\bibitem [{\citenamefont {Yan}\ \emph {et~al.}(2022{\natexlab{a}})\citenamefont {Yan}, \citenamefont {Samajdar}, \citenamefont {Wang}, \citenamefont {Sachdev},\ and\ \citenamefont {Meng}}]{yan2022triangular}%
  \BibitemOpen
  \bibfield  {author} {\bibinfo {author} {\bibfnamefont {Z.}~\bibnamefont {Yan}}, \bibinfo {author} {\bibfnamefont {R.}~\bibnamefont {Samajdar}}, \bibinfo {author} {\bibfnamefont {Y.-C.}\ \bibnamefont {Wang}}, \bibinfo {author} {\bibfnamefont {S.}~\bibnamefont {Sachdev}},\ and\ \bibinfo {author} {\bibfnamefont {Z.~Y.}\ \bibnamefont {Meng}},\ }\bibfield  {title} {\bibinfo {title} {Triangular lattice quantum dimer model with variable dimer density},\ }\href {https://doi.org/10.1038/s41467-022-33431-5} {\bibfield  {journal} {\bibinfo  {journal} {Nature Communications}\ }\textbf {\bibinfo {volume} {13}},\ \bibinfo {pages} {5799} (\bibinfo {year} {2022}{\natexlab{a}})}\BibitemShut {NoStop}%
\bibitem [{\citenamefont {Yan}\ \emph {et~al.}(2023{\natexlab{b}})\citenamefont {Yan}, \citenamefont {Wang}, \citenamefont {Samajdar}, \citenamefont {Sachdev},\ and\ \citenamefont {Meng}}]{ZY2023Emergent}%
  \BibitemOpen
  \bibfield  {author} {\bibinfo {author} {\bibfnamefont {Z.}~\bibnamefont {Yan}}, \bibinfo {author} {\bibfnamefont {Y.-C.}\ \bibnamefont {Wang}}, \bibinfo {author} {\bibfnamefont {R.}~\bibnamefont {Samajdar}}, \bibinfo {author} {\bibfnamefont {S.}~\bibnamefont {Sachdev}},\ and\ \bibinfo {author} {\bibfnamefont {Z.~Y.}\ \bibnamefont {Meng}},\ }\bibfield  {title} {\bibinfo {title} {Emergent glassy behavior in a kagome rydberg atom array},\ }\href {https://doi.org/10.1103/PhysRevLett.130.206501} {\bibfield  {journal} {\bibinfo  {journal} {Phys. Rev. Lett.}\ }\textbf {\bibinfo {volume} {130}},\ \bibinfo {pages} {206501} (\bibinfo {year} {2023}{\natexlab{b}})}\BibitemShut {NoStop}%
\bibitem [{\citenamefont {Feynman}(1982)}]{Feynman1982}%
  \BibitemOpen
  \bibfield  {author} {\bibinfo {author} {\bibfnamefont {R.~P.}\ \bibnamefont {Feynman}},\ }\bibfield  {title} {\bibinfo {title} {Simulating physics with computers},\ }\href {https://doi.org/10.1007/BF02650179} {\bibfield  {journal} {\bibinfo  {journal} {International Journal of Theoretical Physics}\ }\textbf {\bibinfo {volume} {21}},\ \bibinfo {pages} {467} (\bibinfo {year} {1982})}\BibitemShut {NoStop}%
\bibitem [{\citenamefont {Zhou}\ \emph {et~al.}(2022)\citenamefont {Zhou}, \citenamefont {Liu}, \citenamefont {Yan}, \citenamefont {Chen},\ and\ \citenamefont {Zhang}}]{zhou2020quantumstring}%
  \BibitemOpen
  \bibfield  {author} {\bibinfo {author} {\bibfnamefont {Z.}~\bibnamefont {Zhou}}, \bibinfo {author} {\bibfnamefont {C.}~\bibnamefont {Liu}}, \bibinfo {author} {\bibfnamefont {Z.}~\bibnamefont {Yan}}, \bibinfo {author} {\bibfnamefont {Y.}~\bibnamefont {Chen}},\ and\ \bibinfo {author} {\bibfnamefont {X.-F.}\ \bibnamefont {Zhang}},\ }\bibfield  {title} {\bibinfo {title} {Quantum dynamics of topological strings in a frustrated ising antiferromagnet},\ }\href@noop {} {\bibfield  {journal} {\bibinfo  {journal} {npj Quantum Materials}\ }\textbf {\bibinfo {volume} {7}},\ \bibinfo {pages} {60} (\bibinfo {year} {2022})}\BibitemShut {NoStop}%
\bibitem [{\citenamefont {Zhou}\ \emph {et~al.}(2023)\citenamefont {Zhou}, \citenamefont {Liu}, \citenamefont {Liu}, \citenamefont {Yan}, \citenamefont {Chen},\ and\ \citenamefont {Zhang}}]{zhou2020quantum}%
  \BibitemOpen
  \bibfield  {author} {\bibinfo {author} {\bibfnamefont {Z.}~\bibnamefont {Zhou}}, \bibinfo {author} {\bibfnamefont {C.}~\bibnamefont {Liu}}, \bibinfo {author} {\bibfnamefont {D.-X.}\ \bibnamefont {Liu}}, \bibinfo {author} {\bibfnamefont {Z.}~\bibnamefont {Yan}}, \bibinfo {author} {\bibfnamefont {Y.}~\bibnamefont {Chen}},\ and\ \bibinfo {author} {\bibfnamefont {X.-F.}\ \bibnamefont {Zhang}},\ }\bibfield  {title} {\bibinfo {title} {Quantum tricriticality of incommensurate phase induced by quantum strings in frustrated ising magnetism},\ }\href {https://doi.org/10.21468/SciPostPhys.14.3.037} {\bibfield  {journal} {\bibinfo  {journal} {SciPost Phys.}\ }\textbf {\bibinfo {volume} {14}},\ \bibinfo {pages} {037} (\bibinfo {year} {2023})}\BibitemShut {NoStop}%
\bibitem [{\citenamefont {Zhang}\ and\ \citenamefont {Eggert}(2013)}]{zhang01}%
  \BibitemOpen
  \bibfield  {author} {\bibinfo {author} {\bibfnamefont {X.-F.}\ \bibnamefont {Zhang}}\ and\ \bibinfo {author} {\bibfnamefont {S.}~\bibnamefont {Eggert}},\ }\bibfield  {title} {\bibinfo {title} {Chiral edge states and fractional charge separation in a system of interacting bosons on a kagome lattice},\ }\href {https://doi.org/10.1103/PhysRevLett.111.147201} {\bibfield  {journal} {\bibinfo  {journal} {Phys. Rev. Lett.}\ }\textbf {\bibinfo {volume} {111}},\ \bibinfo {pages} {147201} (\bibinfo {year} {2013})}\BibitemShut {NoStop}%
\bibitem [{\citenamefont {Zhang}\ \emph {et~al.}(2016)\citenamefont {Zhang}, \citenamefont {Hu}, \citenamefont {Pelster},\ and\ \citenamefont {Eggert}}]{zhang02}%
  \BibitemOpen
  \bibfield  {author} {\bibinfo {author} {\bibfnamefont {X.-F.}\ \bibnamefont {Zhang}}, \bibinfo {author} {\bibfnamefont {S.}~\bibnamefont {Hu}}, \bibinfo {author} {\bibfnamefont {A.}~\bibnamefont {Pelster}},\ and\ \bibinfo {author} {\bibfnamefont {S.}~\bibnamefont {Eggert}},\ }\bibfield  {title} {\bibinfo {title} {Quantum domain walls induce incommensurate supersolid phase on the anisotropic triangular lattice},\ }\href {https://doi.org/10.1103/PhysRevLett.117.193201} {\bibfield  {journal} {\bibinfo  {journal} {Phys. Rev. Lett.}\ }\textbf {\bibinfo {volume} {117}},\ \bibinfo {pages} {193201} (\bibinfo {year} {2016})}\BibitemShut {NoStop}%
\bibitem [{\citenamefont {Zhang}\ \emph {et~al.}(2018)\citenamefont {Zhang}, \citenamefont {He}, \citenamefont {Eggert}, \citenamefont {Moessner},\ and\ \citenamefont {Pollmann}}]{zhang03}%
  \BibitemOpen
  \bibfield  {author} {\bibinfo {author} {\bibfnamefont {X.-F.}\ \bibnamefont {Zhang}}, \bibinfo {author} {\bibfnamefont {Y.-C.}\ \bibnamefont {He}}, \bibinfo {author} {\bibfnamefont {S.}~\bibnamefont {Eggert}}, \bibinfo {author} {\bibfnamefont {R.}~\bibnamefont {Moessner}},\ and\ \bibinfo {author} {\bibfnamefont {F.}~\bibnamefont {Pollmann}},\ }\bibfield  {title} {\bibinfo {title} {Continuous easy-plane deconfined phase transition on the kagome lattice},\ }\href {https://doi.org/10.1103/PhysRevLett.120.115702} {\bibfield  {journal} {\bibinfo  {journal} {Phys. Rev. Lett.}\ }\textbf {\bibinfo {volume} {120}},\ \bibinfo {pages} {115702} (\bibinfo {year} {2018})}\BibitemShut {NoStop}%
\bibitem [{\citenamefont {Wang}\ \emph {et~al.}(2017)\citenamefont {Wang}, \citenamefont {Qi}, \citenamefont {Chen},\ and\ \citenamefont {Meng}}]{YCWang2017}%
  \BibitemOpen
  \bibfield  {author} {\bibinfo {author} {\bibfnamefont {Y.-C.}\ \bibnamefont {Wang}}, \bibinfo {author} {\bibfnamefont {Y.}~\bibnamefont {Qi}}, \bibinfo {author} {\bibfnamefont {S.}~\bibnamefont {Chen}},\ and\ \bibinfo {author} {\bibfnamefont {Z.~Y.}\ \bibnamefont {Meng}},\ }\bibfield  {title} {\bibinfo {title} {Caution on emergent continuous symmetry: A {M}onte {C}arlo investigation of the transverse-field frustrated {I}sing model on the triangular and honeycomb lattices},\ }\href {https://doi.org/10.1103/PhysRevB.96.115160} {\bibfield  {journal} {\bibinfo  {journal} {Phys. Rev. B}\ }\textbf {\bibinfo {volume} {96}},\ \bibinfo {pages} {115160} (\bibinfo {year} {2017})}\BibitemShut {NoStop}%
\bibitem [{\citenamefont {Wang}\ \emph {et~al.}(2018)\citenamefont {Wang}, \citenamefont {Zhang}, \citenamefont {Pollmann}, \citenamefont {Cheng},\ and\ \citenamefont {Meng}}]{YCWang2018}%
  \BibitemOpen
  \bibfield  {author} {\bibinfo {author} {\bibfnamefont {Y.-C.}\ \bibnamefont {Wang}}, \bibinfo {author} {\bibfnamefont {X.-F.}\ \bibnamefont {Zhang}}, \bibinfo {author} {\bibfnamefont {F.}~\bibnamefont {Pollmann}}, \bibinfo {author} {\bibfnamefont {M.}~\bibnamefont {Cheng}},\ and\ \bibinfo {author} {\bibfnamefont {Z.~Y.}\ \bibnamefont {Meng}},\ }\bibfield  {title} {\bibinfo {title} {{Quantum Spin Liquid with Even Ising Gauge Field Structure on Kagome Lattice}},\ }\href {https://doi.org/10.1103/PhysRevLett.121.057202} {\bibfield  {journal} {\bibinfo  {journal} {Phys. Rev. Lett.}\ }\textbf {\bibinfo {volume} {121}},\ \bibinfo {pages} {057202} (\bibinfo {year} {2018})}\BibitemShut {NoStop}%
\bibitem [{\citenamefont {Wang}\ \emph {et~al.}(2021)\citenamefont {Wang}, \citenamefont {Yan}, \citenamefont {Wang}, \citenamefont {Qi},\ and\ \citenamefont {Meng}}]{YCWang2021vestigial}%
  \BibitemOpen
  \bibfield  {author} {\bibinfo {author} {\bibfnamefont {Y.-C.}\ \bibnamefont {Wang}}, \bibinfo {author} {\bibfnamefont {Z.}~\bibnamefont {Yan}}, \bibinfo {author} {\bibfnamefont {C.}~\bibnamefont {Wang}}, \bibinfo {author} {\bibfnamefont {Y.}~\bibnamefont {Qi}},\ and\ \bibinfo {author} {\bibfnamefont {Z.~Y.}\ \bibnamefont {Meng}},\ }\bibfield  {title} {\bibinfo {title} {Vestigial anyon condensation in kagome quantum spin liquids},\ }\href {https://doi.org/10.1103/PhysRevB.103.014408} {\bibfield  {journal} {\bibinfo  {journal} {Phys. Rev. B}\ }\textbf {\bibinfo {volume} {103}},\ \bibinfo {pages} {014408} (\bibinfo {year} {2021})}\BibitemShut {NoStop}%
\bibitem [{\citenamefont {Motta}\ \emph {et~al.}(2020)\citenamefont {Motta}, \citenamefont {Sun}, \citenamefont {Tan}, \citenamefont {O’Rourke}, \citenamefont {Ye}, \citenamefont {Minnich}, \citenamefont {Brand{\~a}o},\ and\ \citenamefont {Chan}}]{motta2020determining}%
  \BibitemOpen
  \bibfield  {author} {\bibinfo {author} {\bibfnamefont {M.}~\bibnamefont {Motta}}, \bibinfo {author} {\bibfnamefont {C.}~\bibnamefont {Sun}}, \bibinfo {author} {\bibfnamefont {A.~T.}\ \bibnamefont {Tan}}, \bibinfo {author} {\bibfnamefont {M.~J.}\ \bibnamefont {O’Rourke}}, \bibinfo {author} {\bibfnamefont {E.}~\bibnamefont {Ye}}, \bibinfo {author} {\bibfnamefont {A.~J.}\ \bibnamefont {Minnich}}, \bibinfo {author} {\bibfnamefont {F.~G.}\ \bibnamefont {Brand{\~a}o}},\ and\ \bibinfo {author} {\bibfnamefont {G.~K.-L.}\ \bibnamefont {Chan}},\ }\bibfield  {title} {\bibinfo {title} {Determining eigenstates and thermal states on a quantum computer using quantum imaginary time evolution},\ }\href@noop {} {\bibfield  {journal} {\bibinfo  {journal} {Nature Physics}\ }\textbf {\bibinfo {volume} {16}},\ \bibinfo {pages} {205} (\bibinfo {year} {2020})}\BibitemShut {NoStop}%
\bibitem [{\citenamefont {Love}(2020)}]{love2020cooling}%
  \BibitemOpen
  \bibfield  {author} {\bibinfo {author} {\bibfnamefont {P.~J.}\ \bibnamefont {Love}},\ }\bibfield  {title} {\bibinfo {title} {Cooling with imaginary time},\ }\href@noop {} {\bibfield  {journal} {\bibinfo  {journal} {Nature Physics}\ }\textbf {\bibinfo {volume} {16}},\ \bibinfo {pages} {130} (\bibinfo {year} {2020})}\BibitemShut {NoStop}%
\bibitem [{\citenamefont {Nishi}\ \emph {et~al.}(2021)\citenamefont {Nishi}, \citenamefont {Kosugi},\ and\ \citenamefont {Matsushita}}]{Nishi2021}%
  \BibitemOpen
  \bibfield  {author} {\bibinfo {author} {\bibfnamefont {H.}~\bibnamefont {Nishi}}, \bibinfo {author} {\bibfnamefont {T.}~\bibnamefont {Kosugi}},\ and\ \bibinfo {author} {\bibfnamefont {Y.-i.}\ \bibnamefont {Matsushita}},\ }\bibfield  {title} {\bibinfo {title} {Implementation of quantum imaginary-time evolution method on nisq devices by introducing nonlocal approximation},\ }\href {https://doi.org/10.1038/s41534-021-00409-y} {\bibfield  {journal} {\bibinfo  {journal} {npj Quantum Information}\ }\textbf {\bibinfo {volume} {7}},\ \bibinfo {pages} {85} (\bibinfo {year} {2021})}\BibitemShut {NoStop}%
\bibitem [{\citenamefont {Cao}\ \emph {et~al.}(2022)\citenamefont {Cao}, \citenamefont {An}, \citenamefont {Hou}, \citenamefont {Zhou},\ and\ \citenamefont {Zeng}}]{Cao2022}%
  \BibitemOpen
  \bibfield  {author} {\bibinfo {author} {\bibfnamefont {C.}~\bibnamefont {Cao}}, \bibinfo {author} {\bibfnamefont {Z.}~\bibnamefont {An}}, \bibinfo {author} {\bibfnamefont {S.-Y.}\ \bibnamefont {Hou}}, \bibinfo {author} {\bibfnamefont {D.~L.}\ \bibnamefont {Zhou}},\ and\ \bibinfo {author} {\bibfnamefont {B.}~\bibnamefont {Zeng}},\ }\bibfield  {title} {\bibinfo {title} {Quantum imaginary time evolution steered by reinforcement learning},\ }\href {https://doi.org/10.1038/s42005-022-00837-y} {\bibfield  {journal} {\bibinfo  {journal} {Communications Physics}\ }\textbf {\bibinfo {volume} {5}},\ \bibinfo {pages} {57} (\bibinfo {year} {2022})}\BibitemShut {NoStop}%
\bibitem [{\citenamefont {Zhang}\ and\ \citenamefont {Yin}(2023)}]{Zhang2023a}%
  \BibitemOpen
  \bibfield  {author} {\bibinfo {author} {\bibfnamefont {S.-X.}\ \bibnamefont {Zhang}}\ and\ \bibinfo {author} {\bibfnamefont {S.}~\bibnamefont {Yin}},\ }\bibfield  {title} {\bibinfo {title} {{Universal imaginary-time critical dynamics on a quantum computer}},\ }\href@noop {} {\bibfield  {journal} {\bibinfo  {journal} {arXiv:2308.05408}\ } (\bibinfo {year} {2023})},\ \Eprint {https://arxiv.org/abs/arXiv:2308.05408v1} {arXiv:arXiv:2308.05408v1} \BibitemShut {NoStop}%
\bibitem [{\citenamefont {Peruzzo}\ \emph {et~al.}(2014)\citenamefont {Peruzzo}, \citenamefont {McClean}, \citenamefont {Shadbolt}, \citenamefont {Yung}, \citenamefont {Zhou}, \citenamefont {Love}, \citenamefont {Aspuru-Guzik},\ and\ \citenamefont {O'Brien}}]{Peruzzo2014}%
  \BibitemOpen
  \bibfield  {author} {\bibinfo {author} {\bibfnamefont {A.}~\bibnamefont {Peruzzo}}, \bibinfo {author} {\bibfnamefont {J.}~\bibnamefont {McClean}}, \bibinfo {author} {\bibfnamefont {P.}~\bibnamefont {Shadbolt}}, \bibinfo {author} {\bibfnamefont {M.-H.}\ \bibnamefont {Yung}}, \bibinfo {author} {\bibfnamefont {X.-Q.}\ \bibnamefont {Zhou}}, \bibinfo {author} {\bibfnamefont {P.~J.}\ \bibnamefont {Love}}, \bibinfo {author} {\bibfnamefont {A.}~\bibnamefont {Aspuru-Guzik}},\ and\ \bibinfo {author} {\bibfnamefont {J.~L.}\ \bibnamefont {O'Brien}},\ }\bibfield  {title} {\bibinfo {title} {A variational eigenvalue solver on a photonic quantum processor},\ }\href {https://doi.org/10.1038/ncomms5213} {\bibfield  {journal} {\bibinfo  {journal} {Nature Communications}\ }\textbf {\bibinfo {volume} {5}},\ \bibinfo {pages} {4213} (\bibinfo {year} {2014})}\BibitemShut {NoStop}%
\bibitem [{\citenamefont {McClean}\ \emph {et~al.}(2016)\citenamefont {McClean}, \citenamefont {Romero}, \citenamefont {Babbush},\ and\ \citenamefont {Aspuru-Guzik}}]{McClean_2016}%
  \BibitemOpen
  \bibfield  {author} {\bibinfo {author} {\bibfnamefont {J.~R.}\ \bibnamefont {McClean}}, \bibinfo {author} {\bibfnamefont {J.}~\bibnamefont {Romero}}, \bibinfo {author} {\bibfnamefont {R.}~\bibnamefont {Babbush}},\ and\ \bibinfo {author} {\bibfnamefont {A.}~\bibnamefont {Aspuru-Guzik}},\ }\bibfield  {title} {\bibinfo {title} {The theory of variational hybrid quantum-classical algorithms},\ }\href {https://doi.org/10.1088/1367-2630/18/2/023023} {\bibfield  {journal} {\bibinfo  {journal} {New Journal of Physics}\ }\textbf {\bibinfo {volume} {18}},\ \bibinfo {pages} {023023} (\bibinfo {year} {2016})}\BibitemShut {NoStop}%
\bibitem [{\citenamefont {Kitaev}(2003)}]{KITAEV20032}%
  \BibitemOpen
  \bibfield  {author} {\bibinfo {author} {\bibfnamefont {A.}~\bibnamefont {Kitaev}},\ }\bibfield  {title} {\bibinfo {title} {Fault-tolerant quantum computation by anyons},\ }\href {https://doi.org/https://doi.org/10.1016/S0003-4916(02)00018-0} {\bibfield  {journal} {\bibinfo  {journal} {Annals of Physics}\ }\textbf {\bibinfo {volume} {303}},\ \bibinfo {pages} {2} (\bibinfo {year} {2003})}\BibitemShut {NoStop}%
\bibitem [{\citenamefont {Lehtovaara}\ \emph {et~al.}(2007)\citenamefont {Lehtovaara}, \citenamefont {Toivanen},\ and\ \citenamefont {Eloranta}}]{LEHTOVAARA2007148}%
  \BibitemOpen
  \bibfield  {author} {\bibinfo {author} {\bibfnamefont {L.}~\bibnamefont {Lehtovaara}}, \bibinfo {author} {\bibfnamefont {J.}~\bibnamefont {Toivanen}},\ and\ \bibinfo {author} {\bibfnamefont {J.}~\bibnamefont {Eloranta}},\ }\bibfield  {title} {\bibinfo {title} {Solution of time-independent schrödinger equation by the imaginary time propagation method},\ }\href {https://doi.org/https://doi.org/10.1016/j.jcp.2006.06.006} {\bibfield  {journal} {\bibinfo  {journal} {Journal of Computational Physics}\ }\textbf {\bibinfo {volume} {221}},\ \bibinfo {pages} {148} (\bibinfo {year} {2007})}\BibitemShut {NoStop}%
\bibitem [{\citenamefont {McArdle}\ \emph {et~al.}(2019)\citenamefont {McArdle}, \citenamefont {Jones}, \citenamefont {Endo}, \citenamefont {Li}, \citenamefont {Benjamin},\ and\ \citenamefont {Yuan}}]{McArdle2019}%
  \BibitemOpen
  \bibfield  {author} {\bibinfo {author} {\bibfnamefont {S.}~\bibnamefont {McArdle}}, \bibinfo {author} {\bibfnamefont {T.}~\bibnamefont {Jones}}, \bibinfo {author} {\bibfnamefont {S.}~\bibnamefont {Endo}}, \bibinfo {author} {\bibfnamefont {Y.}~\bibnamefont {Li}}, \bibinfo {author} {\bibfnamefont {S.~C.}\ \bibnamefont {Benjamin}},\ and\ \bibinfo {author} {\bibfnamefont {X.}~\bibnamefont {Yuan}},\ }\bibfield  {title} {\bibinfo {title} {Variational ansatz-based quantum simulation of imaginary time evolution},\ }\href {https://doi.org/10.1038/s41534-019-0187-2} {\bibfield  {journal} {\bibinfo  {journal} {npj Quantum Information}\ }\textbf {\bibinfo {volume} {5}},\ \bibinfo {pages} {75} (\bibinfo {year} {2019})}\BibitemShut {NoStop}%
\bibitem [{\citenamefont {Daley}(2014)}]{doi:10.1080/00018732.2014.933502}%
  \BibitemOpen
  \bibfield  {author} {\bibinfo {author} {\bibfnamefont {A.~J.}\ \bibnamefont {Daley}},\ }\bibfield  {title} {\bibinfo {title} {Quantum trajectories and open many-body quantum systems},\ }\href {https://doi.org/10.1080/00018732.2014.933502} {\bibfield  {journal} {\bibinfo  {journal} {Advances in Physics}\ }\textbf {\bibinfo {volume} {63}},\ \bibinfo {pages} {77} (\bibinfo {year} {2014})}\BibitemShut {NoStop}%
\bibitem [{\citenamefont {Yuto~Ashida}\ and\ \citenamefont {Ueda}(2020)}]{doi:10.1080/00018732.2021.1876991}%
  \BibitemOpen
  \bibfield  {author} {\bibinfo {author} {\bibfnamefont {Z.~G.}\ \bibnamefont {Yuto~Ashida}}\ and\ \bibinfo {author} {\bibfnamefont {M.}~\bibnamefont {Ueda}},\ }\bibfield  {title} {\bibinfo {title} {Non-hermitian physics},\ }\href {https://doi.org/10.1080/00018732.2021.1876991} {\bibfield  {journal} {\bibinfo  {journal} {Advances in Physics}\ }\textbf {\bibinfo {volume} {69}},\ \bibinfo {pages} {249} (\bibinfo {year} {2020})}\BibitemShut {NoStop}%
\bibitem [{\citenamefont {Bharti}\ \emph {et~al.}(2022)\citenamefont {Bharti}, \citenamefont {Cervera-Lierta}, \citenamefont {Kyaw}, \citenamefont {Haug}, \citenamefont {Alperin-Lea}, \citenamefont {Anand}, \citenamefont {Degroote}, \citenamefont {Heimonen}, \citenamefont {Kottmann}, \citenamefont {Menke}, \citenamefont {Mok}, \citenamefont {Sim}, \citenamefont {Kwek},\ and\ \citenamefont {Aspuru-Guzik}}]{Bharti2021z}%
  \BibitemOpen
  \bibfield  {author} {\bibinfo {author} {\bibfnamefont {K.}~\bibnamefont {Bharti}}, \bibinfo {author} {\bibfnamefont {A.}~\bibnamefont {Cervera-Lierta}}, \bibinfo {author} {\bibfnamefont {T.~H.}\ \bibnamefont {Kyaw}}, \bibinfo {author} {\bibfnamefont {T.}~\bibnamefont {Haug}}, \bibinfo {author} {\bibfnamefont {S.}~\bibnamefont {Alperin-Lea}}, \bibinfo {author} {\bibfnamefont {A.}~\bibnamefont {Anand}}, \bibinfo {author} {\bibfnamefont {M.}~\bibnamefont {Degroote}}, \bibinfo {author} {\bibfnamefont {H.}~\bibnamefont {Heimonen}}, \bibinfo {author} {\bibfnamefont {J.~S.}\ \bibnamefont {Kottmann}}, \bibinfo {author} {\bibfnamefont {T.}~\bibnamefont {Menke}}, \bibinfo {author} {\bibfnamefont {W.-K.}\ \bibnamefont {Mok}}, \bibinfo {author} {\bibfnamefont {S.}~\bibnamefont {Sim}}, \bibinfo {author} {\bibfnamefont {L.-C.}\ \bibnamefont {Kwek}},\ and\ \bibinfo {author} {\bibfnamefont {A.}~\bibnamefont {Aspuru-Guzik}},\ }\bibfield  {title} {\bibinfo {title} {{Noisy intermediate-scale quantum algorithms}},\ }\href
  {https://doi.org/10.1103/RevModPhys.94.015004} {\bibfield  {journal} {\bibinfo  {journal} {Reviews of Modern Physics}\ }\textbf {\bibinfo {volume} {94}},\ \bibinfo {pages} {015004} (\bibinfo {year} {2022})}\BibitemShut {NoStop}%
\bibitem [{\citenamefont {Cerezo}\ \emph {et~al.}(2021)\citenamefont {Cerezo}, \citenamefont {Arrasmith}, \citenamefont {Babbush}, \citenamefont {Benjamin}, \citenamefont {Endo}, \citenamefont {Fujii}, \citenamefont {McClean}, \citenamefont {Mitarai}, \citenamefont {Yuan}, \citenamefont {Cincio},\ and\ \citenamefont {Coles}}]{Cerezo2021}%
  \BibitemOpen
  \bibfield  {author} {\bibinfo {author} {\bibfnamefont {M.}~\bibnamefont {Cerezo}}, \bibinfo {author} {\bibfnamefont {A.}~\bibnamefont {Arrasmith}}, \bibinfo {author} {\bibfnamefont {R.}~\bibnamefont {Babbush}}, \bibinfo {author} {\bibfnamefont {S.~C.}\ \bibnamefont {Benjamin}}, \bibinfo {author} {\bibfnamefont {S.}~\bibnamefont {Endo}}, \bibinfo {author} {\bibfnamefont {K.}~\bibnamefont {Fujii}}, \bibinfo {author} {\bibfnamefont {J.~R.}\ \bibnamefont {McClean}}, \bibinfo {author} {\bibfnamefont {K.}~\bibnamefont {Mitarai}}, \bibinfo {author} {\bibfnamefont {X.}~\bibnamefont {Yuan}}, \bibinfo {author} {\bibfnamefont {L.}~\bibnamefont {Cincio}},\ and\ \bibinfo {author} {\bibfnamefont {P.~J.}\ \bibnamefont {Coles}},\ }\bibfield  {title} {\bibinfo {title} {Variational quantum algorithms},\ }\href {https://doi.org/10.1038/s42254-021-00348-9} {\bibfield  {journal} {\bibinfo  {journal} {Nature Reviews Physics}\ }\textbf {\bibinfo {volume} {3}},\ \bibinfo {pages} {625} (\bibinfo {year} {2021})}\BibitemShut
  {NoStop}%
\bibitem [{\citenamefont {Li}\ and\ \citenamefont {Benjamin}(2017)}]{Li2017bz}%
  \BibitemOpen
  \bibfield  {author} {\bibinfo {author} {\bibfnamefont {Y.}~\bibnamefont {Li}}\ and\ \bibinfo {author} {\bibfnamefont {S.~C.}\ \bibnamefont {Benjamin}},\ }\bibfield  {title} {\bibinfo {title} {{Efficient Variational Quantum Simulator Incorporating Active Error Minimization}},\ }\href {https://doi.org/10.1103/PhysRevX.7.021050} {\bibfield  {journal} {\bibinfo  {journal} {Physical Review X}\ }\textbf {\bibinfo {volume} {7}},\ \bibinfo {pages} {021050} (\bibinfo {year} {2017})}\BibitemShut {NoStop}%
\bibitem [{\citenamefont {Yuan}\ \emph {et~al.}(2019)\citenamefont {Yuan}, \citenamefont {Endo}, \citenamefont {Zhao}, \citenamefont {Li},\ and\ \citenamefont {Benjamin}}]{Yuan2019theoryofvariational}%
  \BibitemOpen
  \bibfield  {author} {\bibinfo {author} {\bibfnamefont {X.}~\bibnamefont {Yuan}}, \bibinfo {author} {\bibfnamefont {S.}~\bibnamefont {Endo}}, \bibinfo {author} {\bibfnamefont {Q.}~\bibnamefont {Zhao}}, \bibinfo {author} {\bibfnamefont {Y.}~\bibnamefont {Li}},\ and\ \bibinfo {author} {\bibfnamefont {S.~C.}\ \bibnamefont {Benjamin}},\ }\bibfield  {title} {\bibinfo {title} {Theory of variational quantum simulation},\ }\href {https://doi.org/10.22331/q-2019-10-07-191} {\bibfield  {journal} {\bibinfo  {journal} {{Quantum}}\ }\textbf {\bibinfo {volume} {3}},\ \bibinfo {pages} {191} (\bibinfo {year} {2019})}\BibitemShut {NoStop}%
\bibitem [{\citenamefont {Barison}\ \emph {et~al.}(2021)\citenamefont {Barison}, \citenamefont {Vicentini},\ and\ \citenamefont {Carleo}}]{Barison2021z}%
  \BibitemOpen
  \bibfield  {author} {\bibinfo {author} {\bibfnamefont {S.}~\bibnamefont {Barison}}, \bibinfo {author} {\bibfnamefont {F.}~\bibnamefont {Vicentini}},\ and\ \bibinfo {author} {\bibfnamefont {G.}~\bibnamefont {Carleo}},\ }\bibfield  {title} {\bibinfo {title} {{An efficient quantum algorithm for the time evolution of parameterized circuits}},\ }\href {https://doi.org/10.22331/q-2021-07-28-512} {\bibfield  {journal} {\bibinfo  {journal} {Quantum}\ }\textbf {\bibinfo {volume} {5}},\ \bibinfo {pages} {512} (\bibinfo {year} {2021})}\BibitemShut {NoStop}%
\bibitem [{\citenamefont {Benedetti}\ \emph {et~al.}(2021)\citenamefont {Benedetti}, \citenamefont {Fiorentini},\ and\ \citenamefont {Lubasch}}]{Benedetti2021az}%
  \BibitemOpen
  \bibfield  {author} {\bibinfo {author} {\bibfnamefont {M.}~\bibnamefont {Benedetti}}, \bibinfo {author} {\bibfnamefont {M.}~\bibnamefont {Fiorentini}},\ and\ \bibinfo {author} {\bibfnamefont {M.}~\bibnamefont {Lubasch}},\ }\bibfield  {title} {\bibinfo {title} {{Hardware-efficient variational quantum algorithms for time evolution}},\ }\href {https://doi.org/10.1103/PhysRevResearch.3.033083} {\bibfield  {journal} {\bibinfo  {journal} {Physical Review Research}\ }\textbf {\bibinfo {volume} {3}},\ \bibinfo {pages} {033083} (\bibinfo {year} {2021})}\BibitemShut {NoStop}%
\bibitem [{\citenamefont {Lee}\ \emph {et~al.}(2022)\citenamefont {Lee}, \citenamefont {Zhang}, \citenamefont {Hsieh}, \citenamefont {Zhang},\ and\ \citenamefont {Shi}}]{Lee2022z}%
  \BibitemOpen
  \bibfield  {author} {\bibinfo {author} {\bibfnamefont {C.~K.}\ \bibnamefont {Lee}}, \bibinfo {author} {\bibfnamefont {S.-X.}\ \bibnamefont {Zhang}}, \bibinfo {author} {\bibfnamefont {C.-Y.}\ \bibnamefont {Hsieh}}, \bibinfo {author} {\bibfnamefont {S.}~\bibnamefont {Zhang}},\ and\ \bibinfo {author} {\bibfnamefont {L.}~\bibnamefont {Shi}},\ }\bibfield  {title} {\bibinfo {title} {{Variational Quantum Simulations of Finite-Temperature Dynamical Properties via Thermofield Dynamics}},\ }\href {http://arxiv.org/abs/2206.05571} {\bibfield  {journal} {\bibinfo  {journal} {arXiv:2206.05517}\ } (\bibinfo {year} {2022})}\BibitemShut {NoStop}%
\bibitem [{\citenamefont {Wiersema}\ \emph {et~al.}(2020)\citenamefont {Wiersema}, \citenamefont {Zhou}, \citenamefont {de~Sereville}, \citenamefont {Carrasquilla}, \citenamefont {Kim},\ and\ \citenamefont {Yuen}}]{Wiersema2020z}%
  \BibitemOpen
  \bibfield  {author} {\bibinfo {author} {\bibfnamefont {R.}~\bibnamefont {Wiersema}}, \bibinfo {author} {\bibfnamefont {C.}~\bibnamefont {Zhou}}, \bibinfo {author} {\bibfnamefont {Y.}~\bibnamefont {de~Sereville}}, \bibinfo {author} {\bibfnamefont {J.~F.}\ \bibnamefont {Carrasquilla}}, \bibinfo {author} {\bibfnamefont {Y.~B.}\ \bibnamefont {Kim}},\ and\ \bibinfo {author} {\bibfnamefont {H.}~\bibnamefont {Yuen}},\ }\bibfield  {title} {\bibinfo {title} {{Exploring Entanglement and Optimization within the Hamiltonian Variational Ansatz}},\ }\href {https://doi.org/10.1103/PRXQuantum.1.020319} {\bibfield  {journal} {\bibinfo  {journal} {PRX Quantum}\ }\textbf {\bibinfo {volume} {1}},\ \bibinfo {pages} {020319} (\bibinfo {year} {2020})}\BibitemShut {NoStop}%
\bibitem [{\citenamefont {Zhang}\ \emph {et~al.}(2022)\citenamefont {Zhang}, \citenamefont {Hsieh}, \citenamefont {Zhang},\ and\ \citenamefont {Yao}}]{Zhang2020bz}%
  \BibitemOpen
  \bibfield  {author} {\bibinfo {author} {\bibfnamefont {S.-X.}\ \bibnamefont {Zhang}}, \bibinfo {author} {\bibfnamefont {C.-Y.}\ \bibnamefont {Hsieh}}, \bibinfo {author} {\bibfnamefont {S.}~\bibnamefont {Zhang}},\ and\ \bibinfo {author} {\bibfnamefont {H.}~\bibnamefont {Yao}},\ }\bibfield  {title} {\bibinfo {title} {{Differentiable quantum architecture search}},\ }\href {https://doi.org/10.1088/2058-9565/ac87cd} {\bibfield  {journal} {\bibinfo  {journal} {Quantum Science and Technology}\ }\textbf {\bibinfo {volume} {7}},\ \bibinfo {pages} {045023} (\bibinfo {year} {2022})}\BibitemShut {NoStop}%
\bibitem [{\citenamefont {Du}\ \emph {et~al.}(2022)\citenamefont {Du}, \citenamefont {Huang}, \citenamefont {You}, \citenamefont {Hsieh},\ and\ \citenamefont {Tao}}]{Du2020az}%
  \BibitemOpen
  \bibfield  {author} {\bibinfo {author} {\bibfnamefont {Y.}~\bibnamefont {Du}}, \bibinfo {author} {\bibfnamefont {T.}~\bibnamefont {Huang}}, \bibinfo {author} {\bibfnamefont {S.}~\bibnamefont {You}}, \bibinfo {author} {\bibfnamefont {M.-H.}\ \bibnamefont {Hsieh}},\ and\ \bibinfo {author} {\bibfnamefont {D.}~\bibnamefont {Tao}},\ }\bibfield  {title} {\bibinfo {title} {{Quantum circuit architecture search for variational quantum algorithms}},\ }\href {https://doi.org/10.1038/s41534-022-00570-y} {\bibfield  {journal} {\bibinfo  {journal} {npj Quantum Information}\ }\textbf {\bibinfo {volume} {8}},\ \bibinfo {pages} {62} (\bibinfo {year} {2022})}\BibitemShut {NoStop}%
\bibitem [{\citenamefont {Zhang}\ \emph {et~al.}(2021)\citenamefont {Zhang}, \citenamefont {Hsieh}, \citenamefont {Zhang},\ and\ \citenamefont {Yao}}]{Zhang2021z}%
  \BibitemOpen
  \bibfield  {author} {\bibinfo {author} {\bibfnamefont {S.-X.}\ \bibnamefont {Zhang}}, \bibinfo {author} {\bibfnamefont {C.-Y.}\ \bibnamefont {Hsieh}}, \bibinfo {author} {\bibfnamefont {S.}~\bibnamefont {Zhang}},\ and\ \bibinfo {author} {\bibfnamefont {H.}~\bibnamefont {Yao}},\ }\bibfield  {title} {\bibinfo {title} {{Neural predictor based quantum architecture search}},\ }\href {https://doi.org/10.1088/2632-2153/ac28dd} {\bibfield  {journal} {\bibinfo  {journal} {Machine Learning: Science and Technology}\ }\textbf {\bibinfo {volume} {2}},\ \bibinfo {pages} {045027} (\bibinfo {year} {2021})}\BibitemShut {NoStop}%
\bibitem [{\citenamefont {Preskill}(2018)}]{Preskill2018z}%
  \BibitemOpen
  \bibfield  {author} {\bibinfo {author} {\bibfnamefont {J.}~\bibnamefont {Preskill}},\ }\bibfield  {title} {\bibinfo {title} {{Quantum Computing in the NISQ era and beyond}},\ }\href {https://doi.org/10.22331/q-2018-08-06-79} {\bibfield  {journal} {\bibinfo  {journal} {Quantum}\ }\textbf {\bibinfo {volume} {2}},\ \bibinfo {pages} {79} (\bibinfo {year} {2018})}\BibitemShut {NoStop}%
\bibitem [{\citenamefont {McLachlan}(1964)}]{doi:10.1080/00268976400100041}%
  \BibitemOpen
  \bibfield  {author} {\bibinfo {author} {\bibfnamefont {A.}~\bibnamefont {McLachlan}},\ }\bibfield  {title} {\bibinfo {title} {A variational solution of the time-dependent schrodinger equation},\ }\href {https://doi.org/10.1080/00268976400100041} {\bibfield  {journal} {\bibinfo  {journal} {Molecular Physics}\ }\textbf {\bibinfo {volume} {8}},\ \bibinfo {pages} {39} (\bibinfo {year} {1964})}\BibitemShut {NoStop}%
\bibitem [{\citenamefont {Stokes}\ \emph {et~al.}(2020)\citenamefont {Stokes}, \citenamefont {Izaac}, \citenamefont {Killoran},\ and\ \citenamefont {Carleo}}]{Stokes2020quantumnatural}%
  \BibitemOpen
  \bibfield  {author} {\bibinfo {author} {\bibfnamefont {J.}~\bibnamefont {Stokes}}, \bibinfo {author} {\bibfnamefont {J.}~\bibnamefont {Izaac}}, \bibinfo {author} {\bibfnamefont {N.}~\bibnamefont {Killoran}},\ and\ \bibinfo {author} {\bibfnamefont {G.}~\bibnamefont {Carleo}},\ }\bibfield  {title} {\bibinfo {title} {Quantum {N}atural {G}radient},\ }\href {https://doi.org/10.22331/q-2020-05-25-269} {\bibfield  {journal} {\bibinfo  {journal} {{Quantum}}\ }\textbf {\bibinfo {volume} {4}},\ \bibinfo {pages} {269} (\bibinfo {year} {2020})}\BibitemShut {NoStop}%
\bibitem [{\citenamefont {Moessner}\ and\ \citenamefont {Sondhi}(2001)}]{Moessner2001}%
  \BibitemOpen
  \bibfield  {author} {\bibinfo {author} {\bibfnamefont {R.}~\bibnamefont {Moessner}}\ and\ \bibinfo {author} {\bibfnamefont {S.~L.}\ \bibnamefont {Sondhi}},\ }\bibfield  {title} {\bibinfo {title} {{I}sing models of quantum frustration},\ }\href {https://doi.org/10.1103/PhysRevB.63.224401} {\bibfield  {journal} {\bibinfo  {journal} {Phys. Rev. B}\ }\textbf {\bibinfo {volume} {63}},\ \bibinfo {pages} {224401} (\bibinfo {year} {2001})}\BibitemShut {NoStop}%
\bibitem [{\citenamefont {Isakov}\ and\ \citenamefont {Moessner}(2003)}]{Isakov}%
  \BibitemOpen
  \bibfield  {author} {\bibinfo {author} {\bibfnamefont {S.~V.}\ \bibnamefont {Isakov}}\ and\ \bibinfo {author} {\bibfnamefont {R.}~\bibnamefont {Moessner}},\ }\bibfield  {title} {\bibinfo {title} {Interplay of quantum and thermal fluctuations in a frustrated magnet},\ }\href {https://doi.org/10.1103/PhysRevB.68.104409} {\bibfield  {journal} {\bibinfo  {journal} {Phys. Rev. B}\ }\textbf {\bibinfo {volume} {68}},\ \bibinfo {pages} {104409} (\bibinfo {year} {2003})}\BibitemShut {NoStop}%
\bibitem [{\citenamefont {Da~Liao}\ \emph {et~al.}(2021)\citenamefont {Da~Liao}, \citenamefont {Li}, \citenamefont {Yan}, \citenamefont {Wei}, \citenamefont {Li}, \citenamefont {Qi},\ and\ \citenamefont {Meng}}]{YDLiao2021}%
  \BibitemOpen
  \bibfield  {author} {\bibinfo {author} {\bibfnamefont {Y.}~\bibnamefont {Da~Liao}}, \bibinfo {author} {\bibfnamefont {H.}~\bibnamefont {Li}}, \bibinfo {author} {\bibfnamefont {Z.}~\bibnamefont {Yan}}, \bibinfo {author} {\bibfnamefont {H.-T.}\ \bibnamefont {Wei}}, \bibinfo {author} {\bibfnamefont {W.}~\bibnamefont {Li}}, \bibinfo {author} {\bibfnamefont {Y.}~\bibnamefont {Qi}},\ and\ \bibinfo {author} {\bibfnamefont {Z.~Y.}\ \bibnamefont {Meng}},\ }\bibfield  {title} {\bibinfo {title} {Phase diagram of the quantum {I}sing model on a triangular lattice under external field},\ }\href {https://doi.org/10.1103/PhysRevB.103.104416} {\bibfield  {journal} {\bibinfo  {journal} {Phys. Rev. B}\ }\textbf {\bibinfo {volume} {103}},\ \bibinfo {pages} {104416} (\bibinfo {year} {2021})}\BibitemShut {NoStop}%
\bibitem [{\citenamefont {Moessner}\ and\ \citenamefont {Raman}(2011{\natexlab{a}})}]{Moessner2010b}%
  \BibitemOpen
  \bibfield  {author} {\bibinfo {author} {\bibfnamefont {R.}~\bibnamefont {Moessner}}\ and\ \bibinfo {author} {\bibfnamefont {K.~S.}\ \bibnamefont {Raman}},\ }\bibinfo {title} {Quantum dimer models},\ in\ \href {https://doi.org/10.1007/978-3-642-10589-0_17} {\emph {\bibinfo {booktitle} {Introduction to Frustrated Magnetism: Materials, Experiments, Theory}}}\ (\bibinfo  {publisher} {Springer Berlin Heidelberg},\ \bibinfo {address} {Berlin, Heidelberg},\ \bibinfo {year} {2011})\ pp.\ \bibinfo {pages} {437--479}\BibitemShut {NoStop}%
\bibitem [{\citenamefont {Yan}\ \emph {et~al.}(2021{\natexlab{a}})\citenamefont {Yan}, \citenamefont {Zhou}, \citenamefont {Sylju\aa{}sen}, \citenamefont {Zhang}, \citenamefont {Yuan}, \citenamefont {Lou},\ and\ \citenamefont {Chen}}]{yan2019widely}%
  \BibitemOpen
  \bibfield  {author} {\bibinfo {author} {\bibfnamefont {Z.}~\bibnamefont {Yan}}, \bibinfo {author} {\bibfnamefont {Z.}~\bibnamefont {Zhou}}, \bibinfo {author} {\bibfnamefont {O.~F.}\ \bibnamefont {Sylju\aa{}sen}}, \bibinfo {author} {\bibfnamefont {J.}~\bibnamefont {Zhang}}, \bibinfo {author} {\bibfnamefont {T.}~\bibnamefont {Yuan}}, \bibinfo {author} {\bibfnamefont {J.}~\bibnamefont {Lou}},\ and\ \bibinfo {author} {\bibfnamefont {Y.}~\bibnamefont {Chen}},\ }\bibfield  {title} {\bibinfo {title} {Widely existing mixed phase structure of the quantum dimer model on a square lattice},\ }\href {https://doi.org/10.1103/PhysRevB.103.094421} {\bibfield  {journal} {\bibinfo  {journal} {Phys. Rev. B}\ }\textbf {\bibinfo {volume} {103}},\ \bibinfo {pages} {094421} (\bibinfo {year} {2021}{\natexlab{a}})}\BibitemShut {NoStop}%
\bibitem [{\citenamefont {Yan}\ \emph {et~al.}(2019)\citenamefont {Yan}, \citenamefont {Wu}, \citenamefont {Liu}, \citenamefont {Sylju\aa{}sen}, \citenamefont {Lou},\ and\ \citenamefont {Chen}}]{ZY2019}%
  \BibitemOpen
  \bibfield  {author} {\bibinfo {author} {\bibfnamefont {Z.}~\bibnamefont {Yan}}, \bibinfo {author} {\bibfnamefont {Y.}~\bibnamefont {Wu}}, \bibinfo {author} {\bibfnamefont {C.}~\bibnamefont {Liu}}, \bibinfo {author} {\bibfnamefont {O.~F.}\ \bibnamefont {Sylju\aa{}sen}}, \bibinfo {author} {\bibfnamefont {J.}~\bibnamefont {Lou}},\ and\ \bibinfo {author} {\bibfnamefont {Y.}~\bibnamefont {Chen}},\ }\bibfield  {title} {\bibinfo {title} {Sweeping cluster algorithm for quantum spin systems with strong geometric restrictions},\ }\href {https://doi.org/10.1103/PhysRevB.99.165135} {\bibfield  {journal} {\bibinfo  {journal} {Phys. Rev. B}\ }\textbf {\bibinfo {volume} {99}},\ \bibinfo {pages} {165135} (\bibinfo {year} {2019})}\BibitemShut {NoStop}%
\bibitem [{\citenamefont {Yan}\ \emph {et~al.}(2021{\natexlab{b}})\citenamefont {Yan}, \citenamefont {Wang}, \citenamefont {Ma}, \citenamefont {Qi},\ and\ \citenamefont {Meng}}]{yan2020triangular}%
  \BibitemOpen
  \bibfield  {author} {\bibinfo {author} {\bibfnamefont {Z.}~\bibnamefont {Yan}}, \bibinfo {author} {\bibfnamefont {Y.-C.}\ \bibnamefont {Wang}}, \bibinfo {author} {\bibfnamefont {N.}~\bibnamefont {Ma}}, \bibinfo {author} {\bibfnamefont {Y.}~\bibnamefont {Qi}},\ and\ \bibinfo {author} {\bibfnamefont {Z.~Y.}\ \bibnamefont {Meng}},\ }\bibfield  {title} {\bibinfo {title} {Topological phase transition and single/multi anyon dynamics of z 2 spin liquid},\ }\href {https://doi.org/10.1038/s41535-021-00338-1} {\bibfield  {journal} {\bibinfo  {journal} {npj Quantum Materials}\ }\textbf {\bibinfo {volume} {6}},\ \bibinfo {pages} {39} (\bibinfo {year} {2021}{\natexlab{b}})}\BibitemShut {NoStop}%
\bibitem [{\citenamefont {Yan}(2022)}]{yan2020improved}%
  \BibitemOpen
  \bibfield  {author} {\bibinfo {author} {\bibfnamefont {Z.}~\bibnamefont {Yan}},\ }\bibfield  {title} {\bibinfo {title} {Global scheme of sweeping cluster algorithm to sample among topological sectors},\ }\href {https://doi.org/10.1103/PhysRevB.105.184432} {\bibfield  {journal} {\bibinfo  {journal} {Phys. Rev. B}\ }\textbf {\bibinfo {volume} {105}},\ \bibinfo {pages} {184432} (\bibinfo {year} {2022})}\BibitemShut {NoStop}%
\bibitem [{\citenamefont {Yan}\ \emph {et~al.}(2022{\natexlab{b}})\citenamefont {Yan}, \citenamefont {Meng}, \citenamefont {Huse},\ and\ \citenamefont {Chan}}]{ZYhqdm2022}%
  \BibitemOpen
  \bibfield  {author} {\bibinfo {author} {\bibfnamefont {Z.}~\bibnamefont {Yan}}, \bibinfo {author} {\bibfnamefont {Z.~Y.}\ \bibnamefont {Meng}}, \bibinfo {author} {\bibfnamefont {D.~A.}\ \bibnamefont {Huse}},\ and\ \bibinfo {author} {\bibfnamefont {A.}~\bibnamefont {Chan}},\ }\bibfield  {title} {\bibinfo {title} {Height-conserving quantum dimer models},\ }\href {https://doi.org/10.1103/PhysRevB.106.L041115} {\bibfield  {journal} {\bibinfo  {journal} {Phys. Rev. B}\ }\textbf {\bibinfo {volume} {106}},\ \bibinfo {pages} {L041115} (\bibinfo {year} {2022}{\natexlab{b}})}\BibitemShut {NoStop}%
\bibitem [{\citenamefont {Moessner}\ and\ \citenamefont {Raman}(2011{\natexlab{b}})}]{QDMbook}%
  \BibitemOpen
  \bibfield  {author} {\bibinfo {author} {\bibfnamefont {R.}~\bibnamefont {Moessner}}\ and\ \bibinfo {author} {\bibfnamefont {K.~S.}\ \bibnamefont {Raman}},\ }\bibfield  {title} {\bibinfo {title} {Quantum dimer models},\ }in\ \href@noop {} {\emph {\bibinfo {booktitle} {Introduction to Frustrated Magnetism}}}\ (\bibinfo  {publisher} {Springer},\ \bibinfo {year} {2011})\ pp.\ \bibinfo {pages} {437--479}\BibitemShut {NoStop}%
\bibitem [{\citenamefont {Zhang}\ \emph {et~al.}(2023)\citenamefont {Zhang}, \citenamefont {Allcock}, \citenamefont {Wan}, \citenamefont {Liu}, \citenamefont {Sun}, \citenamefont {Yu}, \citenamefont {Yang}, \citenamefont {Qiu}, \citenamefont {Ye}, \citenamefont {Chen}, \citenamefont {Lee}, \citenamefont {Zheng}, \citenamefont {Jian}, \citenamefont {Yao}, \citenamefont {Hsieh},\ and\ \citenamefont {Zhang}}]{Zhang2023tensorcircuit}%
  \BibitemOpen
  \bibfield  {author} {\bibinfo {author} {\bibfnamefont {S.-X.}\ \bibnamefont {Zhang}}, \bibinfo {author} {\bibfnamefont {J.}~\bibnamefont {Allcock}}, \bibinfo {author} {\bibfnamefont {Z.-Q.}\ \bibnamefont {Wan}}, \bibinfo {author} {\bibfnamefont {S.}~\bibnamefont {Liu}}, \bibinfo {author} {\bibfnamefont {J.}~\bibnamefont {Sun}}, \bibinfo {author} {\bibfnamefont {H.}~\bibnamefont {Yu}}, \bibinfo {author} {\bibfnamefont {X.-H.}\ \bibnamefont {Yang}}, \bibinfo {author} {\bibfnamefont {J.}~\bibnamefont {Qiu}}, \bibinfo {author} {\bibfnamefont {Z.}~\bibnamefont {Ye}}, \bibinfo {author} {\bibfnamefont {Y.-Q.}\ \bibnamefont {Chen}}, \bibinfo {author} {\bibfnamefont {C.-K.}\ \bibnamefont {Lee}}, \bibinfo {author} {\bibfnamefont {Y.-C.}\ \bibnamefont {Zheng}}, \bibinfo {author} {\bibfnamefont {S.-K.}\ \bibnamefont {Jian}}, \bibinfo {author} {\bibfnamefont {H.}~\bibnamefont {Yao}}, \bibinfo {author} {\bibfnamefont {C.-Y.}\ \bibnamefont {Hsieh}},\ and\ \bibinfo {author} {\bibfnamefont {S.}~\bibnamefont {Zhang}},\
  }\bibfield  {title} {\bibinfo {title} {Tensor{C}ircuit: a {Q}uantum {S}oftware {F}ramework for the {NISQ} {E}ra},\ }\href {https://doi.org/10.22331/q-2023-02-02-912} {\bibfield  {journal} {\bibinfo  {journal} {{Quantum}}\ }\textbf {\bibinfo {volume} {7}},\ \bibinfo {pages} {912} (\bibinfo {year} {2023})}\BibitemShut {NoStop}%
\bibitem [{Note1()}]{Note1}%
  \BibitemOpen
  \bibinfo {note} {\protect \url {https://github.com/msquare1998/SourceCodes-arxiv2310.04291}}\BibitemShut {NoStop}%
\bibitem [{\citenamefont {Kibble}(1976)}]{T_W_B_Kibble_1976}%
  \BibitemOpen
  \bibfield  {author} {\bibinfo {author} {\bibfnamefont {T.~W.~B.}\ \bibnamefont {Kibble}},\ }\bibfield  {title} {\bibinfo {title} {Topology of cosmic domains and strings},\ }\href {https://doi.org/10.1088/0305-4470/9/8/029} {\bibfield  {journal} {\bibinfo  {journal} {Journal of Physics A: Mathematical and General}\ }\textbf {\bibinfo {volume} {9}},\ \bibinfo {pages} {1387} (\bibinfo {year} {1976})}\BibitemShut {NoStop}%
\bibitem [{\citenamefont {Zurek}\ \emph {et~al.}(2005)\citenamefont {Zurek}, \citenamefont {Dorner},\ and\ \citenamefont {Zoller}}]{PhysRevLett.95.105701}%
  \BibitemOpen
  \bibfield  {author} {\bibinfo {author} {\bibfnamefont {W.~H.}\ \bibnamefont {Zurek}}, \bibinfo {author} {\bibfnamefont {U.}~\bibnamefont {Dorner}},\ and\ \bibinfo {author} {\bibfnamefont {P.}~\bibnamefont {Zoller}},\ }\bibfield  {title} {\bibinfo {title} {Dynamics of a quantum phase transition},\ }\href {https://doi.org/10.1103/PhysRevLett.95.105701} {\bibfield  {journal} {\bibinfo  {journal} {Phys. Rev. Lett.}\ }\textbf {\bibinfo {volume} {95}},\ \bibinfo {pages} {105701} (\bibinfo {year} {2005})}\BibitemShut {NoStop}%
\bibitem [{\citenamefont {Wen}(2017)}]{Wen2017topoZoo}%
  \BibitemOpen
  \bibfield  {author} {\bibinfo {author} {\bibfnamefont {X.-G.}\ \bibnamefont {Wen}},\ }\bibfield  {title} {\bibinfo {title} {Colloquium: Zoo of quantum-topological phases of matter},\ }\href {https://doi.org/10.1103/RevModPhys.89.041004} {\bibfield  {journal} {\bibinfo  {journal} {Rev. Mod. Phys.}\ }\textbf {\bibinfo {volume} {89}},\ \bibinfo {pages} {041004} (\bibinfo {year} {2017})}\BibitemShut {NoStop}%
\bibitem [{\citenamefont {Wen}(2013)}]{Wen2013topoorder}%
  \BibitemOpen
  \bibfield  {author} {\bibinfo {author} {\bibfnamefont {X.-G.}\ \bibnamefont {Wen}},\ }\bibfield  {title} {\bibinfo {title} {Topological order: From long-range entangled quantum matter to a unified origin of light and electrons},\ }\href {https://doi.org/10.1155/2013/198710} {\bibfield  {journal} {\bibinfo  {journal} {ISRN Condensed Matter Physics}\ }\textbf {\bibinfo {volume} {2013}},\ \bibinfo {pages} {198710} (\bibinfo {year} {2013})}\BibitemShut {NoStop}%
\end{thebibliography}%

\end{document}